\documentclass[12pt]{JHEP3}
\usepackage{graphicx,amsmath,amssymb}

\newcommand{\be}{\begin{equation}}
\newcommand{\ee}{\end{equation}}
\newcommand{\ba}{\begin{eqnarray}}
\newcommand{\ea}{\end{eqnarray}}
\newcommand{\barr}{\begin{array}}
\newcommand{\earr}{\end{array}}

\newcommand{\simless}[0]{\mathbin{\lower 3pt\hbox
   {$\rlap{\raise 5pt\hbox{$\char'074$}}\mathchar"7218$}}}
\newcommand{\simgreat}[0]{\mathbin{\lower 3pt\hbox
   {$\rlap{\raise 5pt\hbox{$\char'076$}}\mathchar"7218$}}}
\newcommand{\gta}[0]{\simgreat}
\newcommand{\lta}[0]{\simless}
\newcommand\msun{{\rm M}_\odot}

\title{Clustering of Superstring Loops}
\author{David F. Chernoff\\ Cornell University\\ E-mail: \email{chernoff@astro.cornell.edu}}

\abstract{ Superstring loops formed by intercommutation of low tension
  horizon-crossing superstrings may be captured and accreted by growing
  matter perturbations.

The paper explores the influence of string tension $\mu$ and of the
network formation mechanism on the clustering process.
Galaxy formation and growth is schematically described by a radial
infall model. A fully relativistic treatment of motion in perturbed
Friedmann-Robertson-Walker cosmology is developed and applied to track
the loop center of mass motion.  The local enhancement of loops
(galaxy's loop density divided by universe's loop density) is
calculated and compared to that of cold dark matter for different
$\mu$ in the context of two network formation scenarios, one yielding
large loops (fragmentation near the horizon scale) and the other small
loops (cusp-mediated formation).

The primary physical process that enables capture of a loop by a
growing perturbation is the decrease in peculiar velocity slaved to
the universe's expansion.  The main process that removes the loop from
the galactic potential well is the ``rocket effect'', the recoil of
anisotropic gravitational wave emission.  Quantitative criteria for
capture and detachment are given. There is a critical value of
$\mu$ below which clustering generically occurs
in our own Galaxy.

Fragmentation scenarios for {\it large-scale} loops lead to
significant halo enhancements. With typical model parameters (velocity
dispersion of newly formed loops, loop length distribution, etc.) the
limiting enhancement of loop (energy) density is $\sim 0.25-0.4$ that of
cold dark matter which is fully achieved for $G \mu/c^2 \lta 10^{-13}$ at all
scales $\lta 10^2$ kpc. The fact that the string loop enhancement
roughly tracks that of cold dark matter is a robust result for
small $\mu$ and large-scale loops.

In the radial infall model for the Galaxy the magnitude of the cold
dark matter enhancement is $\sim 10^{5.6}$ at $\sim 10$ kpc.  This
quantitative degree of enhancement, although somewhat model dependent,
implies local loops with $G \mu/c^2 \lta 10^{-13}$ are enhanced
$10^{5}-10^{5.2}$ with respect to the homogeneous universe.  The
degree of enhancement at the same position is also substantial for all
$\mu$ less than the critical value for clustering; it exceeds $10^{3.6}$ for $G
\mu/c^2<10^{-10}$.

The enhancement of the energy density of loops as a function of
galactocentric distance for a range of possible $\mu$ is summarized in
figure \ref{figure-final-avdenL-15-10-0}.  These loops are long-lived
but ultimately transient residents of the galaxy.  Experiments
sensitive to the local Galactic population of loops, especially
microlensing, will enjoy increased detection rates compared to
homogeneous estimates.

{\it Small-scale} loops produced via cusps do not cluster strongly
because they do not live long enough for the universe's
expansion to damp their initial relativistic motions.
The galactic enhancement of loop (energy) density is
$\sim 10^{-5}-10^{-4}$ that of cold dark matter depending upon the
string tension ($G \mu/c^2<10^{-10}$) and some uncertainties in the
velocity dispersion of the newly formed ultra-relativistic loops.  At
$10$ kpc the net enhancement of bound loops is $\lta
1-10$. Experiments searching for evidence of small-scale loops need to be
sensitive to the homogeneous distribution throughout the universe;
the local enhancement is small.

}


\preprint{hep-th/yymmnnn}

\begin{document}

\section{Introduction}

Three recent developments motivate an examination of the clustering of
superstring loops on galactic scales.  First, braneworld cosmological
scenarios that provide a framework for inflation and the big bang,
necessary ingredients in modern cosmology, generically produce
string-like defects.  Second, string theory allows and recent
investigations of warped throat compactification suggest that
superstring tensions can be much smaller than the GUT scale: $G
\mu/c^2 << 10^{-6}$. Third, recent high resolution simulations of
string networks suggest that $\sim 10$\% of the energy available in
horizon-crossing strings is transformed into loops within a few orders
of magnitude of the scale of the horizon.

\def\gutclassic{\cite{kibble_topology_1976,zeldovich_cosmological_1980,vilenkin_cosmological_1981,vilenkin_cosmic_1981}}

In the original GUT scenarios, a phase transition at the GUT energy
scale created string-like defects with tension $G \mu/c^2 \sim
10^{-6}$ whose dynamical motions generated density perturbations
ultimately responsible for large scale structure formation\gutclassic.
Intercommutation chopped long, horizon-crossing strings into loops
moving at relativistic speeds. These loops, distributed in the
universe in an approximately homogeneous fashion, evaporated by
gravitational wave emission within a few Hubble times.

\def\stringsclusteringalaxy{\cite{chernoff_cosmic_2007,depies_harmonic_2009}}

By contrast,
the three advances lead to a qualitatively new and different picture for
the fate of the loops. Large loop size and small string tension
implies that loops survive for many expansion times. As such, they
slow down by cosmic drag and fall into existing matter potential
wells\stringsclusteringalaxy.

This paper shows that low tension string loops cluster and form a halo about the
Galaxy. The enhancement relative to the homogeneous loop distribution
is substantial both in terms of loop numbers and loop energy (total
length).

Superstring loops are roughly analogous to stellar objects in two
respects: they have finite lifetimes as local luminous sources of
gravitons much like nuclear burning stars emit photons over a fixed
main sequence lifetime, and they are massive, compact and optically
dark much like the remnants formed in post main sequence evolution.
It is interesting to consider how loops, like main sequence and post
main sequence stars, may reveal themselves either directly by their
intrinsic emissions or indirectly by altering photon propagation of
background sources. Designing and planning searches for superstring
loops will require a good understanding of the distribution of loops
throughout the universe and especially our own backyard.

\subsection{String Tension}

\def\inflation{\cite{albrecht_cosmology_1982,guth_inflationary_1981,linde_new_1982}}

\def\branes{\cite{buchan_inter-brane_2004,burgess_inflationary_2001,dvali_brane_1999,dvali_d-brane_2001,giddings_hierarchiesfluxes_2002,horava_type_1998,kachru_de_2003,
    kachru_towards_2003,sen_stable_1998,sen_so32_1998,tye_brane_2008,witten_d-branes_1998}}

\def\cosmicstrings{\cite{copeland_cosmic_2004,dvali_formation_2004,
firouzjahi_brane_2005,jackson_collisions_2005,jones_brane_2002,
jones_production_2003,sarangi_cosmic_2002,sen_rolling_2002,sen_tachyon_2002,
shandera_observing_2006}}

\def\cosmicstringnetwork{\cite{albrecht_evolution_1985,allen_cosmic-string_1990,
avgoustidis_effect_2006,bennett_evidence_1988,dubath_cosmic_2008,martins_fractal_2006,polchinski_analytic_2006,
polchinski_cosmic_2007,ringeval_cosmological_2007,sakellariadou_noteevolution_2005,tye_scaling_2005,vanchurin_cosmic_2007,vanchurin_scaling_2006,vilenkin_cosmic_2000}}

Inflation is an essential ingredient in modern cosmology\inflation. In
superstring theory a specific realization is brane inflation and the
simplest example of brane inflation involves the interaction of a
D3-brane moving toward a ${\bar D}3$-brane sitting at the bottom of a
warped throat\branes.  The collision and annihilation of the brane pair
initiates the hot big bang. Cosmic superstrings (F- and D-strings and
their bound states) are produced and stretched to enormous
scales\cosmicstrings. After the epoch of inflation these superstrings evolve by the
processes of intercommutation and gravitational wave emission to yield
a scaling network in which there exists a stable relative distribution
of long, horizon-crossing strings and sub-horizon loops\cosmicstringnetwork.

\def\tensionrange{\cite{firouzjahi_brane_2005,jones_brane_2002,jones_production_2003,
shandera_observing_2006}}

The key property of a cosmic string is the tension $\mu$, or in
dimensionless terms, the string's characteristic gravitational
potential $G \mu/c^2$. Theoretical understanding of the characteristic
tension likely to emerge in a physically realistic string theory
solution is far from complete.  Initial estimates suggested $10^{-11}
\lta G \mu/c^2 \lta 10^{-6}$ \cite{sarangi_cosmic_2002} but recent
analyses of multi-brane, multi-throat scenarios have effectively
removed the lower bound\tensionrange.

\def\lensinglimits{\cite{vilenkin_gravitational_1983,hogan_gravitational_1984,vilenkin_cosmic_1984,de_laix_observing_1997,bernardeau_cosmic_2001,sazhin_csl-1:_2003,sazhin_true_2006,christiansen_search_2008}}

\def\gravitationalwavebackandburst{\cite{hogan_gravitational_1984,vachaspati_gravitational_1985,economou_gravitational_1992,battye_gravitational_1998,damour_gravitational_2000,damour_gravitational_2001,damour_gravitational_2005,siemens_gravitational_2006,hogan_gravitational_2006,siemens_gravitational-wave_2007,abbott_searches_2007,ligo_scientific_collaboration:_b._abbott_first_2009}}

\def\gravitationalwavepulsars{\cite{bouchet_millisecond-pulsar_1990,hogan_gravitational_1984,caldwell_cosmological_1992,depies_stochastic_2007}}

\def\tensionlimits{\cite{smoot_structure_1992,bennett_four-year_1996,pogosian_observational_2003,pogosian_observational_2004,wyman_boundscosmic_2005,pogosian_vector_2006,seljak_cosmological_2006,spergel_three-year_2007,bevis_cmb_2007,fraisse_limitsdefects_2007,bevis_fitting_2008,pogosian_cosmic_2009}}

\def\tensionlimitcmbshape{\cite{wyman_boundscosmic_2005,fraiss_limitsdefects_2007,battye_constraints_2008,bevis_fitting_2008,battye_constaints_2008,urrestilla_cosmic_2008}}
\def\tensionlimitcmbpol{\cite{bevis_cmb_2007,pogosian_b-modes_2008}}
\def\tensionlimitcmbsmall{\cite{fraisse_small-angle_2008,pogosian_cosmic_2009}}

Empirical upper bounds on $G \mu/c^2$ have been derived from null
results for experiments involving lensing\lensinglimits, gravitational
wave background and bursts\gravitationalwavebackandburst, pulsar
timing\gravitationalwavepulsars and cosmic microwave background
radiation\tensionlimits. These may be generally summarized as follows:
(1) Searches for signatures of optical lensing in fields of background
galaxies imply $G \mu/c^2 \lta 3 \times 10^{-7}$. The analysis relies
on the deficit angle geometry of a string in spacetime and the
accurate estimation of survey selection effects. (2) Modeling of the
CMB power spectrum yields $G \mu/c^2 \lta 10^{-7}$. The limit is based
on well-understood properties of large-scale string networks although
the precise quantitative results are sensitive to unknown details of
the spectra of string bound states and the probability of
string-string interactions. (3) Pulsar timing stability gives $G
\mu/c^2 \lta 10^{-9}$. The limit assumes that loops of near-horizon
scale are created by the string network.

In short, cosmic superstrings must have tensions substantially less than the
original GUT-inspired strings and there is no known theoretical
impediment to the magnitude of $G \mu/c^2$ being either comparable to or
much lower than the current observational upper limits.

The lifetime $\tau$ of a loop of size $l$ to emission of gravitational
radiation is, on dimensional grounds, $\tau \sim l c /(G
\mu)$. Smaller tension yields larger $\tau$.  In a cosmology with
power law growth in the scale factor, the scaling solution chops long
strings into loops of size $l \sim
\alpha c t$, i.e. proportional to the size of the horizon. Here, the dimensionless constant $\alpha$ typifies a
characteristic loop size from a broad, possibly multi-peaked,
distribution of string loop scales. A key parameter is the number of
expansion times before the loop evaporates $H \tau \sim \alpha/(G
\mu/c^2)$ where $H$ is the Hubble constant. When $H \tau$ is not big
the loops evaporate before the universe has expanded
significantly. Such is the case for the usual GUT-inspired structure formation
scenarios.

The notions that the tension might be {\it very} small, $G \mu/c^2 <<
10^{-7}$, and that the loop size distribution include some objects
comparable to the scale of the horizon ($10^{-4} < \alpha < 10^{-1}$)
yields qualitatively new cosmological features. If $H \tau >> 1$ the
resultant string network will contain many old loops. Loops born
with relativistic speeds are significantly slowed by cosmic drag
before evaporating. A necessary condition for
clustering on a galactic scale is that the loops damp
to speeds less than the typical speeds within the galaxy: $v/c <
(v/c)_{gal} \sim 10^{-3}$. This is only possible for large $H \tau$.

\def\homogeneousloopslowing{\cite{vachaspati_gravitational_1985,hogan_runaway_1987,durrer_gravitational_1989,kuijken_microlensing_2008,battefeld_magnetogenesiscosmic_2008}}

This paper presents a detailed calculation of the infall of loops in a
{\it schematic} model for growth and formation of a galactic scale cold dark
matter perturbation in the presence of a scaling string network.
\footnote{There have been numerous investigations of loop dynamics in
  {\it homogeneous} background\homogeneousloopslowing. This work
  differs by fully accounting for the presence of growing
  gravitational perturbations which are ultimately responsible for the
  clustering.}  The strings satisfy the Nambu Goto equations of motion
(hereafter NG strings).  The loops accumulate and form a large halo
similar in many respects to the Galaxy's gravitationally dominant dark
matter halo but with several unique features: old captured loops decay
by emission of gravitational radiation and are eventually ejected by
the recoil associated with the anisotropy of their gravitational wave
emission (``rocket effect'') while new loops are added continually
near the Galaxy's turn-around radius. At each epoch the galaxy is
dressed with a long-lived halo of string loops.

\S 2 outlines and summarizes the various calculations and the results. Modeling
details follow in subsequent sections:
\S 3 describes the halo formation model, 
\S 4 sets up the equations of motion for string loops in inhomogeneous 
Friedmann-Robertson-Walker (FRW) cosmology, 
\S 5 characterizes the free motion of loops, accelerated motion and derives an
approximation for the critical acceleration that unbinds a loop,
\S 6 describes the model for the string network,
\S 7 characterizes the clustering for string loops generated by the
network at a given time, with given size, and possessing given tension,
and 
\S 8 calculates the loops currently distributed within the Galaxy
in two different models for loop formation.

Finally, \S 9 discusses some of the implications and outlines future work.

\section{Executive Summary}

\def\selfsimilar{\cite{gott_formation_1975,gunn_massive_1977,fillmore_self-similar_1984,bertschinger_self-similar_1985}}

The background cosmological model is Einstein-de Sitter with a critical
density of non-relativistic matter and scale factor $a(t) \propto
t^{2/3}$. The growing gravitational perturbation
is much smaller in size than the horizon and treated non-relativistically. 
Consider the spherically symmetric infall of cold,
collisionless matter caused by introducing a small overdense top hat
perturbation.  The turn-around radius 
is the point where the Hubble expansion exactly
balances the infall velocity and sets the scale
for the problem. Material well within the turn-around
radius has had sufficient time to collapse, re-expand, recollapse, and
so forth. These motions (``bounces'') rearrange the mass, kinetic
energy and potential energy so as to produce a virialized structure
on scales somewhat less than the turn-around radius.
Material outside the turn-around radius has not yet had time to
bounce. If much more material has bounced than was
initially present in the top hat perturbation then the solution is
self-similar, i.e. all the physical properties can be scaled from one
time to another. The essential character of the self-similar infall model
(i.e. the fully self-consistent density, velocity and gravitational potential
of the collisionless cold dark matter which accretes and forms the
non-linear bound object)
has been spelled out
\selfsimilar. In this paper, the Galaxy's
growing gravitational potential is described in terms of the
gravitational potential $\psi$ of the self-similar infall model.

The background metric for flat FRW is $ds^2 = -c^2 dt^2 + a(t)^2 (dx^2
+ dy^2 + dz^2)$. The perturbed metric is assumed to include only the
scalar terms of the growing, non-relativistic gravitational potential
$ds^2 = -(1 + 2 \psi) c^2 dt^2 + a(t)^2 (1 + 2 \Phi) (dx^2 + dy^2 +
dz^2)$. Anisotropies in the stress tensor are ignored; this implies
$\Phi=-\psi$. There is no back-reaction of the strings on the
spacetime.  In the absence of gravitational wave emission, each string loop
center of mass follows a geodesic in the spacetime;
tidal effects on the loops are ignored.

However, the loop does emit radiation and its impulse alters the loop
trajectories.  Intrinsic variations in the direction of momentum
radiated within the loop center of mass frame would tend to average
out the net impulse given to the loop.  An assumption that no torques
act constitutes the ``worst case'' for binding of loops to the Galaxy
because the direction of the rocket force is influenced only by
special and general relativistic effects which are always active {\it
  not} by additional, intrinsic variations in the loop center of mass
frame.  In this paper no internal or external torques act so that the
direction of the rocket impulse in the center of mass frame behaves
like the spin of a particle. The impulse direction is
Fermi transported along the spacetime trajectory of the loop.

The rates of gravitational wave emission of energy and momentum by
oscillating loops have been previously calculated for a sparse sample
of loop configurations \cite{vachaspati_gravitational_1985,
  burden_gravitational_1985,garfinkle_radiationkinky_1987,allen_analytic_1994,allen_closed-form_1994,allen_closed-form_1995}.
In the center of mass frame the approximate, period-averaged
consequences are (1) a constant rate of change of length (or energy),
$dl/dt = \Gamma_E (G \mu/c^2) c$, and (2) a constant degree of
anisotropy which induces an impulse $a_r = \Gamma_P (G
\mu/c^2)(c^2/l)$. Based on the numerical studies, typically
$\Gamma_E \sim 50$ and $\Gamma_P \sim 10$ but these may
vary by a factor $\sim 2$ for individual loop configurations.
\footnote{This estimate for $\Gamma_P$ may be systematically too big
  \cite{polchinski-private2009}.  Large loops have had their cusps
  excised and the anisotropy of radiation emitted by the kinks that
  remain is smaller than would be generated by cusps. Hence, the force
  of recoil on large loops may be smaller than inferred from the
  calculated examples. The value used is conservative in giving the
  ``worst case'' scenario for binding of loops. }

In summary, a loop undergoes non-geodesic motion in spacetime because
of the rocket effect with Fermi transport of the impulse direction.  In
the center of mass frame, the loop shrinks according to a simple,
approximate description and suffers a recoil based on a fixed degree
of anisotropy which determines the magnitude of the non-gravitational
acceleration (hereafter ``the impulse'').

When loops are chopped off from the string network they are typically
moving at relativistic velocities. Initial conditions are drawn from a
homogeneous distribution in space, with a relativistic center-of-mass
velocity and a range of sizes.  A choice of string tension $\mu$ and
initial conditions for the loop (position, velocity, rocket direction
and size at the time of formation $t_i$) yields a trajectory in the
background spacetime.

The numerical results presented in this paper show how loops
bind to the growing galactic perturbation. The most common scenario is
that a loop born at an early time, slows down by cosmic drag, is
overtaken by the turn-around radius, and accretes. The rocket effect
is initially negligible.  The resultant orbit is very radial
passing back and forth through the galactic center with roughly fixed physical
semi-major axis. Such a loop is easily identified as ``bound to the
perturbation.''  The physical scale of the captured orbit is fixed
even as the perturbation continues to grow in size. Loops captured at
early times end up near the center of the structure, ones added later
at the periphery.

As the loop shrinks, eventually, the acceleration of the rocket
unbinds the loop from the perturbation. Detachment is rather sudden
because the periodic motion in the potential averages the effect of
the impulse, i.e. the orbit is adiabatically invariant. Escape follows
when the rocket's force is large enough to alter the orbital
parameters within a single orbit.

The general trends can be understood by reference to the behavior of
cold dark matter.  Cold dark matter falls into the growing
perturbation and creates a well-defined, universal profile with scale
set by the turn-around radius. The cumulative probability distribution
for a cold dark matter particle to be bound to the perturbation
is a function of a single dimensionless parameter, the ratio of radius
to turn-around radius. The analogous distribution for bound loops is
more complicated since it depends on the epoch of formation, the size
of the loop, and the string tension.

Loops -- formed at early epochs but with small
enough $\mu$ that they have not yet neared the end of their lives --
behave just like cold dark matter. They assume the same universal
profile near the center while at large galactic radii the profile is
truncated due to the rocket effect (``outer cutoff''). 

Dynamical complications ensue for loops formed at much earlier or much
later times. For
old loops that have begun to shrink substantially the importance of
the rocket effect increases; consequently, the outer cutoff in the
Galactic profile of loops becomes more important, i.e. it moves
inward. Every loop is unbound before it completely evaporates because
in that limit the acceleration from the rocket grows large. New loops, on
the other hand, which are typically born with relativistic speeds have
not yet damped sufficiently to allow capture by the potential. A set
of graphs illustrate the constraints on formation size, formation time
and string tension.

An integration over the rate of creation of loops implied by recent
network simulations of Nambu Goto strings yields an estimate of
today's loop profile about the Galactic center. The main part of the
bound population was created by fragmentation of horizon-scale strings
into {\it large} sub-horizon loops. The loops are abundant and overdense
with respect to the universe's average number and energy density of
strings of all sorts. 

The bottom line results for large loops are presented in figures
\ref{figure-final-avdenN-15-10-0} (number density of loops
within the galaxy relative to the homogeneous value) and
\ref{figure-final-avdenL-15-10-0} (energy density of loops relative
to homogeneous value). These figures show a substantial degree of 
enhancement in both measures that depends upon string tension and
spans a large interval of galactocentric radii.  Such
profiles motivate more accurate calculations for
gravitational wave, pulsar timing
and microlensing experiments hunting for evidence of loops within the
Galaxy.  

The radial distribution of loops within the Galaxy is weighted to the
center.  The size distribution of loops is weighted to small scales
with a cutoff corresponding to a loop with evaporation lifetime equal
to the age of the universe.

By contrast, cusp-generated {\it small} loops fail to bind to the
Galaxy. This is not surprising given their ultra-relativistic initial
motions and their reduced lifetimes.  The bottom line results for
small loops are presented in figures
\ref{figure-finalPR-avdenN-15-wwofnof} (number density) and
\ref{figure-finalPR-avdenL-15-wwofnof} (energy density). Little
enhancement is observed within the range of radii at which the radial
infall model is applicable.

These results are broadly suggestive that clustering of large
loops will play an important role in setting microlensing rates
and may also increase the effective sensitivity of gravitational wave
and pulsar timing experiments. On the other hand, experiments sensitive
to small loops will not benefit from significant local enhancement.

\section{Halo Formation Model}
\subsection{Aim}

The role of the halo formation model in this paper is to provide a
dynamical background for the motion and eventual capture of the string
loops generated by the network. The density and potential of the cold
dark matter are determined in a self-consistent fashion and the string
loops moved as test particles in the potential. The distribution of
cold dark matter particles and string loops can then be compared.

\subsection{Self-similar Radial Infall}

Refs. \cite{gott_formation_1975}, \cite{gunn_massive_1977},
\cite{fillmore_self-similar_1984}, and
\cite{bertschinger_self-similar_1985} have analyzed the
spherically symmetric infall of cold, collisionless matter onto
small-scale density enhancements in an Einstein-de Sitter
universe. The solution is self-similar i.e. the form and appearance at
any time is fixed when scaled to a characteristic physical length. The
infall yields power law halo density distribution $\rho \propto
r^{-2.25}$ and rotation curves $\propto r^{-1/8}$.  The
model is physically self-consistent and
simple enough that many difficult aspects of the cosmology plus
network evolution can be handled precisely.

The model solution is exact given the assumptions but the model must
be regarded as a {\it schematic} description of the Galaxy.  Its basic
shortcomings in comparison to a realistic treatment of $\Lambda$CDM
cosmology are the following: (1) it ignores the initial spectrum of
perturbations which span a great range of length scales and it
suppresses the generic asymmetry of a typical perturbation that grows
to encompass a galaxy scale mass, (2) it is only applicable at times
after equipartition and before late-time acceleration.

\def\classiccdm{\cite{peebles_large-scale_1982,blumenthal_formation_1984,davis_evolution_1985,mo_abundance_2002}}

Of these, the more significant issue is the first.  Initial conditions
drawn from a CDM spectrum generate a hierarchy of mergers not a
monolithic infall\classiccdm. Today objects like our Galaxy have dark matter
density profiles that are non-power law \cite{navarro_structure_1996}. The
density $\rho$ varies like $1/r$ at small radii (successive mergers of
small dense objects) and $1/r^3$ at large radii (truncation of
infall). The radial infall model does not capture the behavior at
either extreme: it should be adequate on scales on which the rotation
curve is observed to be flat, $3 < r < 30 $ kpc, and it may be
adequate out to distances where the curve is traditionally assumed or
inferred to be flat, e.g. $\sim 100$ kpc\cite{fich_mass_1991}. The motivation for its use here is that a comparison of the
loop and cold dark matter distributions calculated in the same,
self-consistent time-dependent potential, should yield valid
conclusions of greater generality than might be suggested by a
strict comparison of actual to modeled Galactic profiles.

Of course, $a \propto t^{1/2}$ prior to the time of equipartition
$t_{eq}$ and the perturbation begins to grow only for $t>t_{eq}$. It
is incorrect to use the self-similar form for the potential at early
times. Nonetheless, this paper employs it focusing on clustering after
equipartition.  Since the dynamics of loops formed before
equipartition are an essential part of the story to be set forth one
might worry that this presents an additional important shortcoming.
In the current model, the turn-around radius at $t_{eq}$ is $\sim 15$
pc; the radial profile of cold dark matter and of loops at such small
scales would certainly be inaccurately represented even if the
perturbation were exactly spherical. (Which it is not -- as already
indicated the smallest radius at which the radial infall model applies
is much larger.) On larger scales the distribution of cold dark matter
and loops is not adversely impacted since the potential is
basically flat while cosmic drag and the rocket effect both operate
independently of the potential. A loop slows down first and then binds
to the growing perturbation when the turn-around radius reaches
it. This typically occurs at $t>t_{eq}$.  At times $t>t_{eq}$, the
cold dark matter density and gravitational potential quickly asymptote
to the self-similar form. Computed properties today are
uninfluenced.\footnote{An independent issue relates to the string
  network evolution and the distinction between scaling solutions
  before and after equipartition. This is discussed later.}

Finally, the late-time acceleration of the universe alters the
behavior of the turn-around radius. This certainly changes how loops
might be added to the outermost periphery of the Galaxy.  Since M31
will turn out to be closer than turn-around radius formally inferred
today, it is clear that quantitative agreement
at such large distances is not expected.

Future work will treat structure growth in realistic simulations of
the $\Lambda$CDM paradigm.

\subsection{Model Specifics}
\newcommand{\baroi}{{\bar \Omega}_i}

This exposition follows the mathematical description given by
Ref. \cite{bertschinger_self-similar_1985}.  At time $t_i$ consider a
small top hat density enhancement $\delta_i \equiv (\delta \rho/
\rho_H)_i$ which extends from the origin to physical radius $R_i$;
assume the Hubble constant $H_i$ is independent of radius. All the
material in the universe is bound by the presence of the excess
material and is destined to fall towards it.

The motion of shells depends upon the initial position $r_i$, the
initial velocity $r_i H_i$ and the mass interior to the shell.
Write the initial {\it mean} density within a radius $r_i$
in terms of ${\bar \Omega}_i = 1 + \Delta_i$ where $\Delta_i =
\delta_i \min \left(1 , \left( \frac{R_i}{r_i} \right)^3 \right)$.
Before a shell crosses any other shell, it satisfies the parametric
equations
\ba
r & = & \frac{ r_i \baroi }{2 \left( \baroi - 1 \right) }
\left( 1 - \cos \theta \right) \\
t & = & \frac{ \baroi }{2 H_i\left( \baroi - 1 \right)^{3/2} }
\left( \theta - \sin \theta \right) .
\ea
The initial state at $t=0$ has parameter $\theta=0$.
To lowest order in $\Delta_i << 1$ it turns around ($\theta=\pi$) at
\ba
r_{ta} & = & \frac{r_i}{\Delta_i} \\
t_{ta} & = & \frac{\pi}{2 H_i \Delta_i^{3/2}} .
\ea
Eliminating the occurrences of $r_i$ in favor of $\delta_i^{1/3} R_i$
yields the turn-around radius at any given time $t$ as
\be
r_{ta}  =  \delta_i^{1/3} R_i \left( \frac{2 H_i t}{\pi} \right)^{8/9} .
\ee
The fact that $r_{ta}$ involves the product of 
a single combination of variables
and the characteristic power law $t^{8/9}$ 
is a significant simplification.
The turn-around radius is a suitable length scale with which to
non-dimensionalize the problem. The dimensionless length scale
$\lambda$ is a pure function of the parameter $\theta$:
\ba
\lambda & = & \frac{ r(t) }{r_{ta}(t) } 
         =  \frac{ \sin^2 \left( \frac{\theta}{2} \right) }
{\left( \frac{\theta - \sin \theta}{\pi} \right)^{8/9}} .
\ea
A shell 
starting at a large distance from the top hat has large $\lambda$.
As the turn-around radius increases and the shell falls under the
influence of the perturbation $\lambda$ decreases.

The initial motion is orderly. Before shell-crossing occurs the
dimensionless velocity ${\cal V}$, dimensionless mass ${\cal M}$ and
dimensionless density ${\hat \rho}$ are explicit functions of
$\theta$:
\ba
{\cal V} & = & \frac{\dot r}{ \left( \frac{a_{ta}(t)}{t} \right) } 
           = \frac{ \pi^{8/9} }{2}
        \frac{\sin \theta}{1 - \cos \theta}
        \left( \theta - \sin \theta \right)^{1/9} \\
{\cal M} & = & \frac{ 3 M }{4 \pi \rho_H(t) r_{ta}(t)^3} 
           =   \frac{ 9 \lambda^3}{16}
               \frac{ \left( \theta - \sin \theta \right)^2 }{ \sin^6 \left(
\frac{\theta}{2} \right) } \\
{\hat \rho} & = & \frac{ \rho }{ \rho_H(t) } 
              =   \frac{1}{1 + 3 \chi} 
   \frac{ \left( \frac{3}{4} \left( \theta - \sin \theta \right) \right)^{2} }
        {\sin^6 \left( \frac{\theta}{2} \right) } \\
\chi & = & 1 - \frac{3 V}{2 \lambda} 
\ea
and $\rho_H(t)$ is the density of the background model.

The initial state corresponds to $\theta \to 0$; the asymptotic forms in
this limit are
\ba
\lambda & \to & \frac{ \left( 6 \pi \right)^{8/9} }{ 4 \theta^{2/3} } \\
{\cal V} & \to & \frac{ 2 \lambda}{3} \\
{\cal M} & \to & \lambda^3 \\
{\cal \rho} & \to & 1 .
\ea
The {\it excess} mass associated with the perturbation
\ba
{\cal M}_x & = & \lambda^3 \left(
\frac{ 9}{16}
               \frac{ \left( \theta - \sin \theta \right)^2 }{ \sin^6 \left(
\frac{\theta}{2} \right) } - 1 \right)
\ea
is used to infer the potential that will make its appearance in the
metric in the next section.

Eventually, the infalling shell meets shells that have already
bounced.  The first crossing for a shell occurs at $\lambda_1 =
0.364$. The dimensionless expressions above are exact for $\lambda >
\lambda_1$. Once crossings begin for the shell of interest, 
the mass interior to it varies and must be calculated to find the
acceleration and trajectory. Ref. \cite{bertschinger_self-similar_1985} has solved
the problem numerically and tabulated ${\cal M} (\lambda)$ for $0.02 <
\lambda < \lambda_1$ and also provided a power law approximation for
$\lambda << 1$. This paper uses a combination of asymptotic, tabulated
and analytic expressions to describe ${\cal M}(\lambda)$ and
${\cal M}_x(\lambda)$ over the
complete range of $\lambda$.\footnote{The forms for $\lambda<<1$
  follow the scaling given in
  Ref. \cite{bertschinger_self-similar_1985} with coefficients
  adjusted to fit the last tabulated values; the numerical
  coefficients differs at the 5\% level from limiting analytic expressions
  given elsewhere in the same paper.} The dimensionless cumulative mass is shown in
figure \ref{figure-bert-mass}.

The dimensionless radial component for the force and potential
are deduced from ${\cal M}_x$:
\ba
{\cal F} & = & \frac{f}{\frac{4 \pi}{3} G \rho(t) r_{ta}(t)} 
           =   -\frac{M_x}{\lambda^2} \\
{\hat \psi} & = & \frac{\psi}{\frac{4 \pi}{3} G \rho(t) r_{ta}^2} 
              =  -\int_{r/r_{ta}}^\infty \frac{M_x}{\lambda^2} d\lambda .
\ea
The fact that the force and potential, which are intrinsically
functions of four spacetime variables, are compactly represented in
terms of the one-dimensional functions of the dimensionless radius
$\lambda=r/r_{ta}(t)$ is a great simplification. Test particle
motion in the vicinity of the perturbation depends upon the
these functions. The dimensionless potential is shown
in figure \ref{figure-bert-pot}.

\FIGURE[h]{
\includegraphics{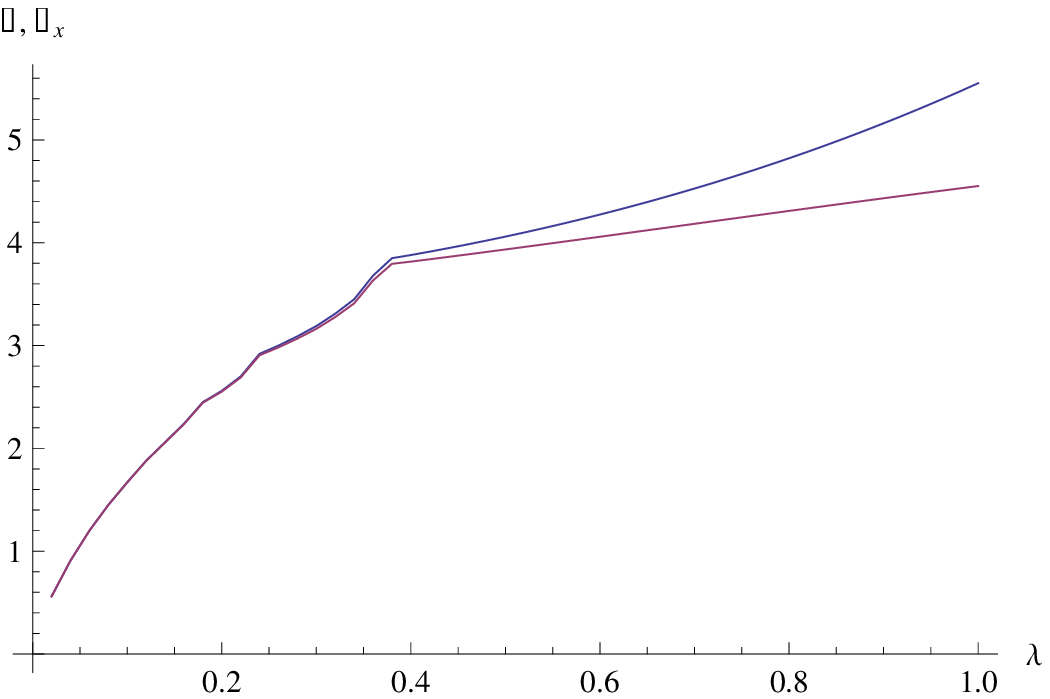}
\caption{\label{figure-bert-mass} The cumulative mass bound to
the perturbation as a function of distance from the
center; $\lambda=r/r_{ta}$ is the scaled radial coordinate;
the mass is expressed in units of
$4 \pi \rho(t) r_{ta}^3/3$ where $\rho(t)$ is the background (critical)
density. The curve is continuous with changes in slope (density
discontinuities) 
at shell-crossings. The upper curve is the entire mass including
that contributed by the mean background; the lower curve is the
excess with respect to the background.}
}

\FIGURE[h]{
\includegraphics{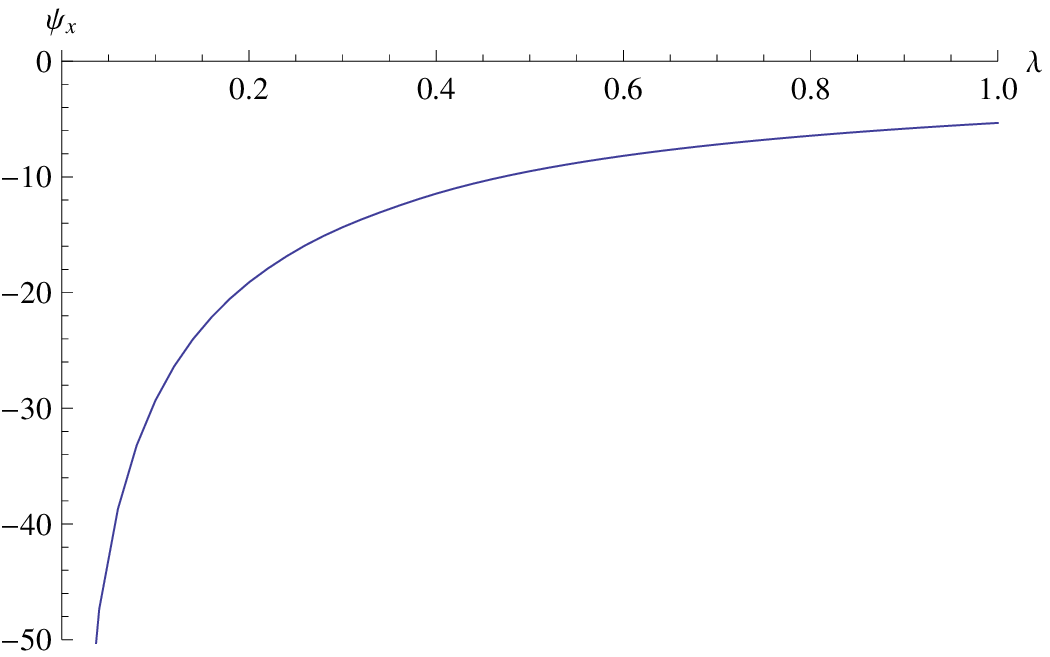}
\caption{\label{figure-bert-pot} The gravitational potential
associated with the excess mass 
as a function of distance from the center; $\psi_x$ is the gravitational
potential measured in units of $4 \pi G \rho(t) r_{ta}^2/3$ (with
value zero at spatial infinity) and $\lambda=r/r_{ta}$.}
}

To apply the infall model to the Galaxy today the turn-around
radius must be specified.
Assume that the
well-virialized part of the halo has a physical scale today $R_g=150$
kpc. The current age is fixed at the concordance value
$t_0 =1.37 \times 10^{10}$ yrs \cite{komatsu_five-year_2009}.  Let the epoch for the turn-around of
the material at $R_g$ be $t_1$ and let the turn-around radius at that time be
$r_{ta,1}$.  Figure \ref{fig-bou} illustrates the history of a
particle as it passes back and forth through the center. Its apocenter
approaches an asymptotic value of $\sim 0.8 r_{ta,1}$. Between 4 and 5
passages, the apocenter has shrunk from $r_{ta,1}$ at time $t_1$ to
$0.845 r_{ta,1}$ at time $8 t_1$.  With some arbitrariness, identify
$R_g = 0.845 r_{ta,1}$ and $t_0 \sim 8 t_1$. Since $r_{ta}
\propto t^{8/9}$ this implies the turn-around radius today is
$r_{ta,0} = 1.1$ Mpc.  

All the quantitative properties of the self-similar
description now follow. Assuming that radial infall continues
to the present, at the Sun's galactocentric radius today
($8.5$ kpc) the rotation velocity is $225$
km s$^{-1}$ and the interior mass is $10^{11} \msun$.
The model has total mass $7.5 \times 10^{11} \msun$ within
$R_g$ today.  Turnaround at $t_1/t_0 \sim 0.12$ occurs before the switch from
power law to exponential expansion in the concordance $\Lambda$CDM model 
($t_\Lambda/t_0 \sim 0.73$) so
the late-time deviations from Einstein-de Sitter should be relatively
unimportant. If infall were to cease at $t>t_1$ the net change in mass
within the solar circle is estimated to be $\sim 4$\%
and within $R_g$ to be $\sim 40$\%.

This completes the specification of the time-dependent Galactic
model.

\FIGURE[h]{
\includegraphics{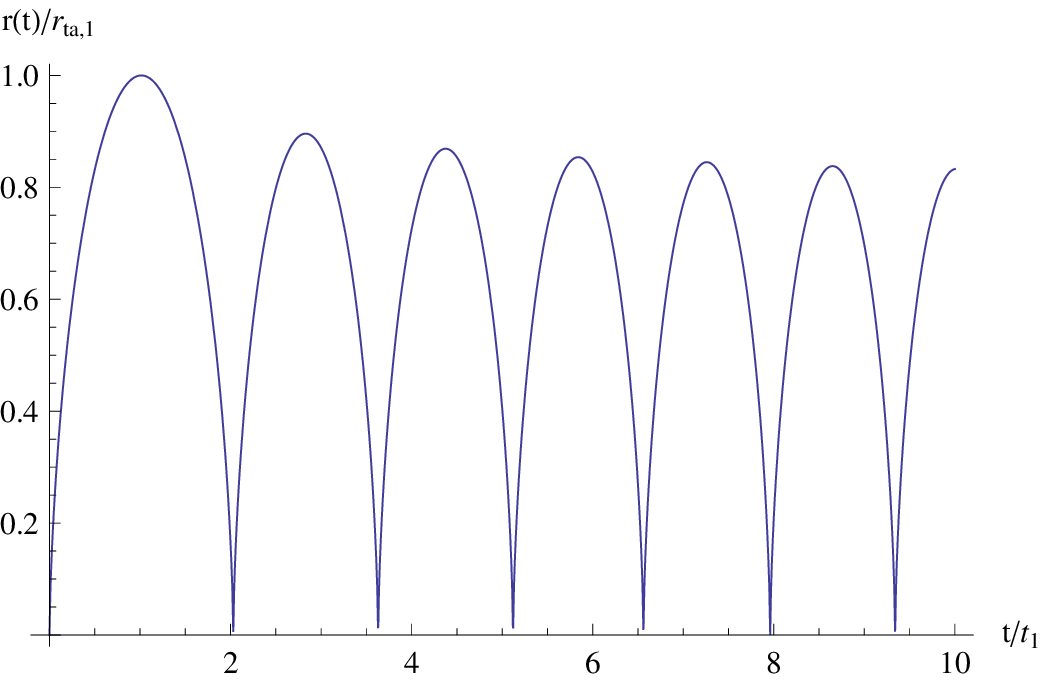}
\caption{\label{fig-bou} The motion of a single particle
in the collisionless infall solution. Time is in units of the
initial turn-around time $t_1$, physical radius in units of the
initial turn-around radius $r_{ta,1}$. The curve is a fit
to numerical results (Table 6 in \cite{bertschinger_self-similar_1985}).}
}

\clearpage

\section{Equations of Motion in Inhomogeneous FRW Cosmology}

The unperturbed background model is flat Friedmann-Robertson-Walker (FRW).
Consider a frame in which the universe appears isotropic and let
$x^\alpha = (t, x^i)$ where $t$ is the global time coordinate and
$x^i$ are the global comoving spatial coordinates. The scale factor is
$a(t)$. Henceforth, adopt units with $c=1$.

The FRW metric with scalar perturbations is
\be
g_{\mu\nu} = \left(
\begin{array}{cccc}
-(1 + 2 \psi) & & & \\
 & a^2 (1 + 2 \Phi) & & \\
 & & a^2 (1 + 2 \Phi) & \\
 & & & a^2 (1 + 2 \Phi)
\end{array}
\right)
\ee
where $\psi(x^\alpha)$ and $\Phi(x^\alpha)$ represent the
effect of inhomogeneities.  Bound structures have $\psi < 0$.

The perturbation potentials are very non-relativistic (rotation
velocity $\sim 225$ km/s for our Galaxy implies $v \sim 10^{-3}$).
The Galaxy is assumed to be at rest in the preferred FRW frame.
Only small errors are made by equating the local time and space
coordinates introduced in the previous section to the
global FRW coordinates ($t, a x^i$). Assume also $\Phi = -\psi$ which
is suitable for small quadrupolar components to the stress energy
tensor for the cold dark matter particles.

In the FRW frame, let a loop's velocity be $V^\alpha$. If there were
no rocket effect the loop would follow a geodesic through spacetime
(ignoring tidal effects). However, the loop does emit radiation and
its impulse alters the loop trajectory. Let the rocket's 4-impulse be
$a_r N^\alpha$ where $a_r$ is the time-varying magnitude of the
impulse.

Consider, first, the direction of the rocket's impulse. In the center
of mass frame, (1) loops formed in cosmological fragmentation
scenarios generally possess net angular
momentum\cite{scherrer_properties_1990}, (2) the angular momentum
radiated over an oscillation period lies parallel to the angular
momentum of the loop\cite{durrer_gravitational_1989}, and (3) the
momentum radiated over an oscillation period lies in a direction
generally different than that of the angular momentum of the
loop. Item (2) implies that loops spin down in a relatively simple
manner, however, item (3) suggests that the net gravitational force
does not act on the center of mass of the loop. The emission of
angular momentum has been studied only for loops of the simplest
complexity. Though there is no evidence to date, more complex loops
might experience more complex dynamics.

Ref. \cite{hogan_runaway_1987} treats the loop as a relativistic
gyroscope and concludes from a dimensional argument that the timescale
for a single precession cycle is $\sim \tau$, i.e. comparable to the
loop lifetime. Since the momentum impulse is not along the angular
momentum direction this argument suggests but does not prove that the
rocket direction is fixed for the life of the loop. On the other hand,
if one treats the loop as a solid body
\cite{durrer_gravitational_1989} the precession time is considerably
shorter $\sim \sqrt{G \mu/c^2} \tau$. A suitable gravitational wave
back-reaction calculation that would definitively address how the
direction of momentum impulse varies is unavailable
(\cite{quashnock_gravitational_1990} did not investigate precession
and \cite{Anderson_self-similar_2005} studied a symmetric loop that
did not radiate momentum).

Intrinsic variations of the rocket direction in the loop center of mass frame
would tend to average out the net impulse given to the loop. A fixed
direction gives the ``most effective'' rocket and presents the
``worst case'' for binding of loops to the Galaxy.

Assume (1) the impulse in
the loop center of mass frame lies along a fixed direction and
(2) no torques act in the loop center of mass frame. Then $N^\alpha$
is simply the Fermi-transported impulse direction of the loop.  The
normalizations are $V^\alpha V_\alpha = -1$ and $N^\alpha N_\alpha =
1$ and orthogonality is $N^\alpha V_\alpha = 0$.  The equations of
motion are
\ba
\frac{d V^\alpha}{d\tau} + \Gamma^\alpha_{\beta\gamma} V^\beta V^\gamma & = & a_r N^\alpha \\
\frac{d N^\alpha}{d\tau} + \Gamma^\alpha_{\beta\gamma}N^\beta V^\gamma & = & 
a_r V^\alpha
\ea
for proper time $\tau$. For the numerical solution in the FRW frame,
the 4-vectors for velocity and for the internally generated impulse
direction are parameterized
\ba
V^\mu & = & \left( \frac{\sqrt{1 + v^2}}{\sqrt{1 + 2 \psi}},
\frac{v {\hat v}^i}{a\sqrt{1 + 2 \Phi}} \right) \\
N^\mu & = & \left( \pm \frac{\sqrt{n^2 -1}}{\sqrt{1 + 2 \psi}},
\frac{n {\hat n}^i}{a\sqrt{1 + 2 \Phi}} \right)
\ea
where $v^2=g_{ij}V^iV^j$, $n^2=g_{ij}N^iN^j$, and ${\hat v}^i$ and
${\hat n}^i$ are 3D-orthonormal unit vectors (${\hat v} \cdot {\hat v}
\equiv \sum_i {\hat v}^i {\hat v}^i = 1$ and ${\hat n} \cdot {\hat n}
= 1$).\footnote{With 
this parametrization, an FRW observer sees an energy per mass
$\sqrt{1+v^2}$ and a momentum per mass $v$. In terms of relativistic
kinematic variables $v = \gamma \beta$. In this paper, $v$ is
called ``velocity'' when the regime is
non-relativistic and ``momentum-per-mass'' for more generality.}
The equations for $dx^i/dt$, $dv/dt$, $d{\hat v}/dt$, $dn/dt$
and $d{\hat n}/dt$ are expressed using the global FRW time as the
independent coordinate.  These equations are applicable to
loops with the whole range of possible velocities from extremely
relativistic to non-relativistic.  The explicit form is given in the
Appendix \ref{appendix-equations}.

Let the total loop energy be $E$ in the FRW frame and, following custom,
denote $E/\mu$ as length $l$. For clarity,
explicitly label quantities in the string's center of mass frame with
``z''. The infinitesimal length (i.e. energy) is
$dl_{(z)} = d\sigma \sqrt{(d{\vec z}/d\sigma)^2/(1-{\dot {\vec z}}^2)}$
where ${\vec z}={\vec z}(\sigma,z^0)$ is the parametric expression
for the string; $l=V^0 l_{(z)}$ and $dt = V^0 dz^0$.

In the loop center of mass frame the rate of energy loss and the magnitude
of the impulse are very simple
\ba
\frac{d l_{(z)}}{dz^0} & = & -\Gamma_E G \mu \\
a_r & = & \Gamma_P \frac{G \mu}{l_{(z)}} .
\ea
The loop lifetime is a fixed increment of time in the center of mass frame.
The 4-impulse in the center of mass frame is
$a_{(z)}^\alpha=(0,a_{(z)}^i)=a_r (0,n_{(z)}^i)$ where $n_{(z)}^i$ is a unit
vector in the direction of the impulse and $a_r$
is the magnitude of the impulse. Since $a_r$ is
a scalar, $a^\alpha a_\alpha = a_{(z)}^\alpha a_{(z)\alpha} = a_r^2$,
write the 4-impulse $a_r N^\alpha$ in the FRW frame.
Since the initial loop configuration ${\vec z}(\sigma,z^0)$ determines
$n_{(z)}^i$ and no torques operate (by assumption)
it is most convenient to find the initial $N^\alpha$ in FRW frame (Appendix
\ref{appendix-acc}) and use Fermi transport to
determine its subsequent evolution in that frame.

In the FRW frame
\ba
\frac{d l}{dt} & = & \frac{l}{V^0} \frac{dV^0}{dt} - \Gamma_E G \mu \\
\frac{d}{dt} \left( \frac{1}{a_r} \right) & = & -\frac{\Gamma_E}{\Gamma_P V^0 } .
\ea
After $a_r N^\alpha$ is initially set the entire calculation can be
carried out in the FRW frame using Fermi transport for $N^\alpha$ and the
above equation for $a_r$.

There are a variety of non-trivial frame transformation effects
that operate in this schematic description of loop evolution.
Ignoring the momentum impulse of the
rocket and the inhomogeneous potential, in the FRW frame
a loop born with length $l_i$ with center of mass
motion $V^0_i=\sqrt{1+v_i^2}$ at time $t_i$ has length
and center of mass momentum
\ba
\label{eqn-loft}
l & = & \sqrt{\frac{1+v^2}{1+v_i^2}} \left( l_i - t_i \frac{\Gamma_E G \mu}{\nu } \sqrt{1+v_i^2}\int_1^{a/a_i}\frac{x^{1/\nu} dx}{\sqrt{x^2 + v_i^2}} \right) \\
v & = & \frac{v_i a_i}{a}
\ea
at scale factor $a=a_i(t/t_i)^\nu$. The initial loop size that just
evaporates at scale $a/a_i$ is explicitly given by setting the 
expression within the
parenthesis to 0. It is clear that complete evaporation occurs in
a finite FRW time. 

The time a loop lives is slightly different than the above result
in a homogeneous universe because of the ever increasing
importance of the rocket effect. The loop length still
vanishes in a finite time. Let the time until evaporation be $\Delta t
= t_{life}-t$ and the ratio of momentum-to-energy loss
$\beta=\Gamma_P/\Gamma_E$. Then the length, acceleration and momentum
parameter vary asymptotically
\ba
l & \propto & \Delta t \\
a_r & \propto & \Delta t^{-1/(1-\beta)} \\
v & \propto & \Delta t^{-\beta/(1-\beta)}
\ea
The difference between the approximate and exact lifetimes
is only $\sim 2.3$\% ($\Gamma_E=50$, $\Gamma_P=10$, $v_i =0.1$)
which will be ignored in subsequent discussion. The acceleration
and the momentum-per-mass both diverge as the evaporation
proceeds to completion.

\section{Characterization of Orbits}

\subsection{Radial Geodesics}
The first calculations illustrate some basic kinematic features for
objects whose initial velocity is very different from Hubble flow. Consider
purely radial geodesics and ignore the rocket effect.  Fix the
magnitude of the initial radial velocity to be $v_{i}=0.1$ (typical
of the largest loops chopped off from the horizon crossing
strings) at the initial time $t_i/t_0 = 10^{-9}$. The results that
follow are for the full, relativistic equations of motion. One
calculation differs from another only in terms of the initial position
of the loop with respect to the spherical center.

Figure \ref{fig-rainbow-comoving-9} shows the comoving trajectories of
a set of loops as a function of $\log t/t_0$. The different colors
label different initial positions and ingoing and outgoing velocities.
The velocity $v$ is damped by the many
decades of expansion (in the absence
of a varying potential $v(t) \sim v_i a_i/a(t)$), a fact made
qualitatively clear from the flattening of all the curves at early
times. Once a loop's motion has been damped, it behaves for all
practical purposes like a cold dark matter particle at the position to
which it has moved. The comoving coordinate is nearly but not exactly
static because every zero-velocity object is bound to the excess
central mass of the perturbation.  In the absence of the rocket
effect, each loop eventually turns around, the comoving coordinate
retreats and the loop oscillates back and forth through the
perturbation center. To avoid clutter, only the first few bounces of
each loop are plotted. The color of the line allows tracing the
epoch of turn-around and recollapse for a loop to a given initial
position. The black line is the comoving turn-around radius in the
radial infall model.

\FIGURE[h]{
\includegraphics{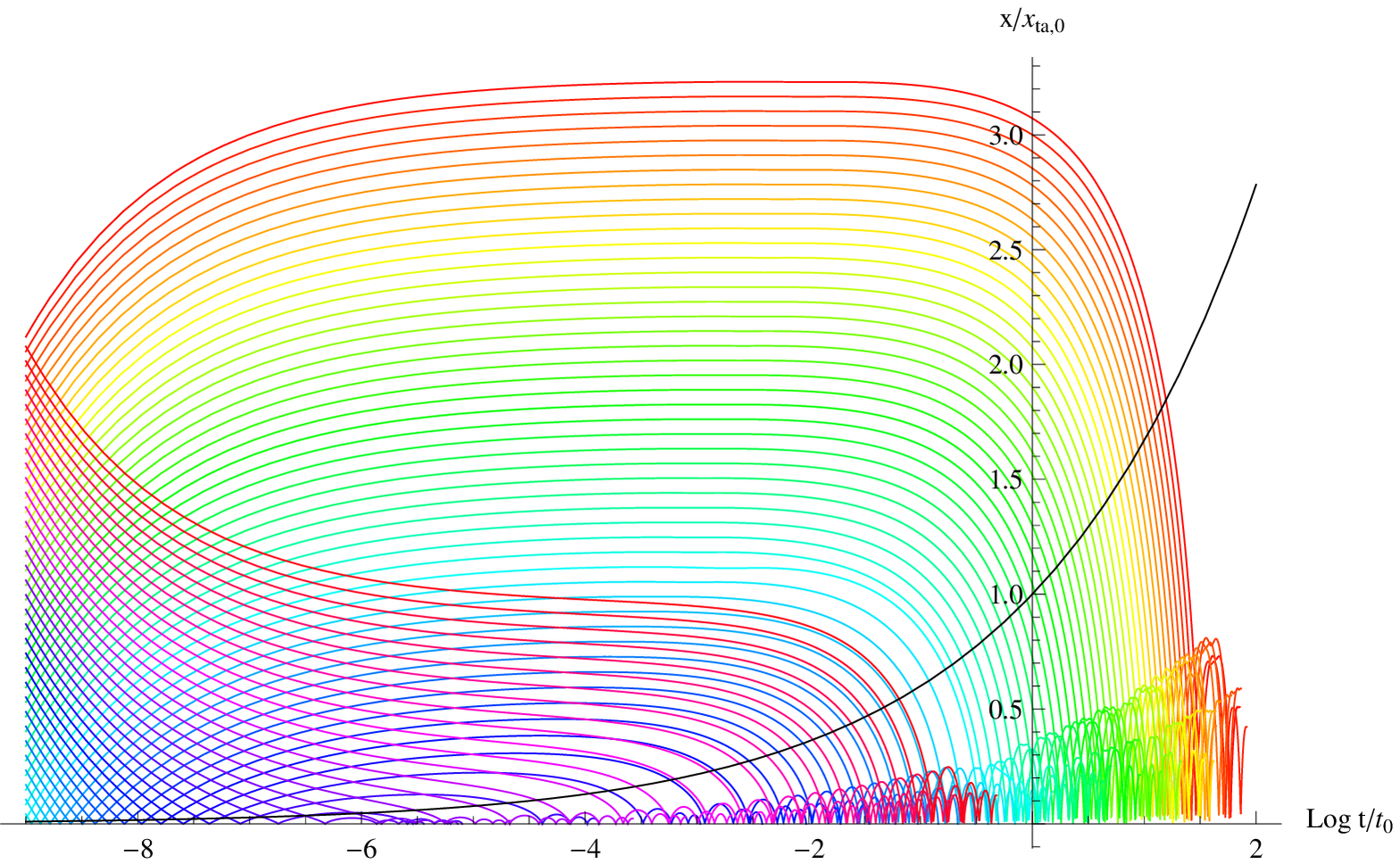}
\caption{\label{fig-rainbow-comoving-9} Comoving position as a function
of $\log t/t_0$. Radial geodesics (no rocket effect) have been
integrated from a set of initial positions with fixed initial radial
velocity $v_i = 0.1$ (either inward or outward) at time $t_i/t_0 =
10^{-9}$. Distance is measured in units of today's comoving
turn-around distance.  The black line shows the comoving turn-around
radius.}
}

Figures \ref{fig-rainbow-captured-comoving-9} and
\ref{fig-rainbow-captured-physical-9} are blow ups in comoving and
physical coordinates respectively for $10^{-3}<t/t_0<1$. They show the
first few bounces after capture of the loops by the
perturbation. These figures as well as the previous one demonstrate
that early (late) turn-around implies small (large) semi-major axes
just as is true for the cold dark matter particles in the radial
infall model. Figure \ref{fig-rainbow-captured-comoving-9} illustrates
that loops starting in different regions of space with different
initial velocities (red and blue lines) can end up with nearly
identical accretion orbits. This is simply the shuffling in position
that occurs during the time it takes for cosmic drag to operate. As a
side note, the absence of red lines is a consequence of the limited
range of initial radii sampled; had larger offsets been plotted such
lines would be present throughout the figure.  Figure
\ref{fig-rainbow-captured-physical-9} makes it clear that there exist
loops that ``turn around'' at the same space time locations as cold
dark matter particles do.

\FIGURE[h]{
\includegraphics{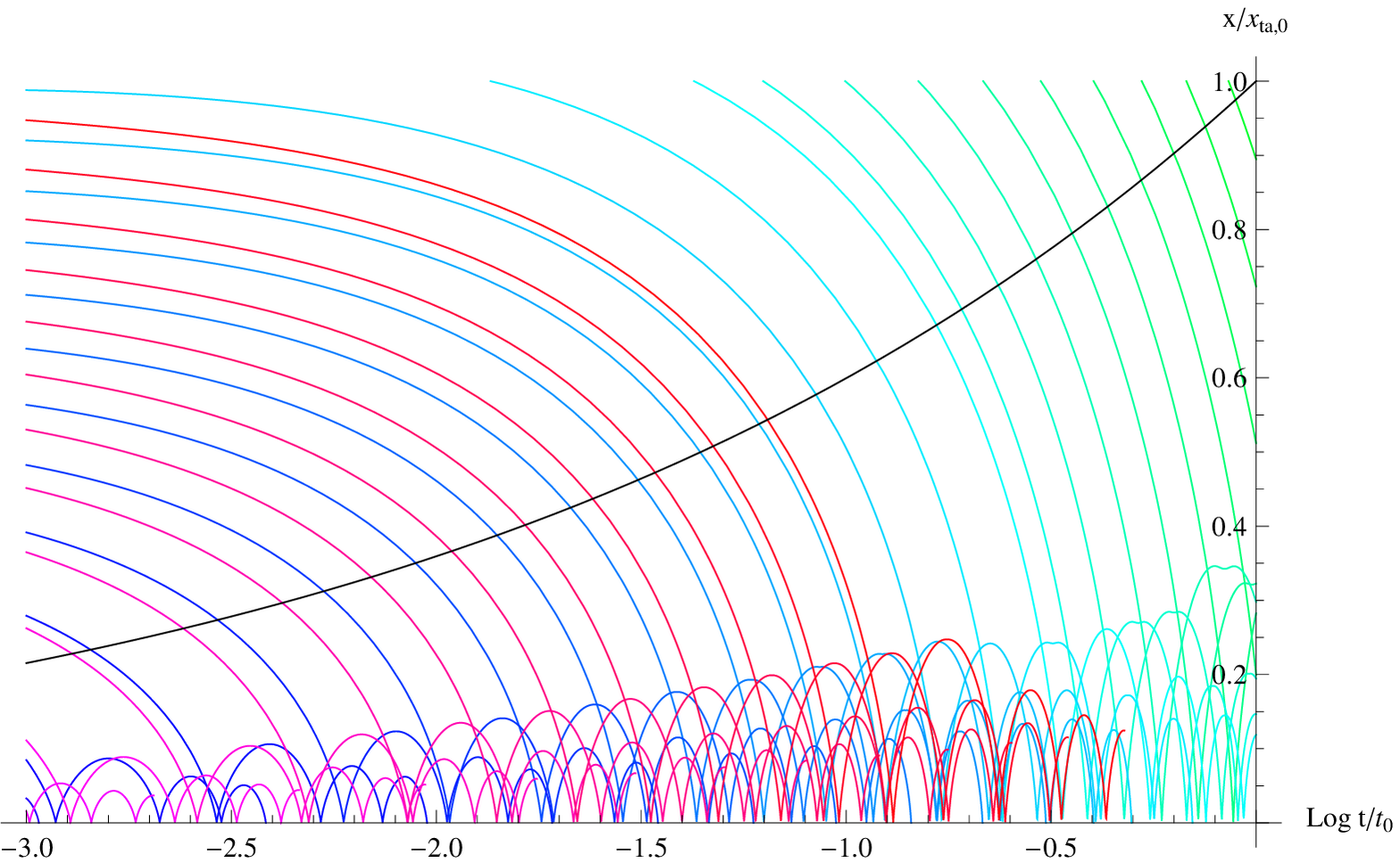}
\caption{\label{fig-rainbow-captured-comoving-9} Detailed view of the
  trajectories in figure \ref{fig-rainbow-comoving-9}. The different
  colored lines indicate different initial conditions. Only the first
  few bounces through the center are plotted. Consequently, the
  increase in amplitude with time highlights a {\it sequence} of loops
  falling into the perturbation having systematically larger orbits
  not that an individual loop's bounces grow in amplitude.  } }

\FIGURE[h]{
\includegraphics{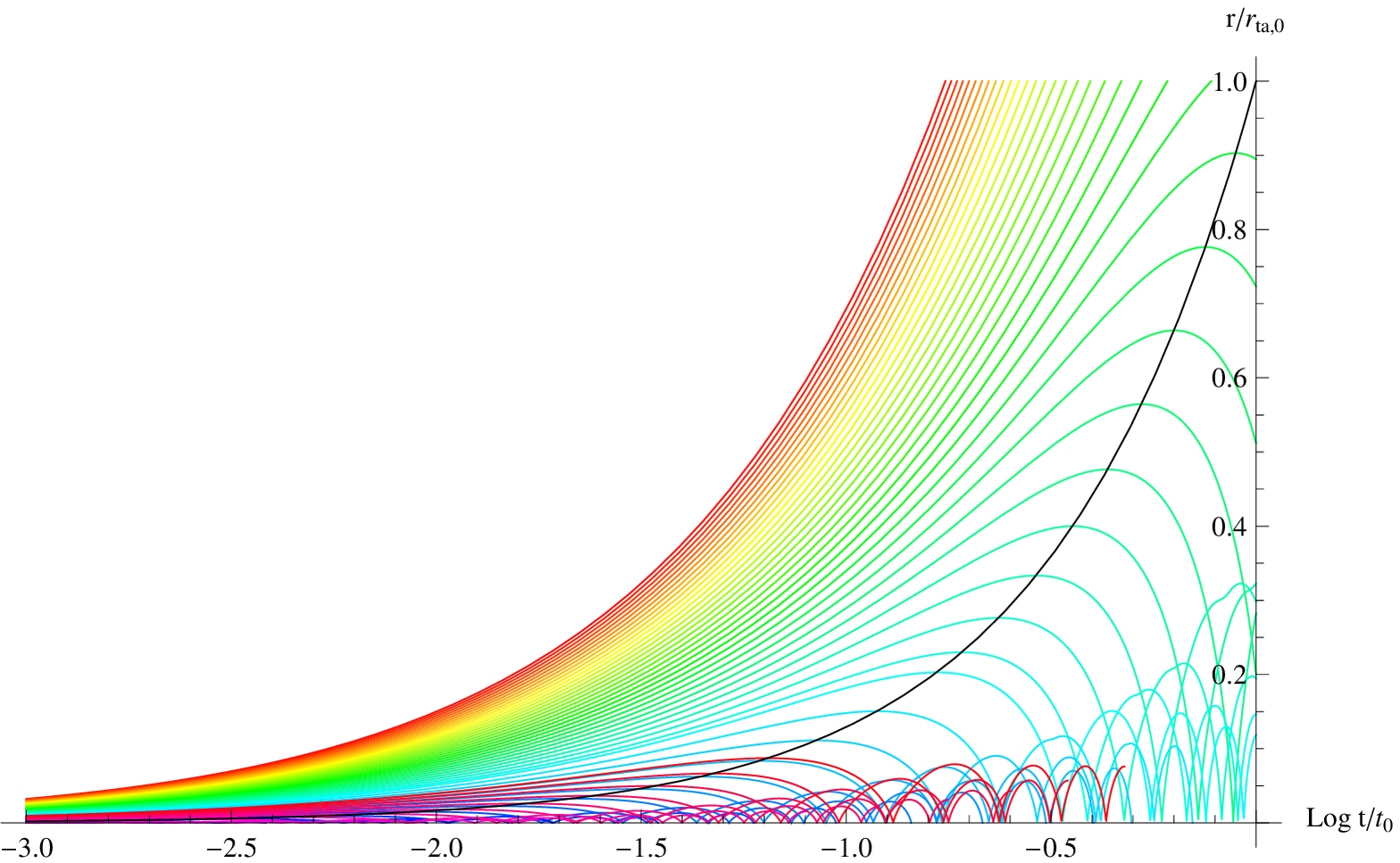}
\caption{\label{fig-rainbow-captured-physical-9} 
Detailed view of physical separation for
trajectories in figure \ref{fig-rainbow-comoving-9}. Only
the first few bounces through the center are plotted. In
physical coordinates, the bounces occur with nearly fixed
amplitude. Same comments as for
figure \ref{fig-rainbow-captured-comoving-9} apply.
}
}

If a loop is young then cosmic expansion may not have had sufficient
opportunity to damp its velocity to allow accretion onto the growing
perturbation at a physical radius of interest.  A simple estimate of
how small $t_i/t_0$ must be for a loop to capture at time $t$ is given
as follows.  Cosmic drag implies the initial velocity decays like
$v(t) = v_i a_i/a(t)$ (flat potentials). A necessary
condition for capture is that $v(t)$ must be less than the escape velocity from
the potential. However, a generally more restrictive condition is that
capture requires $v(t) < H(t) r_{ta}(t)$. Since the turn-around radius
$r_{ta}(t) = r_{ta}(t_0) \left( \frac{t}{t_0} \right)^{8/9}$ the
initial time is constrained to be $t_i/t_0 < \left( \frac{H_0
  r_{ta}(t_0)}{v_i} \right)^{3/2} \left( \frac{t}{t_0}
\right)^{5/6}$. For example, a loop with initial velocity $v_i =0.1$
can be captured today if $t_i/t_0 \lta 7.5 \times 10^{-5}$ and will
have a physical orbit $\sim r_{ta}(t_0)$. Physical radius and
time of capture are inherently linked in the similarity solution.
A loop with physical orbit
$r < r_{ta}(t_0)$ must be accreted at earlier time $t/t_0 =
(r/r_{ta}(t_0))^{9/8}$; the initial time of formation of that loop is
constrained to be $t_i/t_0 < \left( \frac{H_0 r_{ta}(t_0)}{v_i}
\right)^{3/2} \left( \frac{r}{r_{ta}(t_0)} \right)^{15/16}$. For
example, for $r=30$ kpc, sufficient cosmic drag requires the loop be
formed at $t_i/t_0 < 2.5 \times 10^{-6}$ and captured at $t/t_0 \sim
1.7 \times 10^{-2}$.

Figure \ref{fig-rainbow-comoving-4} shows the comoving trajectories of
a set of loops born at $t_i/t_0=10^{-4}$ (the ordinate is greatly
expanded compared to previous figures). The slope for orbits far
away from the perturbation indicates that cosmic drag has not yet
brought the loops to rest. This impedes capture.
Figure \ref{fig-rainbow-captured-physical-4}, a detailed view near the
origin, shows that when it does occur
it does so at large physical separation.

\FIGURE[h]{
\includegraphics{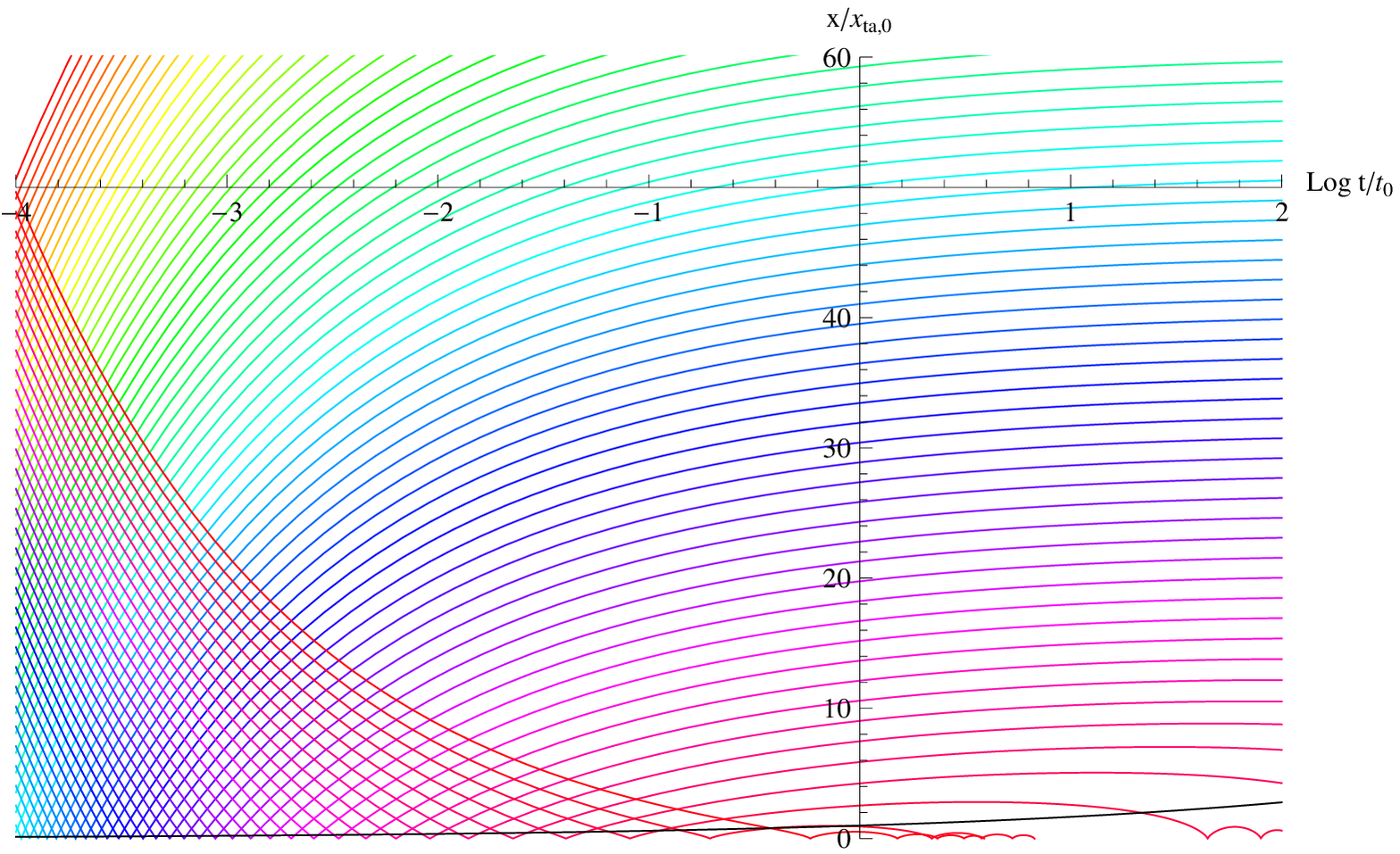}
\caption{\label{fig-rainbow-comoving-4} Same as
figure \ref{fig-rainbow-comoving-9} but with $t_i/t_0=10^{-4}$.
}
}

\FIGURE[h]{
\includegraphics{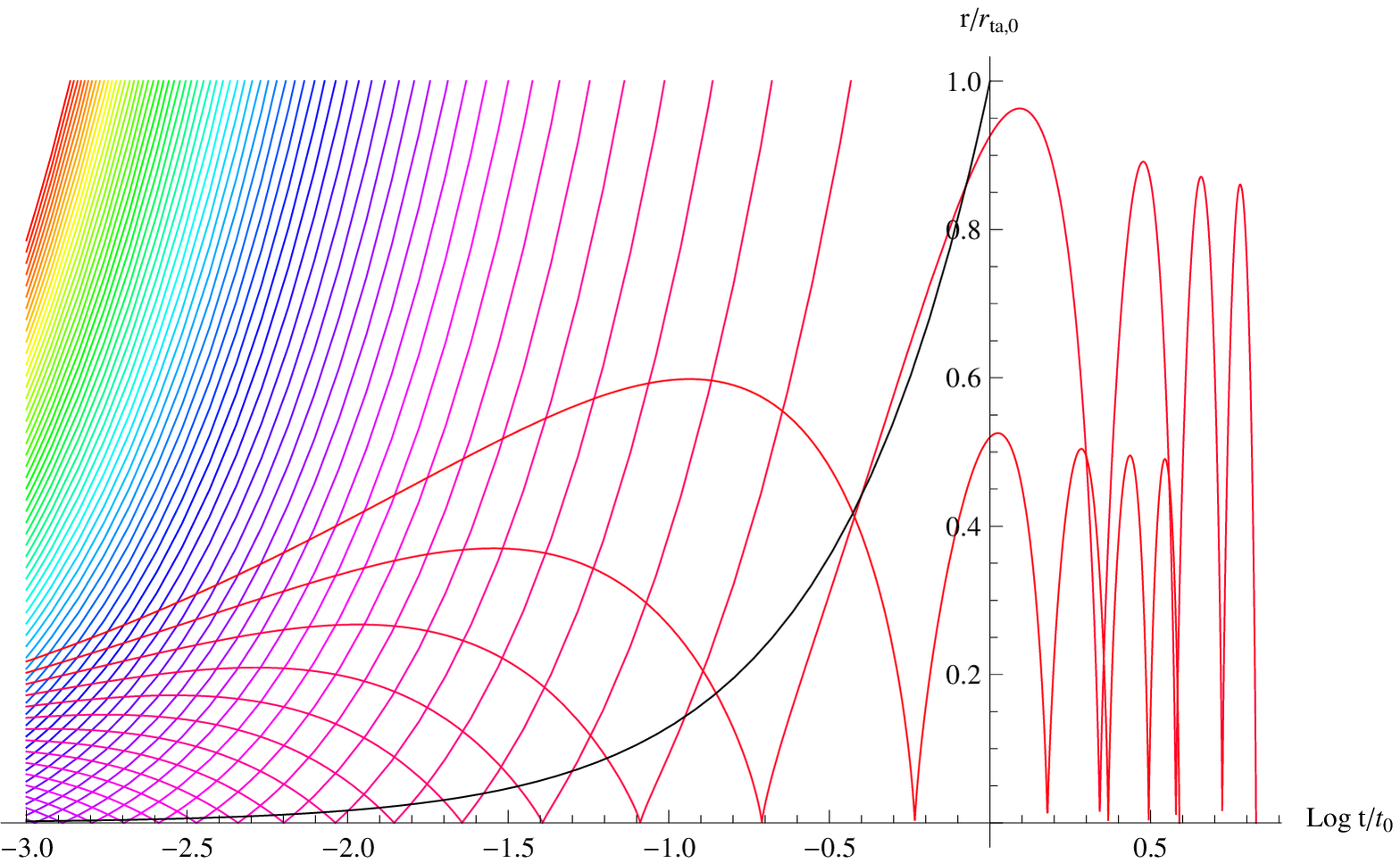}
\caption{\label{fig-rainbow-captured-physical-4} Same as
figure \ref{fig-rainbow-comoving-4} but in physical coordinates.
}
}

\clearpage

\subsection{Rocket Effect}
\label{subsec-rocket}

The rocket effect can inhibit binding of a loop to the
growing perturbation and can unbind a previously captured loop.
This section begins with analytic estimates and follows with
full numerical calculations.

Begin by considering homogeneous FRW with rocket
direction aligned or anti-aligned with initial velocity. The
equations of motion (Appendix \ref{appendix-equations}) reduce to
\ba
\label{eqn-rocketeffect-start}
\frac{dv}{dt} & = & \delta \frac{n a_r}{\sqrt{1+v^2}} - v \frac{\dot a}{a} \\
\frac{dn}{dt} & = & \delta \frac{v a_r}{\sqrt{1+v^2}} - \frac{n v^2 {\dot a}}{a (1+v^2) } \\
\frac{da_r}{dt} & = & \frac{\Gamma_E a_r^2}{\Gamma_P \sqrt{1+v^2}} \\
\label{eqn-rocketeffect-end}
\delta & = & {\hat n} \cdot { \hat v} = \pm 1
\ea
with initial velocity $v_i$ and loop length $l_i = \alpha /H_i = (3/2)
\alpha t_i$ in the matter-dominated, FRW frame 
at time $t_i$. The initial rocket impulse is $a_{r,i} =
\Gamma_P G \mu \sqrt{1+v_{i}^2}/l_i$ and the FRW direction
vector is $n_i=\sqrt{1+v_i^2}$. The equation for $n$
may be solved and its occurrences eliminated. For non-relativistic
motions, the equation for $a_r$ may also be integrated explicitly
to give
\ba
\frac{dv}{dt} & = & \delta \frac{\Gamma_P}{\Gamma_E} \frac{1}{\tau_i + t_i - t} - v \frac{\dot a}{a} \\
\tau_i & = & \frac{l_i}{\Gamma_E G \mu}
\ea
Here $\tau_i$ is the characteristic time for the loop to evaporate
by gravitational wave radiation. If $\tau_i + t_i >> t$ then
the first term is approximately constant and $v$ is a
sum of terms proportional to 
$t$ and $t^{-2/3}$ 
\cite{vachaspati_gravitational_1985,battefeld_magnetogenesiscosmic_2008}.
The limit of interest is $H \tau >> 1$. Specifically, if
$\kappa^{-1} \equiv (5\Gamma_E/(2 \Gamma_P)) v_i H_i \tau_i  = (5 \alpha v_i/(2 \Gamma_P G \mu)) >> 1$ then
\be
v \simeq v_i \left(
\left( \frac{t_i}{t} \right)^{2/3} + \delta \frac{\kappa t}{t_i} 
 \right)
\ee
Capture at radius $r$ of the accelerated trajectory for
a loop with aligned velocity and rocket impulse
($\delta =1$) requires
\be
\kappa < 
\frac{ \frac{H_0 r_{ta,0}}{v_i} 
\left( \frac{t_0}{t_i} \right)^{2/3} 
\left( \frac{r}{r_{ta,0}} \right)^{5/8} - 1 }
{ \left( \frac{r}{r_{ta,0}} \right)^{15/8} 
\left( \frac{t_0}{t_i} \right)^{5/3}
}
\ee
by the same line of argument given in the previous section.
The numerator must be positive and implies an upper
limit on the time of formation
\be
\label{eqn-time-upperlimit}
\left( \frac{H_0 r_{ta,0}}{v_i} \right)^{3/2} 
\left( \frac{r}{r_{ta,0}} \right)^{15/16} 
>
\frac{t_i}{t_0}
\ee
identical to that in the previous section.
This is the effect of cosmic drag and does not depend upon $\mu$.
For loops formed at earlier times, capture requires
\be
\label{eqn-mu-rocket-upperlimit}
G \mu < \frac{5 \alpha}{2 \Gamma_P} H_0 r_{ta,0} 
\left(\frac{r_{ta,0}}{r}\right)^{5/4}
\frac{t_i}{t_0}
\ee
a result that highlights the importance of
the rocket effect.

The condition that the loop not evaporate by
the current epoch requires
\be
\label{eqn-mu-lifetime-upperlimit}
G \mu < \frac{3 \alpha}{2 \Gamma_E} \frac{t_i}{t_0} .
\ee
For a given epoch of formation, the upper limit on tension is
rocket-related for $r/r_{ta,0} > (5 \Gamma_E/(3 \Gamma_P))^{4/5} (H_0
r_{ta,0})^{4/5} \sim 7.6 \times 10^{-3}$ or $r \gta 8.5$ kpc and
age-related at smaller radii.

The maximum tension that permits capture (the intersection of limits
implied by equations \ref{eqn-time-upperlimit} and
\ref{eqn-mu-rocket-upperlimit} or \ref{eqn-mu-lifetime-upperlimit};
this is the rightmost section of the triangular region formed by green
and turquoise lines in figure \ref{fig-summary-ti-mu} ) is
\ba
G \mu |_{critical} & = & \frac{5 \alpha}{2 \Gamma_P} 
v_i \left( \frac{H_0 r_{ta,0}}{v_i} \right)^{5/2}
\left( \frac{r_{ta,0}}{r} \right)^{5/16} 
{\rm min} 
\left( 
1, 
\frac{3 \Gamma_P}{5 \Gamma_E} 
\left( H_0 r_{ta,0} \right)^{-1}
\left( \frac{r}{r_{ta,0}} \right)^{5/4}
\right)
\\
& = & 4.12 \times 10^{-9} 
\left( \frac{\alpha}{0.1} \right)
\left( \frac{ 0.1}{v_i} \right)^{3/2}
\left( \frac{10 {\rm kpc}}{r} \right)^{5/16}
{\rm min} 
\left(
1, \left( \frac{r}{8.5 {\rm kpc} } \right)^{5/4}
\right)
\ea
Above this critical $G \mu/c^2$, loops do not cluster at scale $r$
within the Galaxy. Curiously, the critical $G \mu$ has a maximum close
to the Sun's galactocentric position though the variation from
$r=3-100$ kpc is only about $2.7$ (and with all other parameters
fixed).

All captured loops are eventually stripped from the galaxy by the
rocket effect. A detailed discussion of how removal proceeds
is given in the next section. The result is that $a_r >
0.3 | \nabla \psi |$ leads to detachment. For fixed loop size,
a loop is retained until the current epoch if
\be
G \mu < \frac{0.15 \alpha}{\Gamma_P} H_0 r_{ta,0} 
\frac{{\cal M}_x}{\lambda^2}
\frac{t_i}{t_0}
\ee
For small $\lambda$ the asymptotic form for the mass distribution is
${\cal M}_x \sim 10.5 \lambda^{3/4}$, so that ${\cal M}_x/\lambda^2
\propto \lambda^{-5/4}$ just as in eq. \ref{eqn-mu-rocket-upperlimit}.
This retention criterion is very similar to the capture criterion but
quantitatively a bit stricter (the constant for retention is $\sim
0.6$ that of the capture).  However, the physical interpretations are
very different.  Capture is a statement about the properties of the
loop and turn-around radius at early times whereas retention concerns
all later times up to the current epoch. Because of the self-similar
evolution of the perturbation both criteria vary with $t_0$
identically {\it for fixed loop size}. As the loop shrinks, the
retention criterion becomes more and more strict.  In summary, the
retention criterion is only a bit stricter for fixed loop size but
becomes far stricter once the loop size begins to change.

Figure \ref{fig-summary-ti-mu} provides a graphical summary of some of
the analytic results and a point of reference throughout this section
which describes additional calculations.\footnote{I thank
  Xavier Siemens for his version of this figure.}

The above analysis is now supplemented by a study of the rocket's
effect on loop dynamics via a sequence of trajectory calculations of
increasing complexity. The numerical investigation proceeds along the
following lines. First, locate the specific radial trajectories that
give rise to orbits of fixed physical size today for $\mu=0$,
i.e. without the accelerative force. Second, starting from the
inferred initial conditions, repeat the calculation of the
trajectories including the effect of gravitational wave recoil. The
rocket effect is slight at first but dominates by the end of the loop
lifetime. Comparison of different $\mu$ yields a criterion for
retaining a loop at the radius of interest today.  Figure
\ref{fig-summary-ti-mu} includes some of the numerical results.

\FIGURE[h]{
\includegraphics{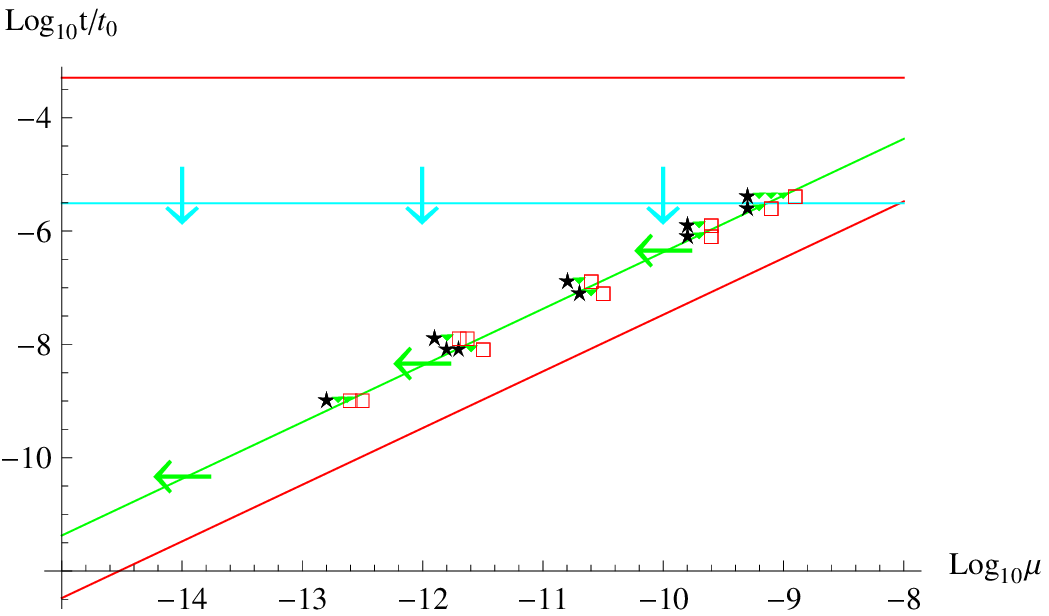}
\caption{\label{fig-summary-ti-mu} Bounds on formation time
and string tension for a loop with initial velocity $v_i=0.1$
to be captured at physical radius $30$ kpc.
{\bf (1)} Upper bounds on the formation time $t_i/t_0$ are given by the
horizontal lines. The condition
that cosmic drag lower the velocity to less than the circular rotation
velocity today is given by the red line. The more stringent condition that
capture occur at $30$ kpc is given by the turquoise line.
{\bf (2)} Upper bounds on the string tension $G \mu/c^2$ are given by the
diagonal lines. The initial loop size is $\alpha/H_i$ and
$\alpha=0.1$. The condition that the loop be younger than its
gravitational wave decay timescale is given by the red line.  The more
stringent condition that the loop not be accelerated out of the Galaxy
is given by the green line ($\chi>\chi_{crit}$). 
The geometric symbols summarize results
from numerical experiments examining the outcome today ($t=t_0$) for
groups of 10-20 loops captured at $30$ kpc with slightly different
string tensions: stars = all loops bound, boxes = all loops ejected,
triangles = some bound and some ejected (for clarity the points are slightly
offset in the vertical but not the horizontal direction).
{\bf (3)} The triangular region encompasses string tensions and formation times
giving bound loops at $30$ kpc for $v_i=0.1$ and $\alpha=0.1$ in a
radial infall model of the Galaxy. The critical value of $\mu$ below
which clustering is possible is determined by the upper right hand corner
of the green and turquoise lines.
Lowering $v_i$ raises the limit on
$t_i/t_0$ (horizontal lines moves upward); lowering $\alpha$ shifts the
bound to smaller $\mu$ (diagonal lines moves leftward). Shifting the
loop orbital scale to smaller values (say the solar position)
requires earlier formation times (horizontal lines shifts down) and
allows larger $\mu$ (green line shifts to the right but is limited by
the red line which is fixed).
} }

Begin by choosing a time for the birth of the loop $t_i/t_0$
($10^{-9}$ to $1$) and fixing the initial radial velocity
($v_i=0.1$). Then find the initial position that ends with an orbit
of the desired apocenter at the current epoch (two specific cases,
$10$ and $30$ kpc, are considered; the apocenter is estimated based on
the last 2 extrema of the orbit prior to $t_0$). This is a
boundary value problem for the equations of motion and is
solved by numerical iteration.

For each case, two qualitatively distinct orbital solutions for small $t_i$ were
found; no attempt was made to find all solutions. As $t_i$ increased
the initial conditions for the individual solutions converged and for
$t$ greater than a critical value no suitable initial conditions could
be found.  This result is consistent with the order-of-magnitude
argument given above.  The variation of the initial
position with $t_i/t_0$ of the two branches
is systematic and shown in figure
\ref{fig-xinit}.

Both branches of $30$ kpc orbits are over plotted in comoving
coordinates in \ref{fig-30kpc-comoving}; they begin at different times
and different velocities but note that all converge to similar oscillatory
solutions. These solutions are the ``baseline'' solutions which are
now perturbed by the rocket effect.

\FIGURE[h]{
\includegraphics{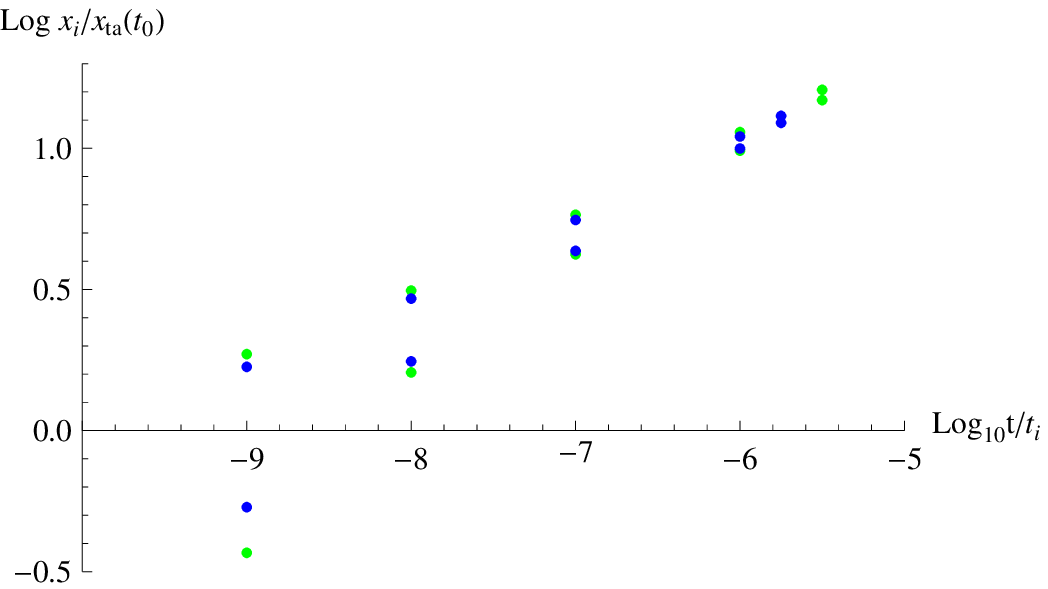}
\caption{\label{fig-xinit}
Initial comoving coordinate positions (ordinate) at a set
of different formation epochs (abscissa) for
loops that evolve to give orbits of fixed physical size today.
The green (blue) points are the initial conditions that give 30 (10) kpc orbits.
The two branches are two qualitatively
different solutions to the boundary value problem described in
the text. The loop in the lower branch passes through the
perturbation center, slows down and is overtaken by the turn-around
radius. The one in the upper branch begins moving outward and is overtaken
by the turn-around radius.
}
}

\FIGURE[h]{
\includegraphics{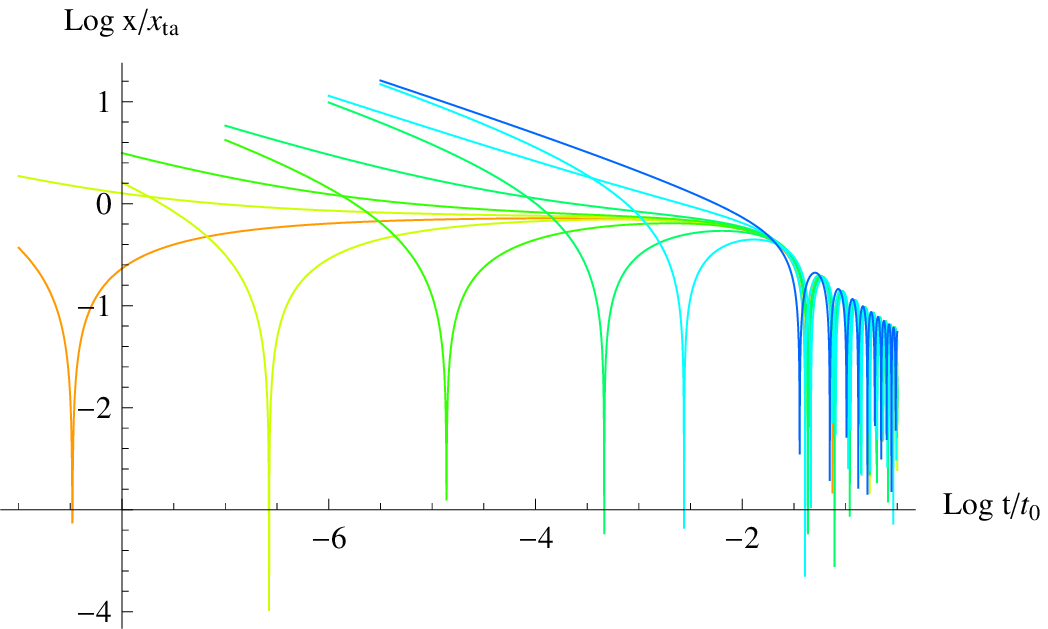}
\caption{\label{fig-30kpc-comoving} 
The different lines illustrate the comoving radial coordinate 
of loops formed at
different times and yielding $30$ kpc orbits today. These are
the explicit solutions to the boundary-value problem whose
initial conditions are displayed in Figure \ref{fig-xinit};
they correspond to the green upper and lower branches.
}
}

Starting from the initial conditions inferred above, repeat the
calculations with non-zero $\mu$ and a set of random orientations for
the rocket in the loop center of mass
frame.  Varying $\mu$ delimits the transition from bound to unbound
orbits at the current epoch. Small (large) $\mu$ implies weak (strong)
acceleration. The transition refers to a specific time, $t=t_0$,
as it is clear that eventually, for long enough integrations, the
orbits of all evaporating loops are unbound.

The initial position and velocity in figures
\ref{fig-physical-ti-9-mu-12.8}-\ref{fig-physical-ti-9-mu-12.6} are
all identical, i.e. that of a loop born at $t_i/t_0=10^{-9}$ which
formed a bound $30$ kpc orbits in the absence of the rocket effect.  A
sequence of calculations with increasing $\log_{10} \mu=-12.8$,
$-12.7$ and $-12.6$ and random rocket orientations unbinds an
increasing fraction of orbits.  Each figure includes the
original unperturbed orbit for comparison.

Besides the fact that the transition occurs over a fairly narrow range
in $\mu$, the figures also illustrate that the apocenter does not
significantly change until just before the orbit is actually
destroyed.

\FIGURE[h]{
\includegraphics{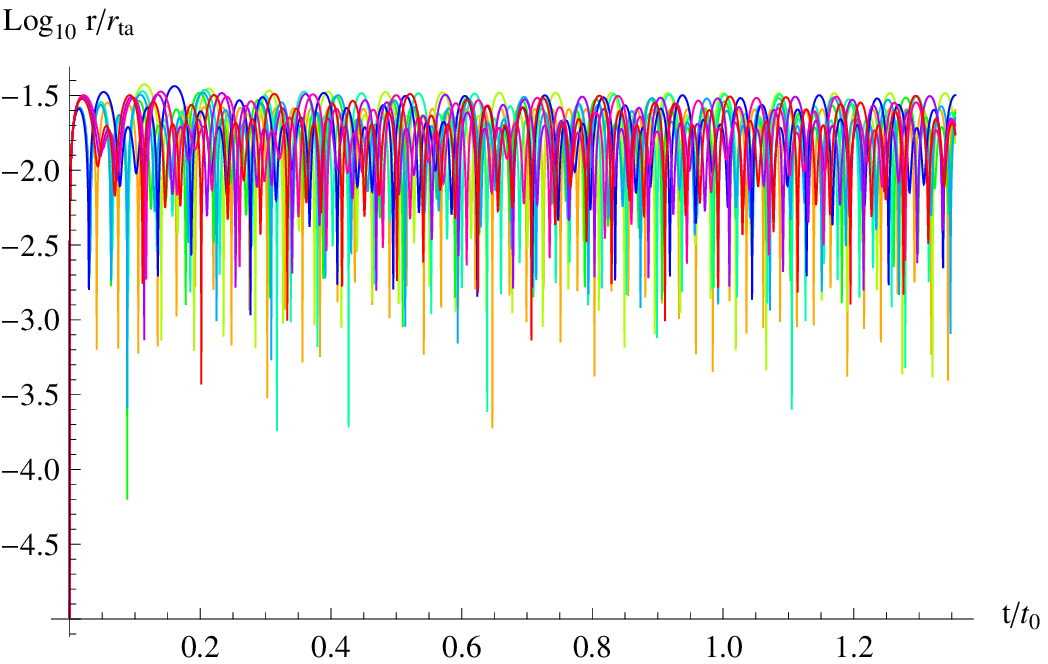}
\caption{\label{fig-physical-ti-9-mu-12.8}
Initial time $t_i/t_0=10^{-9}$, initial velocity $v_i=0.1$, $\mu=10^{-12.8}$.
Eight trajectories with randomly chosen initial momentum directions and
the original unperturbed trajectory (golden yellow color) are
plotted 
}
}

\FIGURE[h]{
\includegraphics{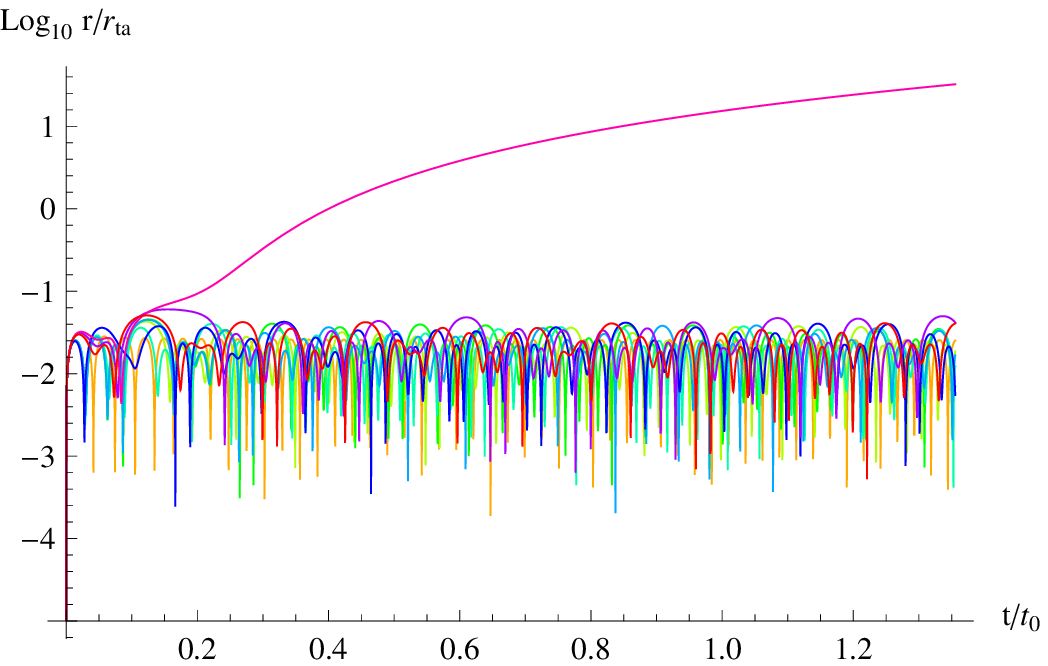}
\caption{\label{fig-physical-ti-9-mu-12.7} Same as \ref{fig-physical-ti-9-mu-12.8} but $\mu =10^{-12.7}$. One trajectory is detached.
}
}

\FIGURE[h]{
\includegraphics{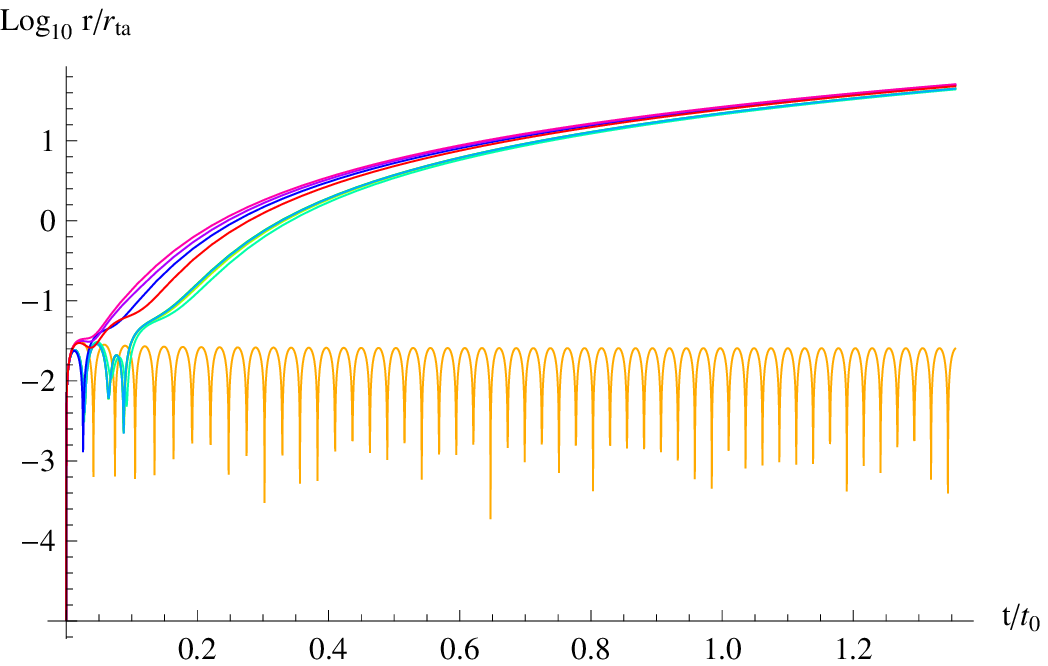}
\caption{\label{fig-physical-ti-9-mu-12.6} Same as \ref{fig-physical-ti-9-mu-12.8} but $\mu =10^{-12.6}$. All the perturbed trajectories escape.
About half the trajectories are captured and later detached while the
rest avoid capture completely. The
unperturbed trajectory is golden yellow.
}
}

A summary of the results for a range of $\mu$ and $t_i/t_0$ for bound
$30$ kpc orbits appears in figure \ref{fig-summary-ti-mu}.  The
geometric symbols characterize the outcomes for groups of orbits:
stars, boxes and triangles mean ``all bound'', ``all ejected'' and
``a mixture'' of both types, respectively.  The green diagonal line is an
analytic estimate for the critical $\mu$ based on
non-relativistic calculations in the following section which imply
that a loop is detached when the magnitude of its rocket impulse $a_r$ satisfies
$a_r \gta 0.3 |\nabla \psi|$. The transition from bound to unbound is
abrupt (a factor of $2$ in $\mu$ encompasses its entire scope)
stemming from the adiabatic invariance associated with averaging weak
forces over periodic orbits.

This figure summarizes the main physical constraints for binding and
residency of loops in the Galactic halo. There are upper bounds on the
formation time $t_i/t_0$ given by the horizontal lines. The condition
that cosmic drag lower the velocity to less than the circular rotation
velocity is given by the red line. The more stringent condition that
capture occur is given by the turquoise line.  There are upper bounds
on the string tension given by the diagonal lines. The condition that
the loop be younger than its gravitational wave decay timescale is
given by the red line.  The more stringent condition that the loop not
be accelerated out of the Galaxy is given by the green line. The
critical $G \mu/c^2$ below which capture and retention is possible is
given by the intersection of the green and turquoise lines. Appendix
\ref{appendix-fig}
includes a graph which replaces the straight lines with
numerically determined conditions for aligned and anti-aligned
rocket orientations.

The triangular region encompasses string tensions and formation times
for loops that are bound to our Galaxy today. The specifics of this
figure refer to loops at radius $30$ kpc, with initial velocity
$v_i=0.1$, and initial length $\alpha/H_i$ where $\alpha=0.1$ in a
radial infall model of the Galaxy. The manner in which the constraints
vary is briefly discussed in the caption.

\clearpage

\subsection{Critical Acceleration}

It is well known from classical mechanics that the action of a simple
harmonic oscillator with frequency $\omega$ is an adiabatic
invariant. Perturbative driving forces having intrinsic frequency
$\Omega$ such that $\Omega << \omega$ produce exponentially small
changes in the action or energy. Here, perturbative means that the
magnitude of the forcing is small, i.e. the instantaneous
change to the coordinate is first order. 
The integrated change of a first order quantity over a full period
is very small.

The rocket acts on the loop's orbit within the Galaxy. For loops which
have slowed enough to bind to the Galaxy, the Fermi transport of the
impulse direction (an analog of Thomas spin precession) has frequency
$\Omega \sim v^2 \omega \sim 10^{-6} \omega$.  Effectively, the
acceleration is in a fixed direction with magnitude $a_r$ governed by
the decrease in length of the loop. The condition for escape is
equivalent to the breakdown in adiabatic invariance of the oscillator
that occurs when the force grows sufficiently large to become
non-perturbative.

The main complication for an actual orbit is that the potential is not
separable and several incommensurate $\omega$'s exist (radial and
angular frequencies), so that one cannot solely focus on the motion in
the coordinate direction defined by the impulse.  However, it is
straightforward to investigate a simple, non-relativistic model having
all the essential features and to infer an approximate criterion for the
transition from perturbative to non-perturbative motion. Consider an
acceleration law of the form 
\be 
{\vec a} = \frac{-{\vec r}}{\left( r^2 +
  r_c^2 \right)^{\frac{n+1}{2}}} + {\vec k} .
\ee 
The interpretation here is
that the first term is the acceleration due to the galactic matter
distribution and the second term is the internal acceleration due to the
rocket.  The constants are core radius $r_c$, internal acceleration $\vec k$
and galactic acceleration power law $n$. For a Keplerian potential $n=2$, for
the radial infall model $n=5/4$ and for galactic potentials the range
of interest is $1<n<2$. To non-dimensionalize, express lengths in
units of $r_c$.

For a numerical investigation, first choose the initial radial
displacement of a zero-velocity particle and the size of the internal
acceleration, and sample random choices of direction $\hat k$.
The unperturbed orbit would remain radial if not for the
internal acceleration which drives it away from that limit.  Define $\chi
\equiv |k| (r^2+r_c^2)^{n/2}$ as a measure of the ratio of internal to
galactic forces and evaluate it along the {\it unperturbed} trajectory.
Next, integrate the actual orbit to determine
whether it remains bound to the center. After many samplings of $\hat k$, one
calculates $f_{bind}(\chi)$ the fraction of bound particles at a fixed
value of $\chi$.

Repeating this procedure for different initial positions ($5$-$20
r_c$) and power law shapes ($n=1$-$2$) allows comparison of the
importance of the various inputs to the calculation.  Figure
\ref{fig-non-relativistic-escape} displays $f_{bind}(\chi)$ for
different $n$. For a fixed $n$, the geometric
orientation of the rocket produces
an intrinsic spread in outcomes for $0.2 \lta \chi \lta 0.8$.  By
comparison, the entire range of $n$ corresponds to relatively small
shifts in $\chi$. This geometric effect is large and
irreducible compared to the uncertainties in the initial
position, power law shape and number of orbital periods.

To help gauge the importance of internal forces on the binding of
orbits, it is useful to define the critical ratio of internal to
galactic forces by $f_{bind}(\chi_{crit}) = 1/2$.  While there is
some variation, $\chi_{crit} \sim 0.3$ is a reasonable estimate. In
applying this result to the loop dynamics evaluate $\chi =
a_r/|\nabla \psi|$ at apocenter and regard the orbit as unbound if
$\chi>\chi_{crit}$.

These results have been used to speed up the large-scale numerical
calculations determining the probability of capture and the halo
profiles in the sections that follow. Essentially, the rocket effect
is ignored before capture and the retention and lifetime criteria at
the epoch of interest are imposed to determine the bound population of
living loops. Since the retention criterion is generally stricter than
the capture criterion little error is made and this allows a single
simulation to be used for multiple values of $G \mu$.

\FIGURE[h]{
\includegraphics{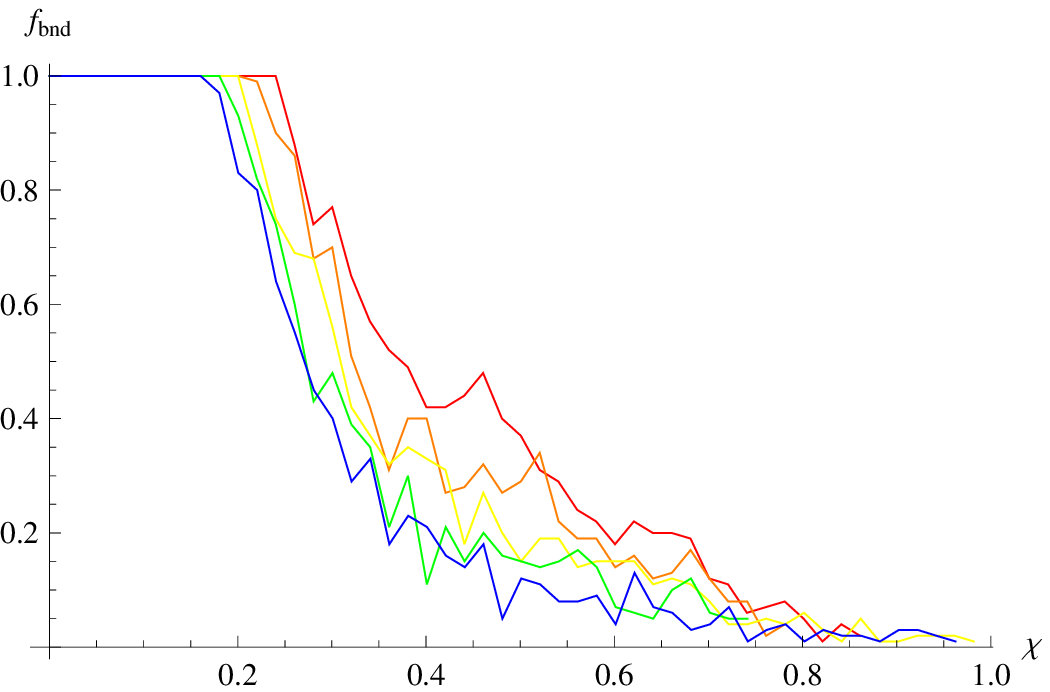}
\caption{\label{fig-non-relativistic-escape} The fraction of
radial orbits that remain bound as a function of $\chi$ for
power law exponents $n=1$, $5/4$, $3/2$, $7/4$ and $2$ (red,
orange, yellow, green and blue, respectively). All orbits began
at an apocenter of $20$ core radii; 100 random orientations
for the internal acceleration were drawn; for each choice the orbit
was integrated many characteristic orbital times; the curves
summarize the bound fraction.
}
}

\section{Network Evolution}

The halo will contain loops born with a variety of positions,
velocities, times of formation and lengths. This section discusses the
birth rate density for loops generated by a scaling network. Succeeding
sections discuss the average probability of capture for loops created
at a given epoch and the net effect when the birth rate density is
integrated over all length and time scales.

\subsection{Status of Large and Small Loops}

\def\stringnetworkearly{\cite{albrecht_evolution_1985,bennett_evidence_1988,albrecht_evolution_1989,bennett_high_1990,allen_cosmic-string_1990}}

Early simulations of Nambu Goto strings\stringnetworkearly 
successfully tackled the large
scale properties of the network, in particular, the relation of
horizon to correlation length, characteristic spacing and persistence
length. They validated the original idea that the network would evolve to
reach a scaling solution in 
simulations\cite{kibble_evolution_1985}. The attraction of arbitrary
initial conditions to a
scaling solution insures that the energy density in long,
horizon-crossing strings never comes to dominate the total energy
density of the universe. This is a necessary but not
sufficient condition to avoid overclosing the universe since loops
formed during the radiation era must decay by {\it some} mechanism
to avoid a monopole-like problem. For strings that couple
only to the gravitational field the decay must involve
gravitational wave emission.

Loops are created when the horizon crossing strings are chopped up. It
was originally thought that loops would form by intercommutations of
the long strings at the characteristic scale of the horizon (the
Kibble one-scale model). Since the early studies found clear signs of
a gas of small loops and dense kink-filled string segments at the
simulations' resolution limit it became
apparent that some basic understanding had yet to be achieved.  In a
perfect scaling solution all the properties of the network, not just
those close to that of the horizon, should scale.  Characterizing
the small scale structure and searching for evidence of its scaling
has been an ongoing effort.

The significance of the small scale structure has now become more
apparent.  It turns out that rather than being a detail to be
disentangled from resolution issues, the small scale structure is
inextricably intertwined with the evolution of the network on all
scales less than the horizon. Small scale structure governs many
potentially observable features of the network and ultimately has a
considerable impact on the expected halo clustering properties of
loops. This section briefly reviews areas of significance for the
problem of loop clustering.

\def\stringnetworkbigloops{\cite{martins_fractal_2006,ringeval_cosmological_2007,olum_cosmic_2007}}

Current numerical simulations of NG strings generate a continuum
distribution of sub-horizon scale loops including some loops that are
much bigger than the resolution limit\stringnetworkbigloops.  Although
non-trivial differences exist between the most recent simulations a
point of common agreement is that the abundance of loops near the
horizon scale is greater than found in the earlier, lower resolution
work. Roughly $\sim 10$\% of the length of long strings ends up in
loops with length $10^{-4}< l/t <
10^{-1}$. For the purposes of this paper, these are all ``large''
loops because they imply $H \tau >> 1$ for $G \mu/c^2<<10^{-7}$.  A
second area of mutual agreement is that the population of these large
loops is judged to be scaling\footnote{The extent of scaling and the criteria for scaling
  differs. See ``Note added'' in
  Ref. \cite{ringeval_cosmological_2007} for a comparison.}
and is not a transient
artifact.  The
presence of a population of large loops hearkens back to the original
expectation that the horizon scale would determine the properties of
newly formed loops.

The traditional interpretation of the large loop part of the
distribution is that it forms via a sequence of multiple encounters in
which a horizon-scale loop is cut into smaller and smaller
progeny. Some of these encounters are with the long, horizon crossing
component, some with other loops and some are self-intersections in
which a loop oscillates so that individual parts collide. Ref
\cite{martins_fractal_2006} indicates that self-intersection of larger
loops is the dominant overall loop production mechanism.  An
assessment of the clustering of such loops is carried out in this
paper.

\def\pol{\cite{polchinski_analytic_2006,polchinski_cosmic_2007,polchinski_cosmic_2007-2,dubath_periodic_2007,rocha_scaling_2008,dubath_cosmic_2008,polchinski_small_2008} }

Both new and old simulations also contain large numbers of small
loops.  Early calculations suggested
\cite{albrecht_smallscale_1990,siemens_cosmic_2003} and recent studies
have demonstrated that tiny loops are produced in great abundance by the
interaction of small scale structure on string segments as those
segments first form a large scale cusp
\cite{polchinski_analytic_2006,dubath_cosmic_2008}.  The small scale
structure today owes its existence to non-linear interactions that the
string experienced in the past at the time it first entered the
horizon. Ref. \cite{polchinski_analytic_2006} quantified the structure
in terms of a correlator $\propto (l/t)^{2 \chi}$ for size scale $l$
at time $t$ and showed that $\chi$ is completely determined by mean
network properties (rate of expansion of the universe and rms velocity
of the strings).  The slope of the spatial correlation function of
simulations agrees well with the theoretically-determined $\chi$ over
the expected range.

Most loops (measured in terms of number or total length) are created
at small physical scale with a cutoff set by
gravitational damping and theoretically derived to be $\alpha \sim (G
\mu/c^2)^{1+2\chi}$ where $\chi=0.1$ or $0.25$ for radiation or matter
respectively\cite{siemens_gravitational_2001,
  siemens_size_2002,polchinski_cosmic_2007}. Numerical simulations do
not include gravitational wave damping so it isn't possible to verify
directly the cutoff prediction. However, the slope of the loop size
distribution depends upon $\chi$, is insensitive above the cutoff
to the actual value of the cutoff and
agrees well with that found in simulation\cite{dubath_cosmic_2008}.

It appears that roughly 90\% of the horizon crossing string goes into
tiny loops \cite{polchinski_cosmic_2007-2,polchinski_small_2008}.
There are many factors which will end up influencing the clustering of
small loops.  While there are more small loops than large ones, small
loops evaporate more quickly than large ones. Loops at the
gravitational wave cutoff are ultra-relativistic
\cite{polchinski_analytic_2006}.  Time-dilation and cosmic
expansion effects (eq. \ref{eqn-loft}) in the ultra-relativistic
limit imply
$H \tau \sim \left( \alpha c^2/G \mu \right)^{1/(1+\nu)} \sim \left( G
\mu/c^2 \right)^{2 \chi/(1+\nu)}$. In the matter era, $\nu=2/3$,
$\chi=0.25$ so $H \tau \sim (G \mu/c^2)^{0.3}$; in the radiation era,
$\nu=1/2$, $\chi=0.1$ and $H \tau \sim (G \mu/c^2)^{0.13}$. All these
results suggest that loops at the cutoff may be unable to cluster but
the small power of $G \mu/c^2$ means
$H \tau$ is never very small and the significance of 
loops larger than the cutoff (which live
longer) is murky. The multiplicity of factors at play suggests that a detailed
calculation of clustering should be carried out.

\subsection{Birth Rate Density}

The long horizon crossing strings are chopped into loops at a
rate
\be
\left( \frac{d \rho}{dt} \right)_{\infty \to {\rm loops}} = \frac{2 \rho_\infty }{t} \left( 1 - \nu \left( 1 + {\bar v}_\infty^2 \right) \right)
\ee
where $\rho_\infty = \mu/(\gamma_s t)^2$ is the energy density
in long strings with typical separation $\gamma_s t$ and mean square
velocity ${\bar v}_\infty^2$ \cite{martins_quantitative_1996}.

As long as a scaling solution is achieved when horizon $\propto t$
(true for power law expansion in radiation and matter dominated eras
but not applicable to the recent $\Lambda$-dominated phase) then the
average birth rate of loops of physical size $l$ per physical volume $V$ at
time $t$
\be
\frac{dN}{dl dt dV} = \frac{f(x)}{t^5}
\ee
where $x = l/t= \alpha/(H t)$ for some function $f(x)$.

The loop formation process involves interactions within the network
and, simultaneously, stretching of the horizon crossing strings and
expansion of the universe.  At least a few expansion times ($\sim
1/H$) are needed for intercommutations to transform long string
segments into sub-horizon loops. Once sub horizon loops are formed the
probability for loop-loop interactions decreases and the loop achieves
a fixed physical size in a few more expansion times. The assumption in
this paper is that during each infinitesimal time interval $(t,t+dt)$
the network produces the loops implied by $dN/dldtdV$. The loops have
physical scale that changes only due to gravitational
radiation; they suffer no further intercommutation. This
prescription provides the distribution of initial conditions for the
clustering calculation.

In principle, the full joint distribution of
$dN/dldtdV$, loop center of mass momentum and rocket direction in the center
of mass frame is needed to realize the initial conditions for the
dynamical motion of a population of loops in the background FRW
cosmology. In lieu of a detailed description, assume a factorized form
for the joint distribution
\ba
\frac{dN}{dldtdVd{\vec v}d{\hat\Omega_r}} 
& = & \frac{dN}{dldtdV} \frac{dP}{d{\vec v}} \frac{dP}{d{\hat \Omega_r}}
\ea
where $\vec v$ is the loop center of mass momentum and
${\hat\Omega}_r$ the direction of the rocket. Here, $dP/d(..)$ means
the differential probability for the initial condition of $(..)$ with
unit normalization. In practice, the correlations between the
magnitude of the momentum $v$ and $l/t$ is retained but all
correlations between the direction of the momentum, the direction of
the impulse and $l/t$ are discarded. 

In the homogeneous limit number $N$ and length $L$ of loops created in
comoving volume $\tilde{V} = V_0/a_0^3$ (i.e. the
comoving volume implied by a physical volume $V_0$ today;
$a_0=a(t_0)$) are
\be
\left(
\begin{array}{c}
N \\
L
\end{array}
\right)
= 
\frac{V_0}{ t_0^3 } 
\int \frac{dy'}{y'^4} \left( \frac{a'}{a_0} \right)^3
\int dx'
\left(
\begin{array}{c}
f(x') \\
f(x') x'
\end{array}
\right)
\ee
where $y=t/t_0$, $x=l/t$ and $a'=a(t')$.
By the current time some loops have evaporated and all have shorter lengths. 
The distribution of loops with length is
\be
\label{eqn-dnadl}
\frac{dN_A}{dl} = 
\frac{V_0}{ t_0^3 } 
\int \frac{dy'}{y'^4} \left( \frac{a'}{a_0} \right)^3
\int dx' f(x') 
\int d{\vec v}' d{\Omega_{{\hat r}'}} \frac{dP}{d {\vec v}'} \frac{dP}{d {\Omega_{{\hat r}'}}} \delta \left( l - L \right) \theta_A
\ee
where $L=L [ {\vec v}', l', {\Omega_{{\hat r}'}}, t' ; t_0]$ is length
at $t_0$ in terms of the initial loop parameters and, similarly,
$\Theta_A = \Theta_A [ {\vec v}', l', {\Omega_{{\hat r}'}}, t' ; t_0]$
is 0 or 1 depending upon whether the loop has reached the end of its
life or not. The dependence of $L$ and $\Theta_A$ on dynamical variables
may be traced to relativistic effects that link FRW and loop center of mass
frames. Quantities $N_A$ and $L_A$ are defined by integrals over $dN_A/dl$.

When the loop's center of mass motion
is only mildly relativistic then
the loop lives until $t_{life} = t' + l'/(G
\mu \Gamma_E)$. 
Hence $\Theta_A=1$ for $t<t_{life}$ and the 
loop length is $L=l' -
\Gamma_E G \mu (t-t')$ independent of the loop dynamics.
To summarize: for non-relativistic kinematics and ignoring the
dynamical influence of the rocket the forms for $L$ and $\Theta_A$
simplify: $L=L[l' , t'; t_0]$ and
$\Theta_A=\Theta_A[l' , t' ; t_0]$ giving
\be
\frac{dN_A}{dl} = 
\frac{V_0}{ t_0^3 } 
\int \frac{dy'}{y'^4} \left( \frac{a'}{a_0} \right)^3
\int dx' f(x')  \delta \left( l - L \right) \theta_A .
\ee
The integration over $x'=l'/t'$ follows directly since
occurrences of $l'$ are now easily rewritten in
terms of $l$, $t_0$ and $t'$.

\subsection{Fragmentation and Cusp-Mediated Loop Formation}

Until this point, the network evolution has been treated in an
agnostic fashion with respect to the scaling solutions for matter and
radiation eras. For the purposes of presenting numerical results, the
focus tightens to matter era network evolution.  This choice avoids
any potential inconsistency with respect to the growing galactic
perturbation for which matter era dynamics are most appropriate. It
also simplifies and streamlines the presentation. However, most loops
in the halo today were born before equipartition and the absolute
numbers of such loops depend upon the radiation era expansion dynamics
\cite{chernoff_cosmic_2007}.  This paper concentrates on {\it
  enhancement}, i.e. the ratio of clustered to homogeneously
distributed loops which is expected to be less sensitive to expansion
dynamics.

Large loops are created by hierarchical fragmentation; small loops by
cusps interacting with pre-existing small scale structure. The processes
are both active at the same time.
The description for each mechanism begins with the power law form
\be
f(x) =  \left\{
\begin{array}{cc}
A x^{-\beta} & {\rm ~if} ~~x_L< x < x_U \\
0 & {\rm else}
\end{array}
\right.
\ee
which depends upon $A$, $\beta$, $x_L$ and $x_U$. 
Individual $f$'s for individual mechanisms are
weighted by the fraction $\delta$ of the total
energy loss rate by horizon crossing string 
$(d\rho/dt)_{\infty \to {\rm loops}}$ that ends up in
loops created by a given mechanism. Energy balance gives
\be
A = \delta \frac{2 (2 - \beta)}
{\gamma_s^2 \left( x_U^{2-\beta}-x_L^{2-\beta} \right) }
\left( 1 - \nu \left( 1 + {\bar v}^2_\infty \right) \right)
\ee
In the non-relativistic matter era, $\nu=2/3$,
$\gamma_s^2 = 10^{-1}$ and ${\bar v}^2_\infty = 0.35$.

Refs. \cite{polchinski_analytic_2006,dubath_cosmic_2008} 
explored the cusp-mediated mechanism and
combined theoretical arguments and simulation-derived
average quantities to infer in the matter era
\ba
\chi & = & 0.25 \\
\beta & = & 3 - 2 \chi \\
x_L & = & \frac{l_{gw}}{t} = 20. \left(G \mu \right)^{1+2 \chi} .
\ea
For the cusp-mediated mechanism $f$ is strongly tilted to small scales;
the upper cutoff $x_U$ has little effect.
Ref. \cite{polchinski_analytic_2006} derived the means square
velocity distribution for newly formed loops
\be
\label{eqn-vrms}
\langle \beta^2 \rangle = 
1 - \frac{2 B}{(2 \chi+1)(2 \chi + 2) } \left( \frac{l}{t} \right)^{2 \chi}
\ee
where $B=0.61$.

In this paper, the theoretically derived $f$ is adopted to describe
cusp-mediated loop production.  The cutoff $x_U$ is adjusted
freely. Ref. \cite{polchinski_analytic_2006} noted that a puzzling
discrepancy exists between the above expression for $\langle \beta^2
\rangle$ and previous, simulation-derived dispersions
\cite{bennett_two-point_1989} with the
theoretical result being larger. Since loops with higher velocity must
experience more damping before 
they are able to cluster adopting the theoretical
expression gives the ``worst case'' scenario for small loop
clustering. The implications of a reduced $\langle \beta^2 \rangle$ will
be investigated as well.

Ideally a similar approach for the hierarchical formation mechanism
should be followed.  Fits for $f$ for large loops in an expanding
cosmology are not generally available and a systematic comparison of
existing simulation results is lacking.  One network simulation
\cite{olum_cosmic_2007} gave a scaling, unimodal distribution at $x
\sim 0.1$ for $f$ but another \cite{ringeval_cosmological_2007} lacked
the peak and generated an approximate power law form for $f$ at
smaller $x$.

A potential practical complication is the extent the cusp-mediated contribution
interferes in fits to $f$ designed to characterize the fragmentation
mechanism. Figure 3 in \cite{dubath_cosmic_2008} showed that
cusp-mediated loop production traces $x^2 f(x)$ from the simulation
\cite{olum_cosmic_2007} over the range $x < 10^{-2}$. This
subsumes a significant part of the loop range termed ``large'' here.

In sum, the form for $f$ for the fragmentation mechanism for large
loops has not yet been sufficiently well-characterized to yield
specific values for $\delta$, $\beta$, $x_L$ and $x_U$.\footnote
{Nor is it clear that a truncated power law form will ultimately be
  sufficient to trace $f$. The segment of curve $10^{-2}<x<10^{-1}$ in
  Figure 3 (\cite{dubath_cosmic_2008}) may not be well fit with a
  simple power law.  } The values of $\delta$ and $x_U$
are the most important and the most uncertain input for many
purposes.

In this paper $\delta$, $\beta$, $x_L$ and $x_U$ are regarded as free parameters
for assessing the impact on loop clustering.  Simulation-based results
for $\langle \beta^2 \rangle$ as a function of $l/t$
\cite{bennett_two-point_1989} were fit using the
same form as eq. \ref{eqn-vrms} in a purely empirical manner. In the
matter era the results are $B=1.46$ and $\chi=0.114$. This fit
for large loops formed by fragmentation will
also be employed as an alternative to the theoretically derived
$\langle \beta^2 \rangle$ for small loop motions.

For all mechanisms, the loop momentum distribution ($v$)
is assumed to be thermal in the FRW frame, i.e.
\ba
\frac{dP}{d{\vec v}} & = & \frac{dP}{dv} \frac{dP}{d\Omega_{\hat v}} \\
\frac{dP}{dv} & = & \frac{v^2 e^{-\kappa E}}{\int dv' v'^2 e^{-\kappa E'}}  \\
E(v) & = & \sqrt{1+v^2} \\
\frac{dP}{d\Omega_{\hat v}} & = & \frac{1}{4 \pi}
\ea
where $\kappa=\kappa(x)$ is set by
requiring agreement with the fits to the dispersion $\left< \beta^2 \right>$.

The direction of rocket impulse is assumed to be isotropic in the loop
center of mass frame.

\section{Loop Clustering}
\label{sec-loop-clustering}

This section describes the halo profile formed by loops which are born
at a single epoch and with a fixed $l/t$. 

\subsection{Probability of Capture}

Consider the probability that a single loop formed at time $t_i$ in a
large but arbitrary comoving volume $\tilde{V}$ ends up
today bound to the galaxy with physical semi-major axis $r$.  Let
$\Delta N$ be the number of loops formed in infinitesimal time
interval $t'$ to $t'+dt'$ for $t'=t_i$.

First, write out the formal probability that the loop
has not evaporated and is bound
with position $\vec x$, momentum $\vec v$ and length $l$ today
\ba
\frac{dP_{AB}}{d{\vec x}d{\vec v}d{l}} & = & 
\frac{ \frac{d}{d{\vec x}d{\vec v}d{l}}{\Delta N_{AB}}}{ \Delta N} \\
 & = &
\frac{1}{\tilde{V}}\int d{\vec x'} d{\vec v'} d{\Omega_{{\hat r}'}}
\frac{dP}{dl'} \frac{dP}{d\vec v'} \frac{dP}{d \Omega_{{\hat r}'}}
\delta^3({\vec x} - {\vec X}) \delta^3({\vec v} - {\vec V}) \delta(l-L) \theta_{AB} \\
\frac{dP}{dl'}          & = & \frac{f}{\int f dl} = \frac{f}{t' F} \\
F                       & = & \int f(x) dx
\ea
The initial variables are position ${\vec x'}$ (assumed homogeneous),
velocity ${\vec v'}$, length $l'$. Here ${\vec X}={\vec X}({\vec x'},
{\vec v'}, l', {\Omega_{{\hat r}'}}, t'; t_0)$ is the formal
time-dependent solution for position, likewise for ${\vec V}$ and
$L$. The function $\theta_{AB}$ is 1 if the loop is alive (has not yet
evaporated) and bound to the perturbation and 0 otherwise. The
probability that the loop has initial length $l'$ is $dP/dl'$ and
similarly for the other initial variables. Here, $F$ is
a normalization constant.

Evaluate this integral by Monte-Carlo methods for $t'=t_i$: first,
sample $l'$, ${\vec x}'$ and
${\vec v}'$ and then use direct numerical integration to evaluate
the final positions and momenta.  Finally, marginalize the
multi-dimensional distribution and focus solely on semi-major axes $r$
and current loop length $l$:
\ba
\frac{dP_{AB}}{dr dl} & = & \int d{\vec x}d{\vec v} \frac{dP_{AB}}{d{\vec x}d{\vec v}d{l}} \delta(r - R) \\
\frac{dP_{AB}}{dr} & = & \int dl \frac{dP_{AB}}{dr dl} 
\ea
where $R=R({\vec x},{\vec v},t_0)$ is the formal expression for the
semi-major axis in terms of the current phase space coordinates.  A
more detailed description is given in Appendix
\ref{appendix-monte-carlo}.

The total probability the loop is bound today
is proportional to $1/\tilde{V}$. As a basis for comparison, consider
the case of a cold dark matter 
particle: it is judged to be bound if it lies
within today's comoving turn-around volume $\tilde{V}_{ta}$.
Scale the differential and cumulative forms by the same factor:
\ba
\frac{dQ_{AB}}{dr} & = & \frac{\tilde{V}}{\tilde{V}_{ta}} \frac{dP_{AB}}{dr} \\
\label{eqn-q}
Q_{AB}(<r) & = & \int_0^r \frac{dQ_{AB}}{dr} dr
\ea
The combination $n V_{ta} Q_{AB}(<r)$ is the expected number of bound objects
today within distance $r$ for a mean homogeneous density $n$
and turn-around volume $V_{ta}$. 
By construction, cold dark matter has $Q_{AB}(<r_{ta}) = {\cal
  M}(\lambda = 1) = 9 \pi^2/16 = 5.55$. The fact that $Q_{AB}(<r_{ta}) > 1$
shows that the perturbation attracts distant particles so that the total
within today's comoving turn-around volume is larger than the 
number in an equivalent volume far from the perturbation center.

The mean interior density is
\be
{\bar n_{AB}} = \frac{Q_{AB}(<r)}{\lambda^3}
\ee
also a useful measure of the clustering and highlights the
part of the distribution near the center.

\subsection{Halo Profiles}

Integrating from $t'=t_i$ to the present gives the
expected differential
number of loops in comoving volume $\tilde{V} = V_0/a_0^3$ today 
\ba
\frac{dN_{AB}}{d{\vec x} d{\vec v} dl} & = & 
V_0 F \int \frac{dt'}{(t')^4} \left( \frac{a'}{a_0} \right)^3
\frac{dP_{AB}}{d {\vec x} d{\vec v} dl } \\
& = & 
V_{ta} F \int \frac{dt'}{(t')^4} \left( \frac{a'}{a_0} \right)^3
\frac{dQ_{AB}}{d {\vec x} d{\vec v} dl } \\
\frac{dN_{AB}}{dr dl} & = & 
V_{ta} F \int \frac{dt'}{(t')^4} \left( \frac{a'}{a_0} \right)^3
\frac{dQ_{AB}}{dr dl }
\ea
The differential probabilities include a factor $1/F t'$.

Objects must not have fully evaporated ($\Theta_A=1$) and must be bound 
($\Theta_B=1$) to the perturbation to contribute to these
distributions. The length is integrated as a variable so its straightforward
to evaluate $\Theta_A$. In practice, deciding whether a loop is bound
amounts to checking whether the orbit has experienced multiple passages through
the perturbation center (capture). If captured, the integration is suspended but
it must still be determined when the rocket effect detaches it. If
the loop is captured and not detached at the epoch of interest it is
called bound.

\subsection{Truncation by Rocket}

A loop remains bound with approximately fixed semi-major axis until
the internal acceleration exceeds that due to the gravitational
potential. The condition for escape is $\chi > \chi_{crit}$ at which
point the loop leaves quickly on an orbital timescale.

When the rocket is ignored, the geodesic trajectory is independent of
$l$ and $\mu$ i.e. just the motion of a test object.\footnote{ The
  probability distributions $dP_{AB}/drdl$ and $dQ_{AB}/drdl$ depend
  upon $l$ not just because of $f$ but also
because of the correlation between the initial velocity and
  loop length.}  As the loop shrinks, the rocket acceleration grows
monotonically $\propto \mu/l$ while the acceleration at apocenter for
a captured loop is constant. Consequently, eventually
$\chi>\chi_{crit}$. For loops with non-relativistic velocities
\ba
\chi     & = & \frac{a_R}{a} \\
a_R      & = & \Gamma_P \frac{G \mu}{l} 
           =   \frac{\Gamma_P}{\Gamma_E} \frac{1}{t_{life}-t} \\
a        & = & \frac{G M_x}{r_{ap}^2} 
           =   \frac{4 \pi}{3} G \rho_H(t_0) r_{ta} \frac{{\cal M}_x}{\lambda^2}
\ea
where $r_{ap}$ is the physical apocenter, the loop lives until
$t_{life}  = t + l/(G \mu \Gamma_E)$ and
$\lambda = {r_{ap}}/{r_{ta}}$. In practice, $t$, $r_{ap}$ and $l$ are
determined at the instant the loop is judged as bound so that the
non-relativistic approximation is good.

The acceleration $a \propto {\cal M}_x/\lambda^2$ is monotonically
decreasing with $\lambda$ or $r_{ap}$. For bound
loops with given $t_{life}$ (i.e. time of formation, size and tension)
there is a single value for $r_{ap}$ today which
satisfies $\chi=\chi_{crit}$.  Denote the solution $r_{ap,cr}
(t',l',\mu)$.

Assume an ejected loop instantaneously leaves the galaxy.  Loops with
$r_{ap}>r_{ap,cr}$ have been lost; loops with $r_{ap}<r_{ap,cr}$ are
still in the halo. To summarize: $\Theta_B$ contains a
factor $\Theta\left( r_{ap} < r_{ap,cr} \right)$ which accounts for
loop loss by the rocket.

\subsection{Results}

The cumulative distribution $Q_{AB}(<r)$ was calculated for specific
choices of $\mu$, formation time $t'$ and length $l' =\alpha c/H'$
and results are displayed in figure \ref{fig-dqdr-1}-\ref{fig-dqdr-3}.  Recall
that $Q_{AB}(<r)$ measures the expected number of bound objects within
distance $r$ for a mean homogeneous density equal to one object per
turn-around volume $V_{ta}$. 

The figure's black line is $Q$ for collisionless cold dark matter as
calculated in the self-similar radial infall model. It serves as a
standard of comparison for the more complicated infall scenario in
which loops need to slow down to be captured and eventually are
ejected by the rocket effect.  The colored lines give $Q$'s for loops
with different formation epochs.  All have string tension $G \mu
=10^{-13}$ and length $l' = \alpha /H'$ for fixed
$\alpha=10^{-1}$. Loops are a fixed fraction of the horizon size at
birth; older loops are smaller. The initial velocities were drawn from
a fixed thermal distribution with $\langle \beta^2 \rangle = 0.2$ (this is
arbitrary
and not directly tied to any of the string network estimates).

Early formation gives a profile that closely matches the cold dark
matter one at small $r/r_{ta}$ (leftmost lines). Such loops have had
plenty of opportunity to damp by cosmic drag so they cluster just
like cold dark matter. Note the empty circle at which some curves
end. The profile is truncated at larger radii because the rocket
effect is able to unbind orbits at apocenter. To summarize: for fixed
$\alpha$, the oldest loops are smallest, closest to the end of their
lives, feel the largest rocket effect and may be retained only by the
centermost parts of the potential.

For loops that are not as old, outer regions of $r/r_{ta}$ are
accessible. Note that many curves end near $r/r_{ta} \sim 0.1$ {\it
  without} an empty circle. The endpoint is not a
consequence of physical ejection but of the minimum time needed to
bind an infalling loop to the perturbation.  Unlike a cold dark matter
particle which is known {\it a priori} to be bound, a loop is judged bound
only after it has passed back and forth several times through the
center.

With a limited number of particles the Monte-Carlo calculation always
has an innermost radii. However, arbitrarily small velocities are
present in the initial conditions, so arbitrarily small turn-around
radii are possible.

\FIGURE[h]{
\includegraphics{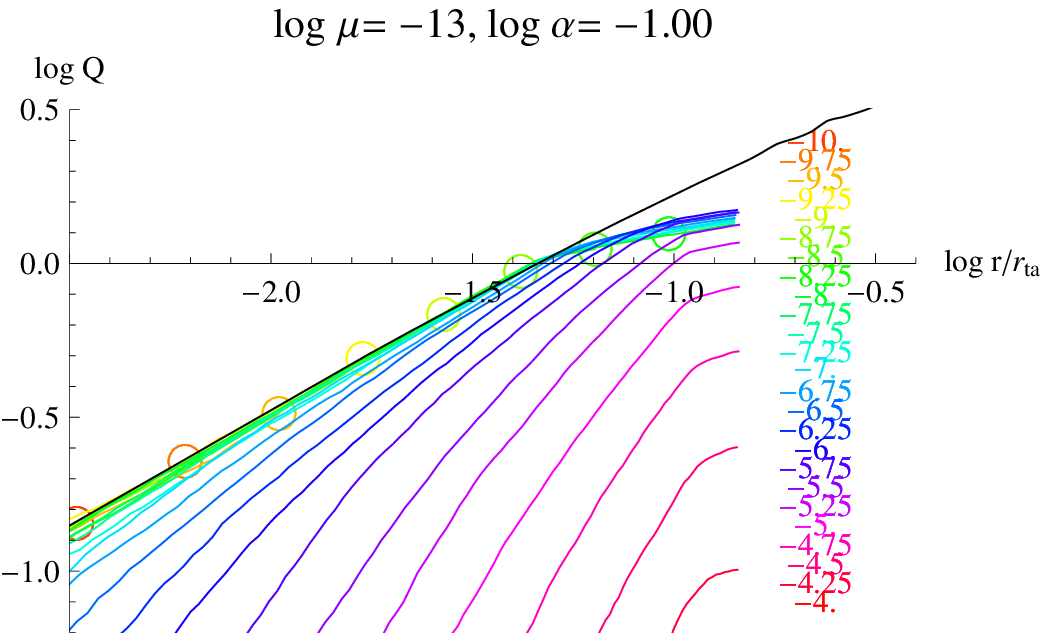}
\caption{\label{fig-dqdr-1} The cumulative distribution 
$Q$ as a function of $r/r_{ta}$. 
The black line is $Q$ for collisionless cold dark matter according to
the radial infall model. It includes material infalling for the first
time as well as regions of multi-streamed flow ($r/r_{ta}<0.36$).
The colored lines show $Q$ for loops formed at different
epochs which are known to be bound to the perturbation. 
All have $G \mu=10^{-13}$ and $\alpha=10^{-1}$
and formation lengths $l' = \alpha /H'$. The time of formation
varies $t'/t_0 = 10^{-10}-10^{-4}$; the color key is
$\log_{10} t'/t_0$ written on the right. Note, for example, that
the orange line closest to the center is formation at the
earliest epoch while the
red line at larger $r/r_{ta}$ is the most recent.
These indicate the degree of clustering in
loops compared to the cold dark matter case.
(1) Old loops closely track the cold dark matter.
(2) Empty circles indicate the truncation of the halo's loop profile
because $r_{ap} > r_{ap,cr}$, i.e. the rocket effect strips
the loops further out.
(3) $Q$ is not plotted beyond $r/r_{ta} \sim 0.1$ -- not enough
time has passed to satisfy
the criteria that the loop is bound (that it pass through
the perturbation center several times). 
(4) Profiles which terminate at small $r/r_{ta}$ do so
only because of the limited
number of particles used in the calculation or the limited extent of
the figure; actual profiles extend to the center.
}
}

Compared to figure \ref{fig-dqdr-1}, figure \ref{fig-dqdr-2} has
smaller $\alpha=10^{-2}$ while figure \ref{fig-dqdr-3}
has larger $G \mu=10^{-12}$. To the extent that the rocket
effect is ignored prior to ejection the $Q$'s are the same (all were
constructed from the same data). The only impact of these changes is
to shorten the loop lifetime, so removing some of the lines, and to
decrease $r_{ap,cr}$, truncating the profile at a smaller radius.

Figure \ref{fig-drhodr-1} presents the mean interior density ${\bar n}$
for the string loops compared to that of the cold dark matter.
Evidently, old loops with small $\mu$ have sufficient time to
cluster strongly.

\FIGURE[h]{
\includegraphics{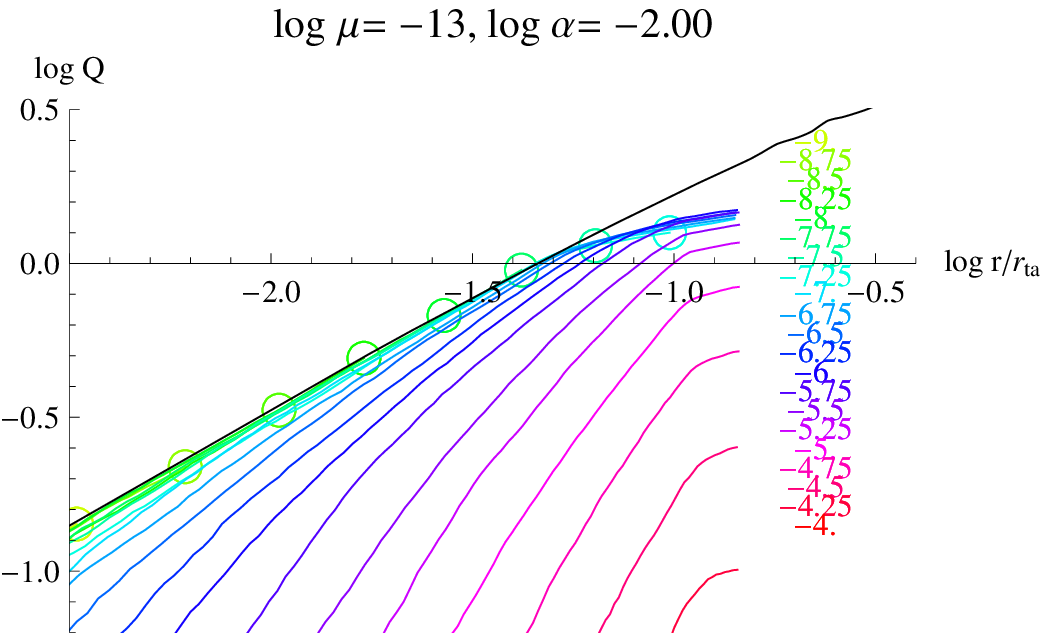}
\caption{\label{fig-dqdr-2} The cumulative distribution 
$Q$ as a function of $r/r_{ta}$. Same as \ref{fig-dqdr-1}
except $\alpha =10^{-2}$.
}
}

\FIGURE[h]{
\includegraphics{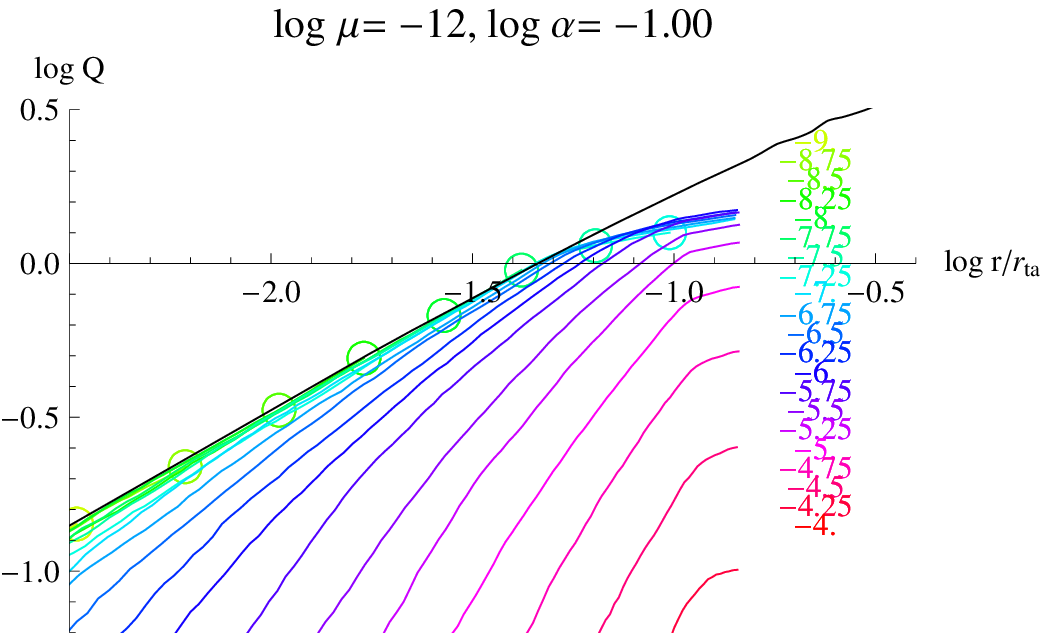}
\caption{\label{fig-dqdr-3} The cumulative distribution 
$Q$ as a function of $r/r_{ta}$. Same as \ref{fig-dqdr-1}
except $G \mu =10^{-12}$.
}
}

\FIGURE[h]{
\includegraphics{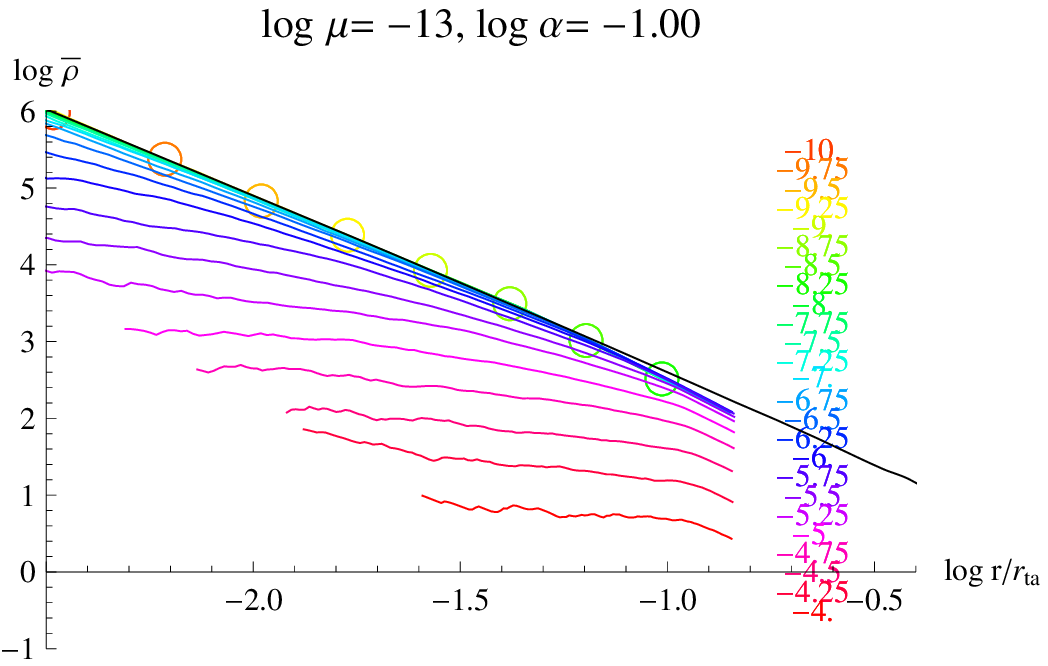}
\caption{\label{fig-drhodr-1} The number density within
a radius $r/r_{ta}$ for $G \mu =10^{-13}$ and $\alpha=10^{-1}$
(same as Figure \ref{fig-dqdr-1}).
}
}

\clearpage

\section{Current Halo Profile}
\label{sec-halo-profile}
The actual halo profile is more complicated than the examples
constructed in the previous section because it involves
loops of many sizes created over a range of epochs. Results for
two specific formation scenarios are illustrated in this section.

\subsection{Measures of the Loop Distribution}

The cumulative distribution $N_{AB}(<r) = \int dl \int_0^r dr
dN_{AB}/dl dr$ and the length-weighted cumulative distribution
$L_{AB}(<r) = \int l dl \int_0^r dr dN_{AB}/dl dr$ provide summary
information about the number and total length of all loops
bound to the galaxy. These quantities are normalized
with respect to $N_A$ and $L_A$, the equivalent quantities
expected to be present in homogeneous space (eq. \ref{eqn-dnadl}).
A measure of the cumulative number of loops bound to the galaxy
compared to the total within the turn-around volume is 
\be
{\cal Q}(<r) = \frac{N_{AB}(<r)}{N_A(<r_{ta})}
\ee
If all loops behaved dynamically like cold dark matter
particles then one would expect ${\cal Q}(<r)$ to resemble
$Q_{AB}(<r)$ (eq. \ref{eqn-q}). 

In a similar manner, start with the average number density of
alive, bound loops within radius $r$
\be
{\bar n_{AB}}(r) = \frac{N_{AB}(<r)}{V(r)} 
\ee
and the average number density of alive loops 
in homogeneous space ${\bar n_A} = N_A(<r)/V(r)$.
A measure of the average overdensity of loops bound
to the galaxy is
\be
{\bar {\cal N}}(<r) = \frac{{\bar n_{AB}}(<r)}{\bar n_A}
\ee
To the extent that loops behave like cold dark
matter then one expects ${\bar {\cal N}} (<r)$ to vary
like $Q_{AB}(<r) (r_{ta}/r)^3$.

Let $X$ be the average over the length distribution of some
function; denote by $X^{(i)}$ the length-weighted
moment of the function times $l^i$ for $i=1$, $2$,... The
first moment
corresponds to total length or energy.

The substitutions $N_{AB} \to N_{AB}^{(1)}=L_{AB}$ and $N_A \to N_A^{(1)}=L_A$
leads to cumulative and density with respect to energy instead
of numbers. In an obvious fashion, let
${\cal Q}(<r) \to {\cal Q}^{(1)}(<r) = \frac{L_{AB}(<r)}{L_A(<r_{ta})}$,
${\bar n_{AB}}(r) \to {\bar n_{AB}^{(1)}}$,
${\bar n_{A}} \to {\bar n_{A}^{(1)}}$,
and ${\bar {\cal N}}(<r) \to {\bar {\cal N}^{(1)}}(<r)$.

\subsection{Point: Large Loops from Fragmentation Model}

A model in which all the long string length goes into
sub-horizon-scale loops will be discussed first. The model parameters
are $\delta_{frag}=1$, $\alpha_L=10^{-3}$, $\alpha_U=10^{-1}$, $\mu =
10^{-15}$-$10^{-9}$ with $f(x) \propto x^{-\beta}$.  
Let $g(x,t)$ be integrated over the birth rate density:
\ba
I[g] & = & \int dl \int dt \frac{f\left( \frac{l}{c t} \right) }{c^4 t^5} a(t)^3 g(x,t) \\
      & \propto & \int dx x^{-\beta} \int dt t^{3 \nu - 4} g(x,t)
\ea
For $g=1$ the integral $I[g]$ is proportional to total number
of loops born. For matter (radiation) era $\nu=2/3$ ($1/2$) 
the result varies like $1/t$ ($1/t^{3/2}$)
and the number of loops is dominated by the earliest epochs. For
$g=l= x t$, $I[g]$ is proportional to total length of loops born
and the result varies like $\log t$ ($1/t^{1/2}$).
For $\beta<1$ large $x$ loops dominate as measured by
number and by length.

The cumulative number $\cal Q$ is shown in figure
\ref{figure-final-cdfN-15-0wwo} for $\mu=10^{-15}$ with and without
the cutoff imposed by the gravitational wave recoil. $Q_{AB}$ for cold
dark matter (the black line) provides a point of reference. All three
cumulative distributions are close at small radii ($r/r_{ta} <
10^{-2.5}$ or approximately $<3$ kpc). The rocket is effective at
depleting the old and hence small loops that would otherwise be
present at large radii.  Since loop numbers are dominated by formation
at the earliest epochs, the characteristic signature of the recoil
effect is the depletion of large numbers of loops at large radii.  The inner
regions are not immune to the depletion but it is less severe.

\FIGURE[h]{
      \includegraphics[width=0.8\textwidth]{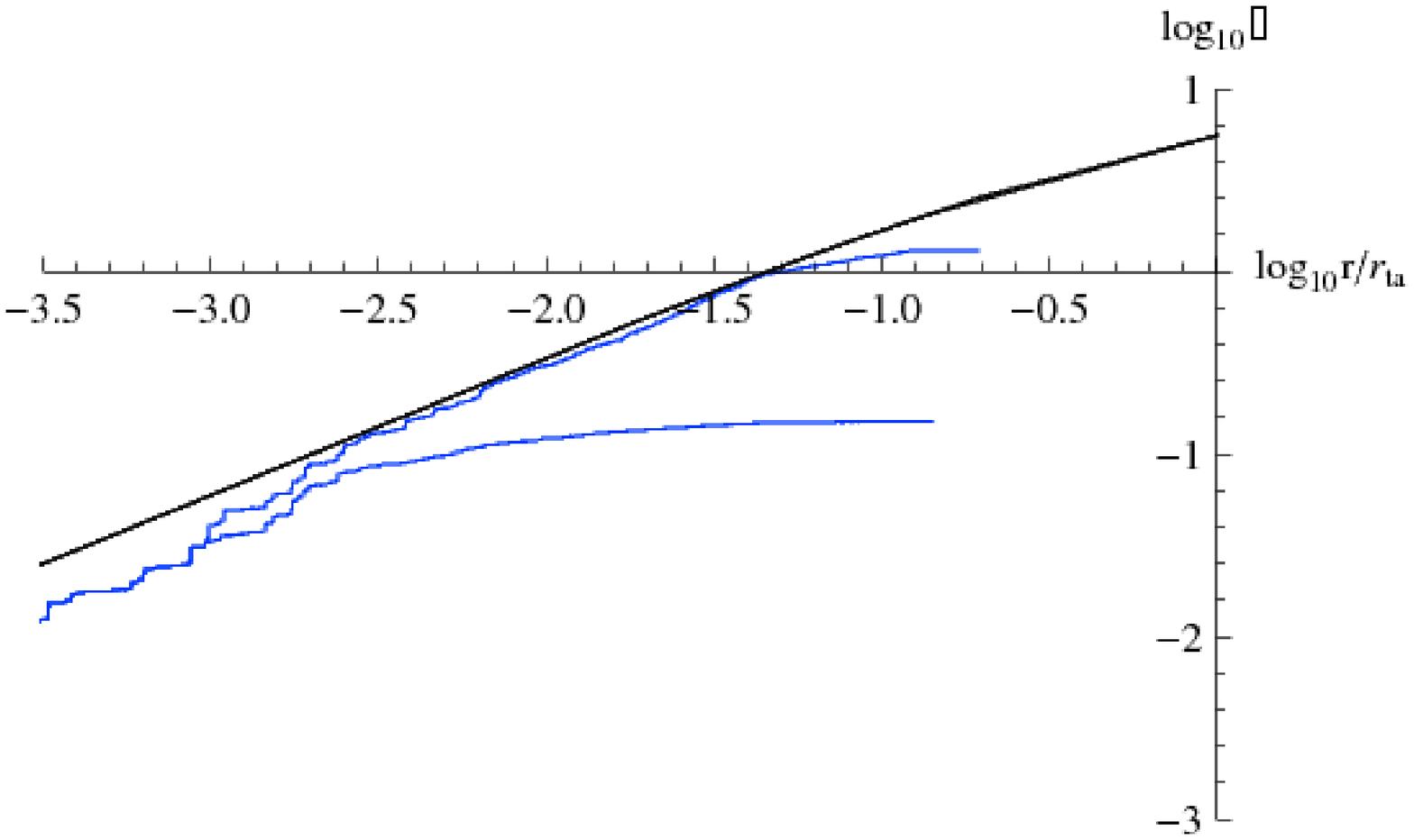}
\caption{\label{figure-final-cdfN-15-0wwo} Clustering
for $G \mu=10^{-15}$ 
by {\it number} of loops for $f$ constant ($\beta=0$),
loop size $l=\alpha/H$, $(\alpha_L,\alpha_U) = (10^{-3},10^{-1})$,
matter era expansion dynamics ($a \propto t^\nu$ with $\nu=2/3$).
The black line is $\log Q_{AB}(<r)$ for cold dark matter, the blue
lines are $\log {\cal Q}$ for loops with (lower) and without
(upper) the gravitational wave recoil (the rocket effect).
}
}

By contrast ${\cal Q}^{(1)}$ weights loops by today's length. This
distribution samples a broader range of times.  The three cumulative
distributions (with and without rocket and cold dark matter) now have
a very different set of relationships. There are two qualitative
observations: recoil makes little apparent difference and the
amplitude of the
cumulative loop distribution lies below the cold dark matter one.

\FIGURE[h]{
      \includegraphics[width=0.8\textwidth]{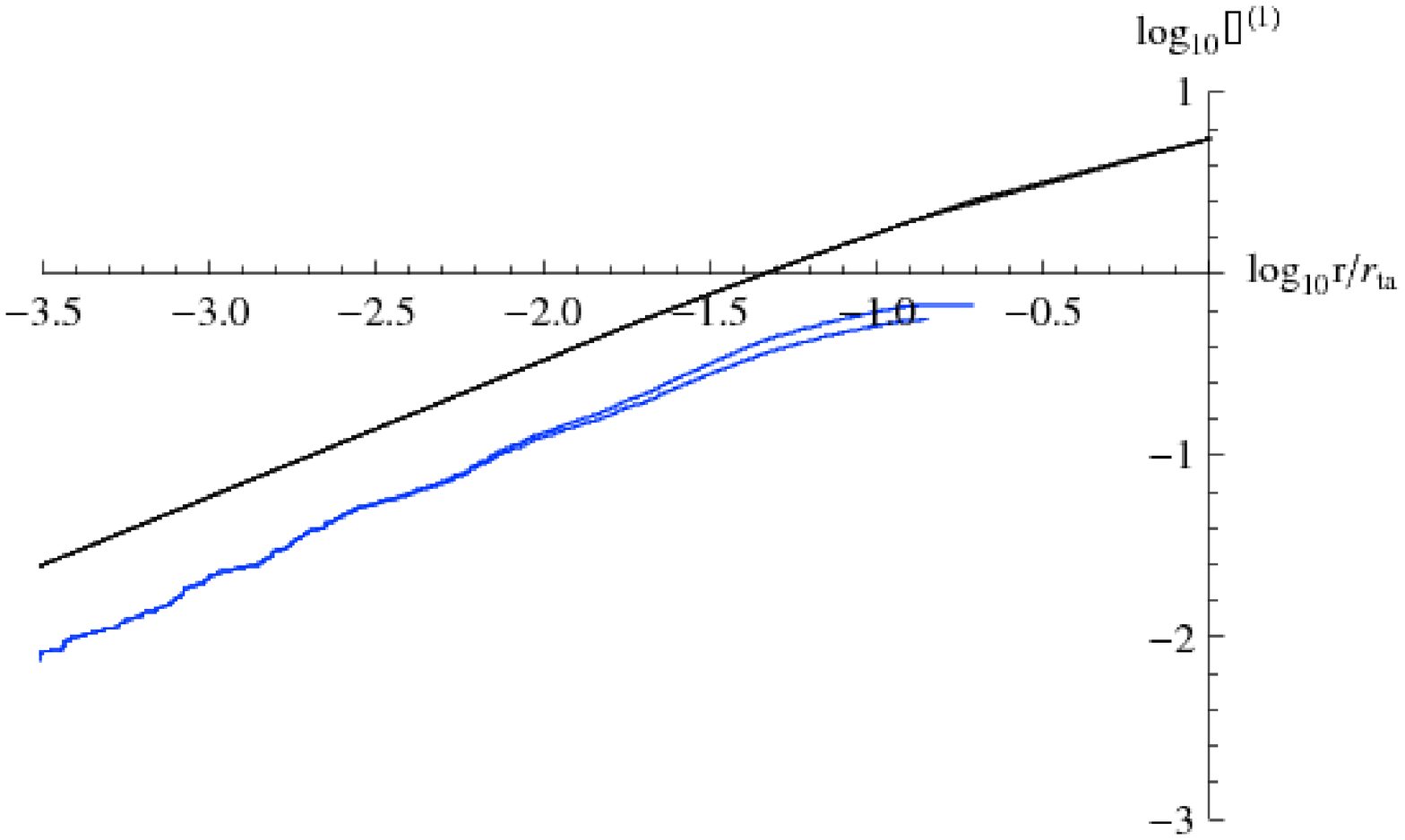}
\caption{\label{figure-final-cdfL-15-0wwo} Clustering
by {\it energy} of loops (same parameters as figure 
\ref{figure-final-cdfN-15-0wwo}).
The black line is $\log Q_{AB}(<r)$ for cold dark matter, the blue
lines are $\log {\cal Q}^{(1)}$ for loops with (lower) and without
(upper) the gravitational wave recoil (the rocket effect).
}
}

Recoil is not dramatic in ${\cal Q}^{(1)}$ because only a small
contribution to total length is made by loops near the end of their
lifetimes.  This can be understood by revisiting figure
\ref{fig-summary-ti-mu} which displays the bounds on formation time
and string tension. At fixed tension the logarithmic interval between
the lifetime (red diagonal) line and the current epoch is proportional
to total loop length created and present in homogeneous space. Only
part binds to the galaxy, the interval between the lifetime
line and the capture time (turquoise horizontal) line.  The rocket
cuts out the space below the acceleration condition (green diagonal).
For small $G\mu$ many decades lie above the acceleration line and
below the capture time so it is difficult to see the effect of the
rocket on the length-weighted loop distribution.  

Insensitivity of ${\cal Q}^{(1)}$ is not a prerequisite for
clustering but an interesting consequence of the weighting
and scale factor.  One expects the rocket to have a more visible influence
on ${\cal Q}^{(1)}$ in the radiation era when the distribution is
$\propto t^{-1/2}$ not $\log t$.

The second observation, that the amplitude of the cumulative
distribution lies below the cold dark matter comparison, is related to
the necessity of slowing down enough for capture. Again, referring to figure
\ref{fig-summary-ti-mu}, at fixed $G \mu$ loops born between the
capture line and the current epoch are present in homogeneous space
but cannot bind to the galaxy today.  The effect lowers
the amplitude relative
to the cold dark matter scenario where there is no
such constraint.  At first glance, the figure would suggest that
the amplitude be diminished by $\sim 2$ (i.e. at $G \mu =
10^{-15}$ about half the decades lie above the capture line and half
below). The figure illustrates $v_i =0.1$ whereas the simulation
velocities are larger (rms dispersions are $0.5-0.9$).  The
excluded region in the figure increases and accounts for
the factor of $3-4$ diminution in amplitude observed in
the simulation.

As the tension increases the natural expectation is that the curves
should fall away from the cold dark matter analog since $H \tau$
diminishes and less damping will be possible. In figure
\ref{fig-summary-ti-mu} this corresponds to trying to move to the
right and the available phase space shrinks.  The cumulative number
$\cal Q$ are shown in figure \ref{figure-final-cdfN-15-10-0} for range
of string tensions $10^{-15}$-$10^{-10}$. 
The lowermost profiles have tensions $G \mu=10^{-10}$
and $10^{-11}$ and fulfill this
expectation.  However, a striking feature is that all curves with $G \mu
\lta 10^{-12}$ are bunched together.  They track the cold dark matter
profile at small radii and are stripped beyond a characteristic
radius. Without the rocket effect the same subset of curves closely
tracks the cold dark matter profile to about $r/r_{ta} \sim 0.1$ (at
which point deciding whether or not a loop is bound is problematic).

Since most loops are created at early times, in the absence of
dynamics most have ages $\sim t_0$. The gravitational wave
acceleration of such loops is $a_R \sim \Gamma_P c/\Gamma_E
t_0$. Setting this equal to $\chi_{crit} | \nabla \phi |$ gives the
characteristic radial scale $ \chi_{crit} v_c^2 \Gamma_E t_0/(c \Gamma_P)
\sim 2.3$ kpc or $\log r/r_{ta} \sim -2.7$, approximately what is
observed in the simulations and independent of $G \mu/c^2$.

\FIGURE[h]{
      \includegraphics[width=0.8\textwidth]{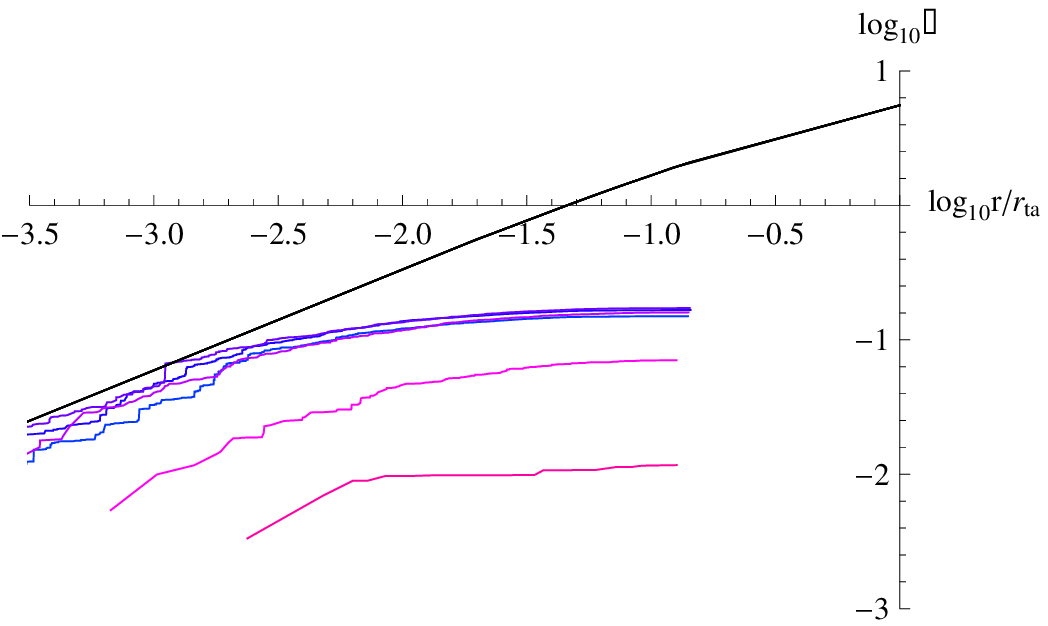}
\caption{\label{figure-final-cdfN-15-10-0}
String tensions $G \mu = 10^{-10}$ to $10^{-15}$ in powers
of ten, bottom to top with all other parameters the same as figure 
\ref{figure-final-cdfN-15-0wwo}).
The black line is $\log Q_{AB}(<r)$ for cold dark matter, the 
colored lines are $\log {\cal Q}$ (all with rocket effect).
For $r/r_{ta} < 10^{-2.5}$
the group with $G \mu <10^{-12}$ are numerically indistinguishable
given the finite Monte-Carlo sample.
}
}

Profiles with loops weighted by length are shown in
\ref{figure-final-cdfL-15-10-0}. Because the loop distribution is
logarithmic, the measures are not dominated by the loops formed at the
earliest times nor is the rocket effect a dominant influence on the
shape. As tension decreases, the cumulative approaches the cold dark
matter limit albeit slowly.

\FIGURE[h]{
      \includegraphics[width=0.8\textwidth]{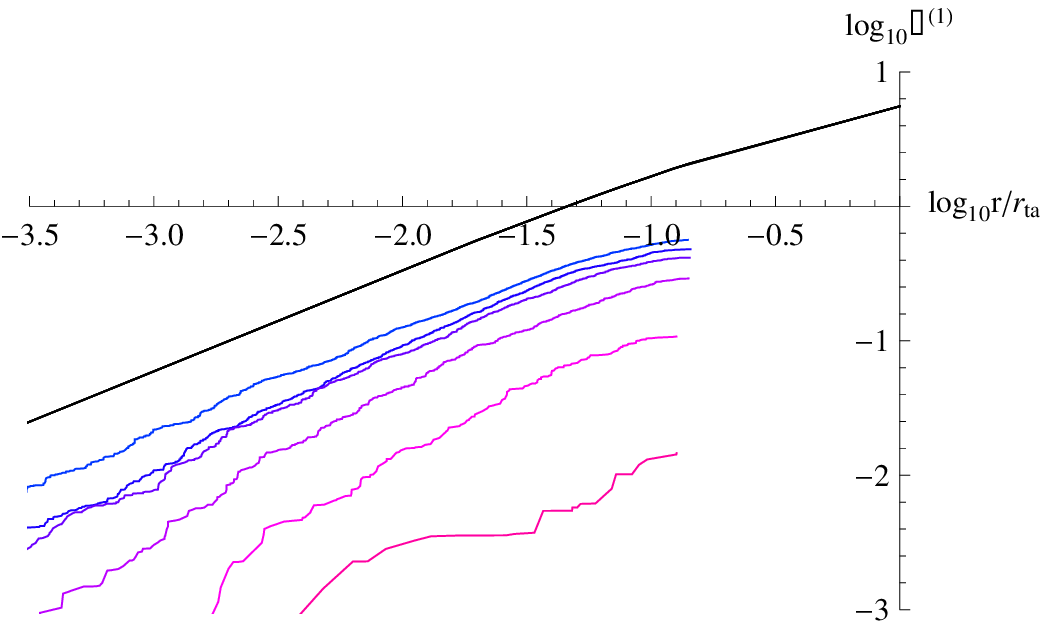}
\caption{\label{figure-final-cdfL-15-10-0} Clustering
by {\it energy} of loops with
string tensions $G \mu = 10^{-10}$ to $10^{-15}$ in powers
of ten, bottom to top (colored lines)
and all other parameters the same as figure 
\ref{figure-final-cdfN-15-0wwo}).
The black line is $\log Q_{AB}(<r)$ for cold dark matter, the colored
lines are $\log {\cal Q}^{(1)}$ for loops (all with rocket effect).
}
}

The average interior density within a radius is displayed in figures
\ref{figure-final-avdenN-15-10-0} (number) and 
\ref{figure-final-avdenL-15-10-0} (length). They illustrate that
the loop clustering follows that of the dark matter.

\FIGURE[h]{
      \includegraphics[width=0.8\textwidth]{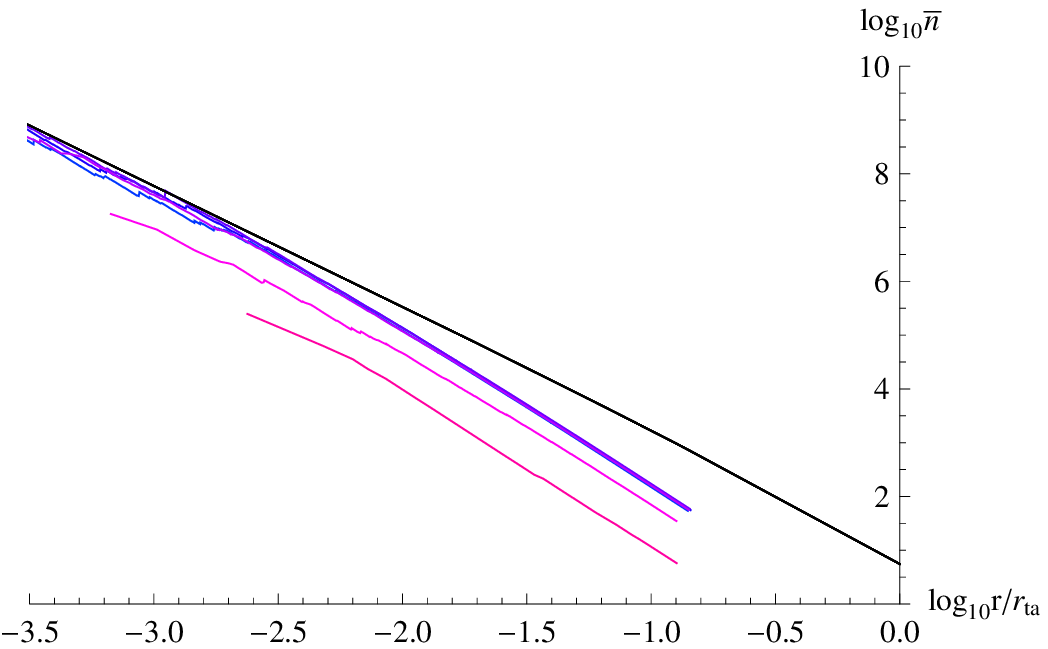}
\caption{\label{figure-final-avdenN-15-10-0} Mean interior {\it
    number} density of loops $\log_{10} {\bar {\cal N}}(<r)$
  (relative to unclustered loop population,
including rocket effect).  String tensions $G \mu = 10^{-10}$ to
  $10^{-15}$ in powers of ten, bottom to top (colored lines) and all
  other parameters the same as figure
  \ref{figure-final-cdfN-15-0wwo}).  The black line is $\log
  Q_{AB}(<r)(r_{ta}/r)^3$ for cold dark matter.
The infall model has turn-around radius $r_{ta} = 1.1$ Mpc.
}
}

\FIGURE[h]{
      \includegraphics[width=0.8\textwidth]{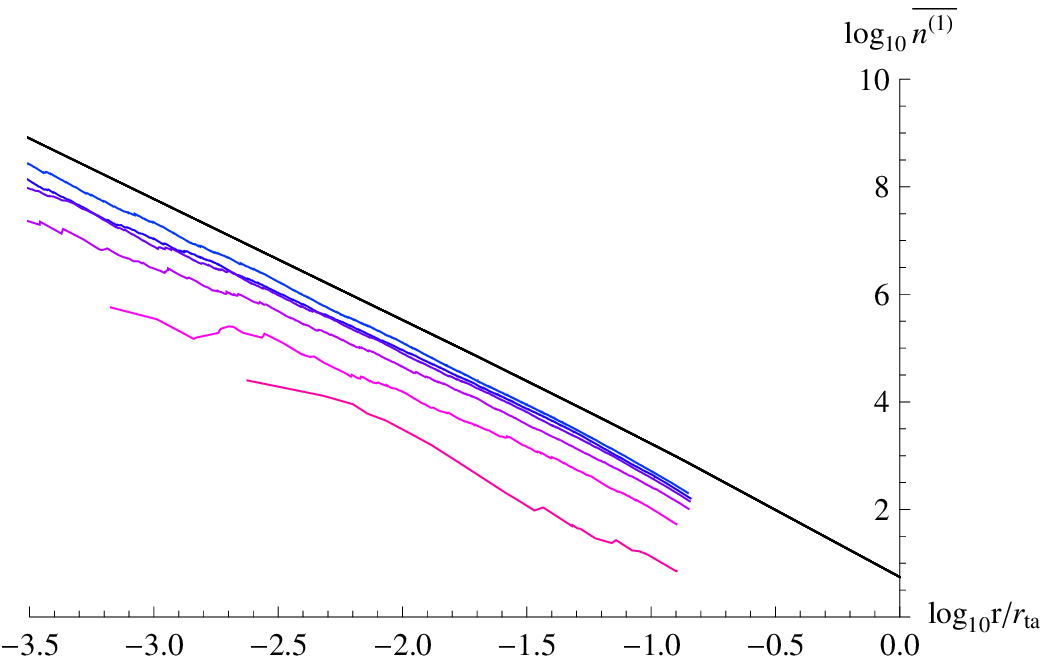}
\caption{\label{figure-final-avdenL-15-10-0} Mean interior
{\it energy} density of loops $\log_{10} {\bar {\cal N}^{(1)}}(<r)$
(relative to unclustered loop population, including rocket effect). 
String tensions $G \mu = 10^{-10}$ to
  $10^{-15}$ in powers of ten, bottom to top (colored lines) and all
  other parameters the same as figure
  \ref{figure-final-cdfN-15-0wwo}).  The black line is $\log
  Q_{AB}(<r)(r_{ta}/r)^3$ for cold dark matter.
The infall model has turn-around radius $r_{ta} = 1.1$ Mpc.
}
}

Figure \ref{figure-final-cdfNandL-15-0-1-1.63} shows that varying $\beta$,
the slope of the loop distribution function, has little effect. The
significance of this observation
is not that the change in $\beta$ is ignorable but that
it is subsumed by scaling with respect to the homogeneous results.

\FIGURE[h]{
      \includegraphics[width=0.8\textwidth]{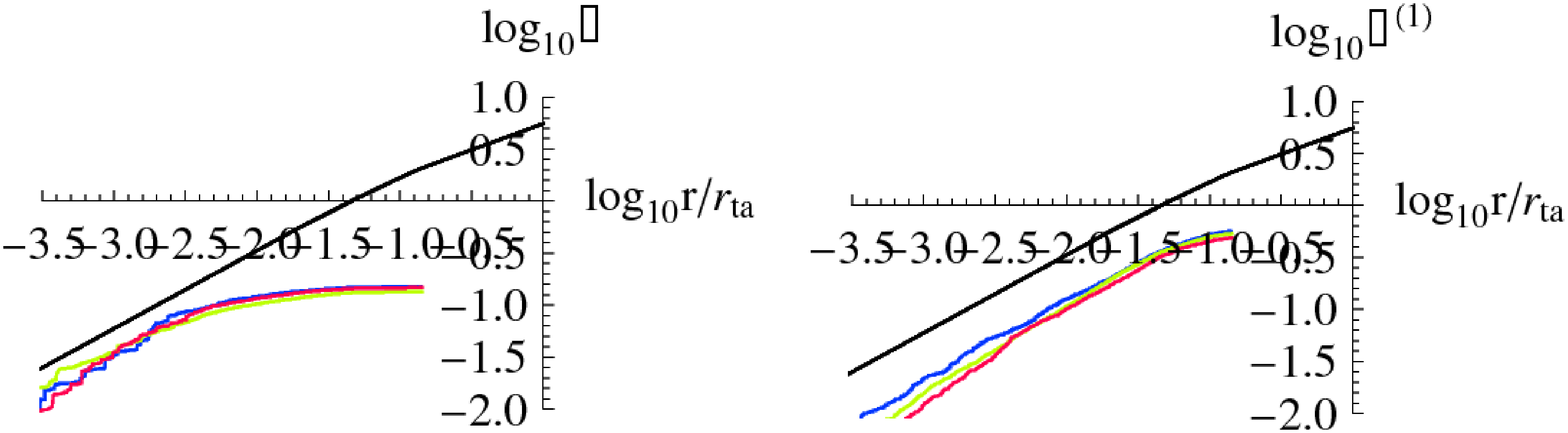}
\caption{\label{figure-final-cdfNandL-15-0-1-1.63} Cumulative
distributions for $\beta$=$0$ (blue), $1$ (green) and $1.63$ (red) for
for $G \mu =10^{-15}$ 
for {\it number} (left) and {\it energy} (right)
of loops. The remaining parameters are the same as
figure \ref{figure-final-cdfN-15-0wwo}.
}
}

All the calculations in this section will be altered by treating
the loop production during the radiation era with the correct
scale factor $a(t) \propto t^{1/2}$. The weighting of the
birth rate in the comoving volume will shift from a
logarithmic distribution to one that $\propto t^{-1/2}$. New results
will be calculated in the future but one can infer that
this new ${\cal Q}^{(1)}$ will have properties
intermediate between ${\cal Q}$ (dominated by early time production)
and ${\cal Q}^{(1)}$ (logarithmic)
calculated thus far.

\clearpage

\subsection{Counterpoint: Small Loops by Cusp Formation}
According to current estimates 80-90\% of the horizon-crossing string
ends up forming small loops. They are unlikely to cluster well because
they are born moving fast and they have intrinsically small
lengths. However, they have a broad range in $x=l/t$ and the
relativistic kinematic effects are non-trivial (time dilation, energy
shifts). So it is of interest to investigate the degree of clustering
within the galaxy. For a simplified, canonical treatment assume 100\%
efficiency ($\delta=1$) for chopping long strings into loops characterized by $f(x)
\propto x^{-\beta}$ over the range $x=(x_{GW},1)$. Here, the power law
index $\beta=3-2 \chi$ and $\chi=0.25$ in the matter era. The loops
are weighted in number and mass to the low cutoff at the gravitational
wave damping scale $x_{GW}=20. \left( G \mu \right)^{1+2 \chi}$. The
velocity dispersion of newly formed loops is given by the
ultra-relativistic theoretical result eq. \ref{eqn-vrms}.

Results for $G \mu = 10^{-15}$ are presented in figure
\ref{figure-finalPR-cdfN-15-wwofnof} for the cumulative number
distribution ${\cal Q}$ and the cold dark matter behavior $Q_{AB}$. It
is apparent that only a very small fraction of the loops generated by
cusp formation are able to live long enough to slow down and bind to
the galaxy. Recall that homogeneous space has ${\cal Q}=1$; the
majority of small living loops in proximity with the galaxy are not
bound to it.  The figure also illustrates the effect of turning off
the rocket effect and starting from a less-relativistic velocity
distribution function. These change ${\cal Q}$ in the manner expected
but still leaves the relative numbers small. A similar picture emerges
from the cumulative length distribution ${\cal Q}^{(1)}$ in figure
\ref{figure-finalPR-cdfL-15-wwofnof}.

\FIGURE[h]{
      \includegraphics[width=0.8\textwidth]{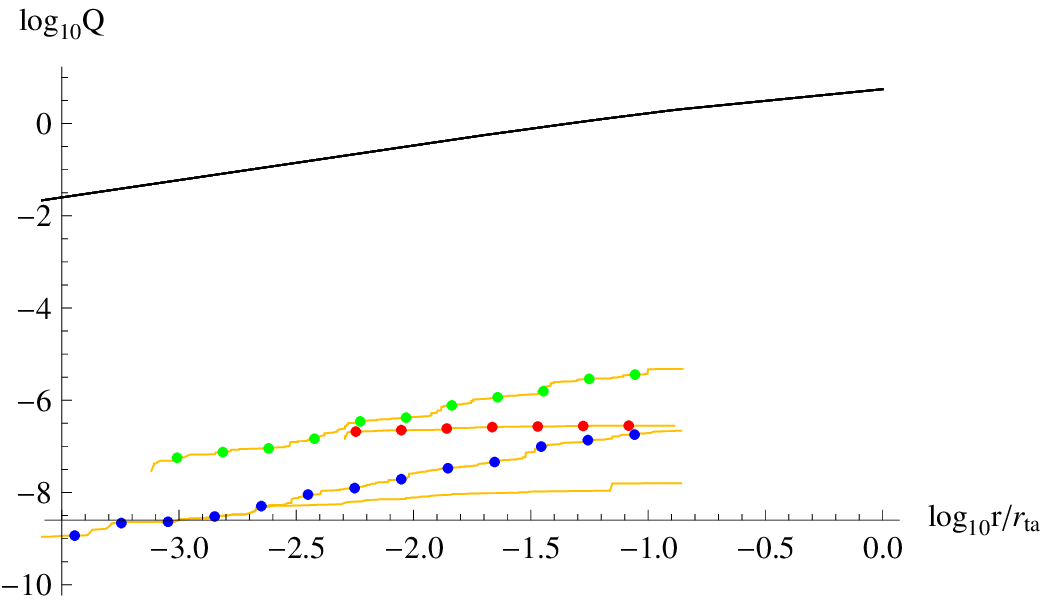}
\caption{\label{figure-finalPR-cdfN-15-wwofnof}The yellow unadorned line is the
cumulative {\it number} distribution of loops ($\log {\cal Q}$)
for $G \mu = 10^{-15}$ in cusp-generated loop formation
during the matter era ($\nu=2/3$, $\chi=0.25$, ${\bar {v_\infty^2}} = 0.35$,
$1/\gamma_s^2 = 10$, $(x_L,x_H) = \left( x_{GW}, 1 \right)$ with
$x_{GW} = 20. \mu^{1+2 \chi}$ and with theoretically determined
ultra-relativistic velocity distribution eq. \ref{eqn-vrms}).
The black line is $\log Q_{AB}(<r)$ for cold dark matter.
The lines with dots represent altering the standard model
assumptions: suppress rocket effect (blue), adopt less relativistic
initial velocity distribution (red), and combined (green).
}
}
\FIGURE[h]{
      \includegraphics[width=0.8\textwidth]{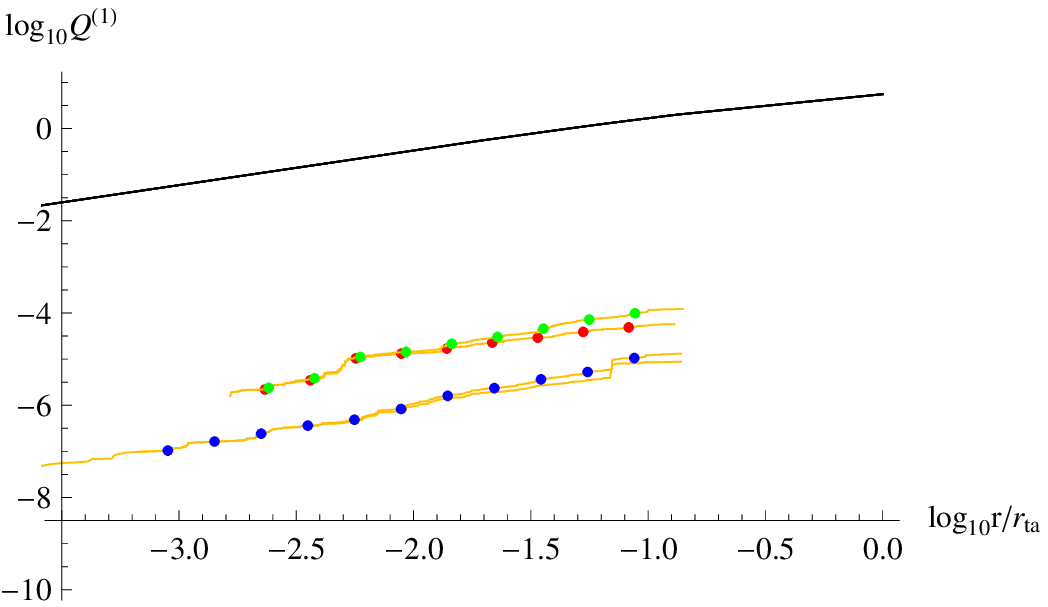}
\caption{\label{figure-finalPR-cdfL-15-wwofnof}
Cumulative {\it energy} distribution of loops ($\log {\cal Q}^{(1)}$);
identical parameters as figure \ref{figure-finalPR-cdfN-15-wwofnof}.
}
}

The average interior number and energy densities are shown in figures
\ref{figure-finalPR-avdenN-15-wwofnof} and 
\ref{figure-finalPR-avdenL-15-wwofnof}. The canonical model reaches
the background number (energy) density at $r/r_{ta} = 10^{-2.9}$ ($10^{-2}$).
The smallest $r/r_{ta}$  
at which the infall model will be a reasonably accurate physical description
of the galaxy is $\sim 10^{-3}$.

\FIGURE[h]{
      \includegraphics[width=0.8\textwidth]{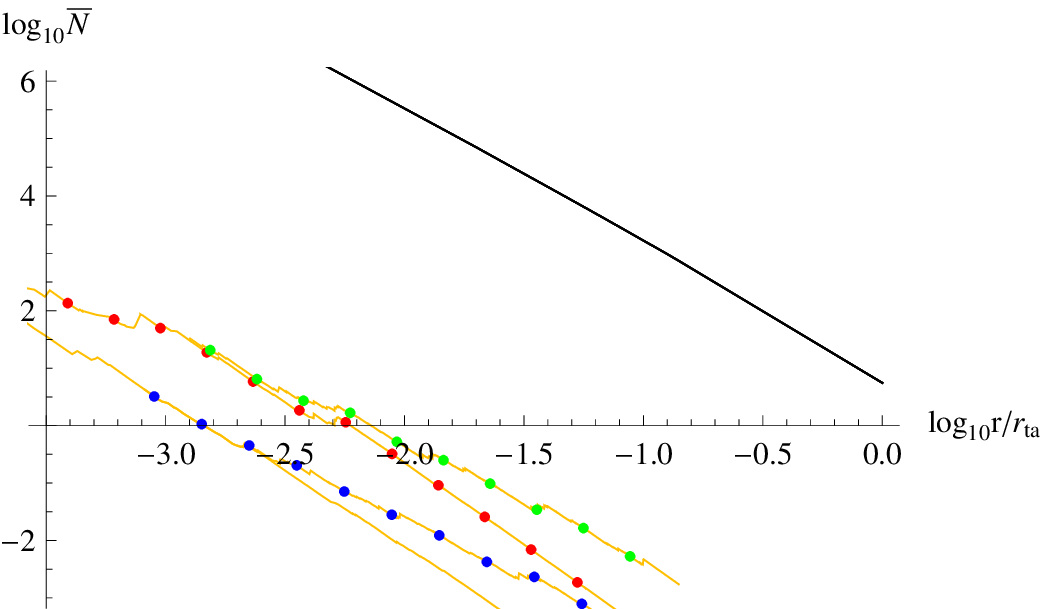}
\caption{\label{figure-finalPR-avdenN-15-wwofnof}
Average interior {\it number} density of loops ($\log {\bar {\cal N}}$);
identical parameters as figure \ref{figure-finalPR-cdfN-15-wwofnof}.
}
}
\FIGURE[h]{
      \includegraphics[width=0.8\textwidth]{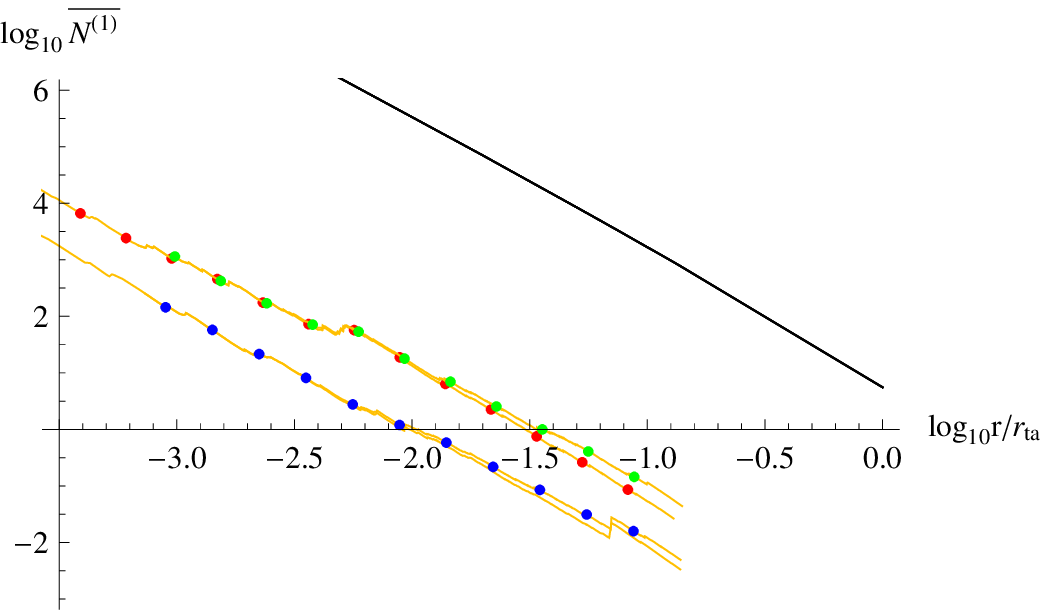}
\caption{\label{figure-finalPR-avdenL-15-wwofnof}
Average interior {\it energy} density of loops ($\log {\bar {\cal N}^{(1)}}$);
identical parameters as figure \ref{figure-finalPR-cdfN-15-wwofnof}.
}
}

A comparison of ${\cal Q}^{(1)}$ for tensions $G
\mu=10^{-15}$-$10^{-12}$ shows that increasing $\mu$ increases the
${\cal Q}^{(1)}$. Essentially, the gravitational wave cutoff $x_{GW}
\propto \mu^{1.5}$ increases and each horizon crossing string makes more
large loops which live longer and move more slowly. The range of
uncertainty is shown by comparing the results for velocities drawn
from the less relativistic Ref. \cite{bennett_two-point_1989}.

\FIGURE[h]{
      \includegraphics[width=0.8\textwidth]{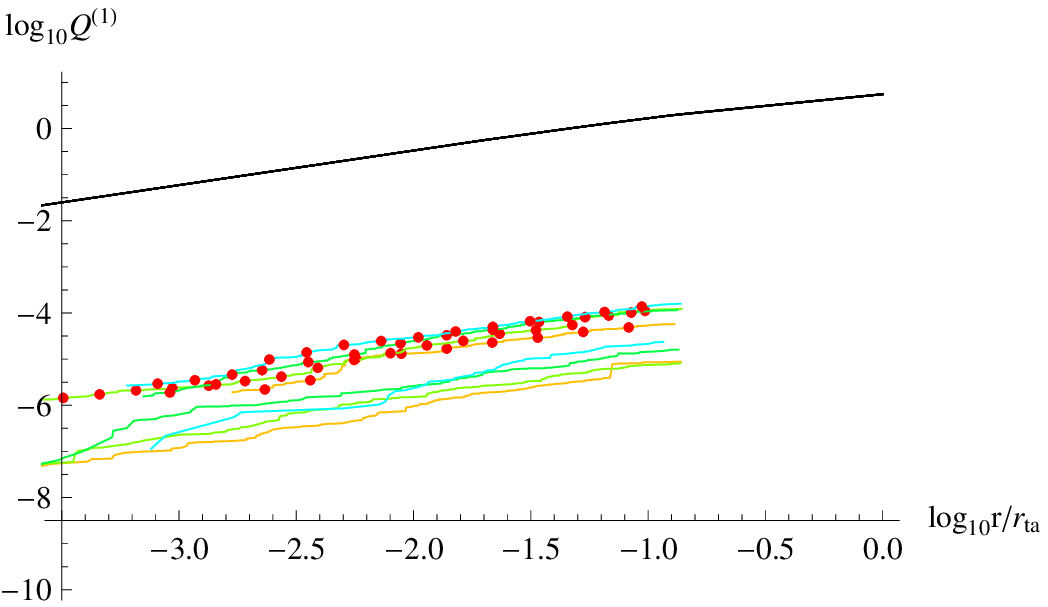}
\caption{\label{figure-finalPR-cdL-15-12-fnof} The cumulative {\it
    energy} distribution of loops ($\log {\cal Q}^{(1)}$) for $G
  \mu = 10^{-12}-10^{-15}$ (yellow to blue) in cusp-generated loop
  formation during the matter era ($\nu=2/3$, $\chi=0.25$, ${\bar
    {v_\infty^2}} = 0.35$, $1/\gamma_s^2 = 10$, $(x_L,x_H) = \left(
  x_{GW}, 1 \right)$ with $x_{GW} = 20.\ \mu^{1+2 \chi}$ and with
  theoretically determined ultra-relativistic velocity distribution
  eq. \ref{eqn-vrms}).  The black line is $\log Q_{AB}(<r)$ for cold
  dark matter.  The lines with red dots show the effect of altering
  the standard model assumptions to adopt less relativistic initial
  velocity distribution. All lines include rocket effect.}  }

\clearpage

\section{Conclusions}

\def\smallmulargealpha{\cite{depies_stochastic_2007,chernoff_cosmic_2007}}

Cosmic superstrings are produced towards the end of brane inflation, a
scenario in modern superstring theory.  As string tension $\mu$
decreases and/or formation size $\alpha$ increases (where $\alpha$ is
roughly the fraction of the horizon), then the timescale for the loop
to evaporate by gravitational emission $\tau \propto \alpha/\mu$
increases. This has important consequences for the abundance of
strings as potentially observable astrophysical
objects\smallmulargealpha.  Although born moving at close to the speed
of light, long-lived loops damp by the expansion of the universe and
fall into potential wells created by cold dark matter and baryons
after equipartition\stringsclusteringalaxy.

This paper explores the dynamics of loops in the vicinity of the
galaxy. The model used to represent the growing galaxy is a
self-similar, radial infall model for the cold dark matter
component. The form for the perturbed FRW metric then provides the
playground for the fully relativistic treatment of string loop
dynamics.

A succession of increasingly complex dynamical investigations
addresses the basic question ``how did the halo get its
loops?'' These include: capture of fast moving objects by the growing
perturbation, recoil by radiated gravitational waves, protection of
tightly bound orbits by adiabatic invariance and a critical transition
from confined to free, liberated motion.

The basic picture that emerges is straightforward.  Loops damp by
cosmological expansion coming nearly to rest and some find themselves
in the vicinity of a growing matter perturbation.  These fall into the
perturbation and, like cold dark matter, acquire radial orbits with
scale comparable to the turn-around radius.  They linger at roughly
fixed physical radius while both the galaxy and the turn-around radius
grow with age. A snapshot of the galaxy would reveal the oldest loops
near the center and the newest ones on the periphery. The oldest loops
are also the smallest loops because the horizon size $\propto t$.

All loops shrink by emission of gravitational wave energy and are
subject to significant recoil because the antenna pattern is highly
anisotropic. The length decreases and the non-gravitational
acceleration increases in tandem. The specific assumptions about
recoil made in this paper are conservative in the sense that they
maximize the importance of the rocket effect and minimize the
possibility of loop clustering.  

For loops of a given size and age the galaxy is stripped from the
outside in. Conversely, these loops are retained latest in the central
parts of the potential well where the binding is greatest. Eventually
as the length vanishes and the acceleration diverges all will be
removed.

The halo is grown from loops of different sizes accreted over a range
of times subject to the dynamical capture process. Since small loops
move faster than large ones at birth two different sized loops born at
the same time will generally not accrete at the same time. The halo is
a mix of loops with a range of sizes and ages at a given apocenter.

Two broad classes for loop formation were characterized.  Loops
directly generated by fragmentation (``large'' loops) and those
mediated by cusp formation (``small'' loops). It was anticipated and
verified that only the large loops cluster about the galaxy to a
significant extent.

The bottom line results for {\it large} loops are presented in figures
\ref{figure-final-avdenN-15-10-0} (density of the number of loops
within the galaxy relative to the unclustered, homogeneous value) and
\ref{figure-final-avdenL-15-10-0} (density of length or energy of
loops). There is a substantial degree of 
enhancement in both number and energy density at a broad
range of galactocentric radii that depends upon string tension.

The bottom line results for {\it small} loops are presented in
figures \ref{figure-finalPR-avdenN-15-wwofnof} (number density) and
\ref{figure-finalPR-avdenL-15-wwofnof} (energy density). Little
enhancement is observed and that only at radii $\lta 1$ kpc where
the radial infall model is not applicable.

Some essential input physics needs improvement:
\begin{itemize}
\item The loop production function needs to be characterized
  accurately, especially the fraction of horizon-crossing strings that
  form large loops ($\delta_{frag}$) and the characteristic scale of
  the large loops ($x_U$).
\item The secular changes to the loop parameters as the loop shrinks
  need to be calculated. This determines the timescale for intrinsic
  variation of the recoil direction, the possibility of loop
  precession (angular momentum changes not along the angular momentum
  direction), and the propensity to convert non-self-intersecting to
  self-intersecting loops.
\end{itemize}
These have direct implications for the number and size of the loops
that are formed, the efficacy of the rocket effect and
the loop lifetime.

From the viewpoint of clustering dynamics these areas need
attention:
\begin{itemize}
\item The string network and background cosmology need to be
  calculated in the context of $\Lambda$CDM cosmology. All
  calculations here use the Einstein-de Sitter non-relativistic matter
  model.
\item The galaxy formation model should be made more realistic and
  consistent with structure formation in $\Lambda$CDM cosmology.
\item Dissipative effects may be important in two contexts: dynamical
  friction can slow down loops before clustering begins and can remove
  energy from loops orbiting within the galaxy.
\item Loop-loop interactions may develop at the galactic center as
  nearly radial orbits converge. If such interactions do occur then
  the small loops formed by intercommutation may acquire relativistic
  velocities and be ejected.
\end{itemize}
The first two items should generally improve the treatment of the loop
distribution and bring it up to date vis-a-vis modern cosmology. The
third and fourth items are unique to the loop dynamics. They may
increase and decrease, respectively, the theoretically calculated halo
densities. A paper on extending the current results to loops
born in the radiation era is being prepared.

From the perspective of detectable astrophysical signatures a
clustered loop halo is a natural source of signals for the following
experiments: direct detection of gravitational wave emission,
pulsar timing variation, and microlensing.  Experiments sensitive to
the local population, especially microlensing of Galactic stars, will
see the most significant impact of clustering.

\acknowledgments 
This work was carried out at Cornell University and on a Sabbatical
visit to KITP at the University of California at Santa Barbara and
Caltech. I appreciate and acknowledge the support provided by KITP and
Caltech.

I thank Sergei Dyda, Vicky Kaspi, Shri Kulkarni, Eran Ofek, Sterl
Phinney, Joe Polchinski, Xavier Siemens, Henry Tye, and Ira Wasserman
for instructive conversations and comments on the manuscript.

\clearpage

\appendix
\section{Equations of Motion in Inhomogeneous FRW}
\label{appendix-equations}
The 4-velocity and direction of the
4-impulse are parameterized in terms of $v$, ${\hat v}$, $n$,
${\hat n}$ according to
\ba
V^\mu & = & \left( \frac{\sqrt{1 + v^2}}{\alpha_\psi},
\frac{v {\hat v}^i}{a\alpha_\Phi} \right) \\
N^\mu & = & \left( \pm \frac{\sqrt{n^2 -1}}{\alpha_\psi},
\frac{n {\hat n}^i}{a\alpha_\Phi} \right) \\
\alpha_{\Phi} & = & \sqrt{1 + 2 \Phi} \\
\alpha_{\psi} & = & \sqrt{1 + 2 \psi}
\ea
where $c=1$ and hat quantities are unit vectors. The sign
of $N^0$ is the sign of ${\hat n} \cdot {\hat v}$.
The equations of motion are
\ba
\frac{\text{dv}}{\text{dt}} & = & \frac{\left(\hat{n}\cdot \hat{v}\right) a_r n \alpha _{\psi }}{\sqrt{1+v^2}}-\frac{\left(\hat{v}\text{$\cdot
\nabla \psi $)}\right. \sqrt{1+v^2}}{a \alpha _{\Phi } \alpha _{\psi }}-\frac{v \dot{a}}{a}-\frac{v \dot{\Phi }}{\alpha _{\Phi }^2} \\
\frac{d\hat{v^i}}{\text{dt}} & = & \frac{a_r
n \alpha _{\psi } (\hat{n^i}-\left(\hat{n}\cdot \hat{v}\right) \hat{v^i})}{v \sqrt{1+v^2}}-\frac{\alpha _{\psi } v (\left(\hat{v}\text{$\cdot \nabla
\Phi $)}\right. \hat{v^i}-\Phi _{,i})}{a \alpha _{\Phi }^3 \sqrt{1+v^2}}+ \\
& & \frac{\sqrt{1+v^2} (\left(\hat{v}\text{$\cdot \nabla \psi $)}\right. \hat{v^i}-\psi
_{,i})}{a \alpha _{\Phi } \alpha _{\psi } v} \nonumber\\
\frac{\text{dn}}{\text{dt}} & = & -\frac{\left(\hat{n}\text{$\cdot \nabla \psi $)}\right. \left(\hat{n}\cdot
\hat{v}\right) n v}{a \alpha _{\Phi } \alpha _{\psi } \sqrt{1+v^2}}+\frac{\left(\hat{n}\cdot \hat{v}\right) a_r \alpha _{\psi } v}{\sqrt{1+v^2}}-\frac{\left(\hat{n}\cdot
\hat{v}\right)^2 n v^2 \dot{a}}{a \left(1+v^2\right)}-\frac{\left(\hat{n}\cdot \hat{v}\right)^2 n v^2 \dot{\Phi }}{\alpha _{\Phi }^2 \left(1+v^2\right)} \\
\frac{d\hat{n^i}}{\text{dt}} & = & -\frac{a_r
\alpha _{\psi } v (\left(\hat{n}\cdot \hat{v}\right) \hat{n^i}-\hat{v^i})}{n \sqrt{1+v^2}}+\frac{\left(\hat{n}\cdot \hat{v}\right) v^2 (\left(\hat{n}\cdot
\hat{v}\right) \hat{n^i}-\hat{v^i}) \dot{a}}{a \left(1+v^2\right)}+ \\
 & &
\frac{v \left(\left(\hat{n}\text{$\cdot \nabla \psi $)}\right. \left(\hat{n}\cdot
\hat{v}\right) \hat{n^i} \alpha _{\Phi }^2-\left(\hat{n}\text{$\cdot \nabla \Phi $)}\right. \alpha _{\psi }^2 \hat{v^i}+ \left(\hat{n}\cdot \hat{v}\right)
\alpha _{\psi }^2 \Phi _{,i}-\left(\hat{n}\cdot \hat{v}\right) \alpha _{\Phi }^2 \psi _{,i}\right)}{a \alpha _{\Phi }^3 \alpha _{\psi } \sqrt{1+v^2}}+ \nonumber\\
& & \frac{\left(\hat{n}\cdot
\hat{v}\right) v^2 (\left(\hat{n}\cdot \hat{v}\right) \hat{n^i}-\hat{v^i}) \dot{\Phi }}{\alpha _{\Phi }^2 \left(1+v^2\right)} \nonumber\\
\frac{\text{dx}^i}{\text{dt}} & = & \frac{\alpha
_{\psi } v \hat{v^i}}{a \alpha _{\Phi } \sqrt{1+v^2}} \\
\frac{\text{da}_r}{\text{dt}} & = & \frac{\Gamma_E a_r{}^2 \alpha _{\psi }}{\Gamma_P \sqrt{1+v^2}}
\ea
Terms have been organized into 3-vector dot products ($(d \cdot e)
\equiv \sum_{i=1,3} d^i e^i$), the explicit index $i$ runs $1-3$ and
time derivatives are indicated ${\dot e} \equiv \frac{de}{dt}$.

\section{Relating Center of Mass and FRW Accelerations}
\label{appendix-acc}

Let $V^\mu$ be the initial 4-velocity of the string center of mass 
in the FRW frame. For clarity, let $V_{(x)}^\alpha$ be
the components in the FRW frame (``x'') and first define
a transformation to an orthonormal frame (``y'') by
\ba
dy^0 & = & \sqrt{1 + 2 \psi} dx^0 \\
dy^i & = & a \sqrt{1 + 2 \Phi} dx^i \\
V_{(y)}^\alpha & = & \frac{ \partial y^\alpha}{\partial x^\beta} V_{(x)}^\beta
\ea
Second, introduce a Lorentz boost to the rest frame (``z''; designated
$\Lambda^\alpha_\beta$) i.e. $\Lambda V_{(y)} \to V_{(z)}=(1,0,0,0)$.
Explicitly,
\ba
\Lambda^\alpha_\beta & = & \frac{ \partial z^\alpha }{\partial y^\beta} \\
& = & \left(
\begin{array}{cc}
V_{(y)}^0 & -V_{(y)}^i \\
-V_{(y)}^i & 1 + \frac{(V_{(y)}^0-1) V_{(y)}^i V_{(y)}^j}{\sum_k (V_{(y)}^k)^2}
\end{array}
\right)
\ea
The impulse in the string center of mass frame is
$a_{(z)}^\alpha=(0,a_{(z)}^i)=a_r (0,n_{(z)}^i)$ where $n_{(z)}^i$ is a
unit-vector. The 4-impulse in the
FRW frame is 
\ba 
a_{(x)}^\alpha & = & \frac{\partial x^\alpha}{\partial y^\beta} \frac{\partial y^\beta}{\partial z^\gamma} a_{(z)}^\gamma \\
             & = & a_r \left(
\begin{array}{c}
a \sqrt{\frac{1 + 2 \Phi}{1 + 2 \psi}} V_{(x)} \cdot n_{(z)} \\
\frac{1}{a \sqrt{1 + 2 \Phi}} \left(
n_{(z)}^i + \left( V_{(x)}^0 \sqrt{1 + 2 \psi} - 1 \right) V_{(x)}^i \frac{ V_{(x)} \cdot n_{(z)} }{ V_{(x)} \cdot V_{(x)} } \right)
\end{array}
\right) \\
& = & a_r N^\alpha.
\ea
where $V_{(x)} \cdot n_{(z)} = \sum_{k=1,3} V_{(x)}^k n_{(z)}^k$ and
$V_{(x)} \cdot V_{(x)} = \sum_{k=1,3} V_{(x)}^k V_{(x)}^k$.  This specifies
the initial $N^\alpha$.

It remains to determine the scalar $a_r$.
In the center of mass of the loop the rates of energy and momentum loss are
\ba
\frac{d l_{(z)}}{dz^0} & = & -\Gamma_E \left( G \mu \right) \\
a_r & = & \Gamma_P \left( G \mu \right) \frac{1}{l_{(z)}}
\ea
Since, $dz^0 = \frac{dx^0}{V^0_{(x)}}$ the length may be straightforwardly
expressed in terms of the FRW time
\be
l_{(z)}=l_{(z),init} - \Gamma_E \left( G \mu \right) \int_{x^0_{init}} \frac{dx^0}{V^0_{(x)}}
\ee
Loops at highly relativistic speeds in the FRW
frame have extended lifetimes because their center of mass clocks
advance more slowly. It is convenient
to express the evolution of $a_r$ directly in the FRW frame using the
global time coordinate
\be
\frac{d}{dx^0} \left( \frac{1}{a_r} \right) = -\frac{\Gamma_E}{\Gamma_P V_{x}^0 }
\ee
After $a_r N^\alpha$ is initially set the entire calculation can be
carried out in the FRW frame.

\section{Variants of Figure 9}
\label{appendix-fig}

Figure \ref{fig-summary-ti-mu-numerical} shows
the bounds on the formation time and string tension for a loop
with initial velocity $v_i=0.1$ captured and retained
at physical radius 30 kpc. It is essentially the same as
\ref{fig-summary-ti-mu} but the black lines are based on
$v$ found by numerical calculations using
eqs. (\ref{eqn-rocketeffect-start})-(\ref{eqn-rocketeffect-end}) for
aligned and anti-aligned rockets rather than by the analytic
expressions in \S \ref{subsec-rocket}. It generally validates
the analytic approximations, however, it shows the existence of
some additional phase space for capture when the
relative directions of impulse and initial velocity vary.
The green line is the retention criterion based
on the critical acceleration at the current epoch.

Figure \ref{fig-summary-ti-mu-numerical2} 
is similar to the one above but applies to
capture at a smaller physical radius 10 kpc. The formation
time must be earlier and the acceleration limit approaches the
loop lifetime limit.

\clearpage

\FIGURE[h]{
\includegraphics[width=0.8\textwidth]{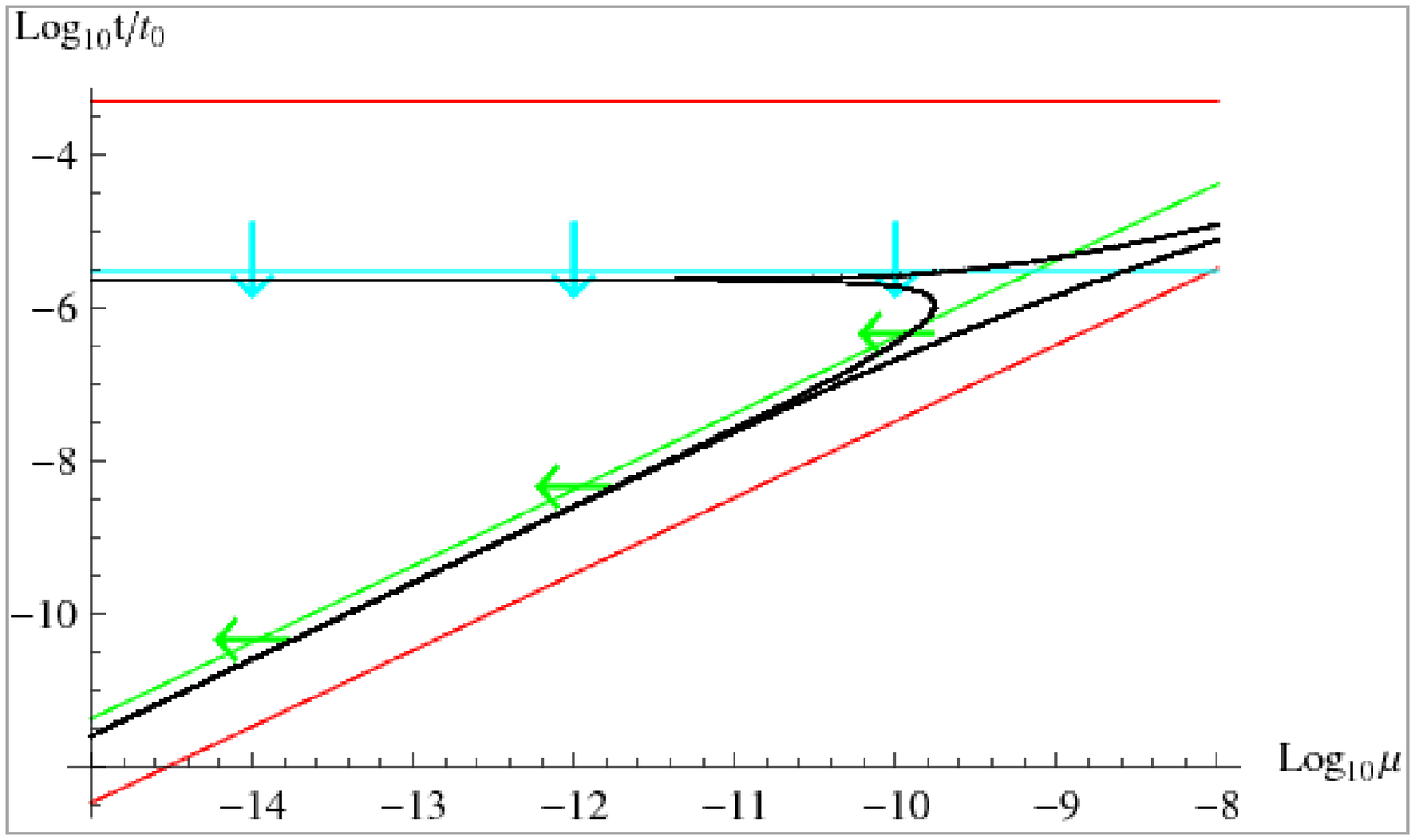}
\caption{\label{fig-summary-ti-mu-numerical} Same as
\ref{fig-summary-ti-mu} except that $v$ is calculated with the full
relativistic equations of motion for aligned and anti-aligned
rockets to determine capture and the numerical
experiments for randomly oriented rockets are omitted.
} }

\FIGURE[h]{
\includegraphics[width=0.8\textwidth]{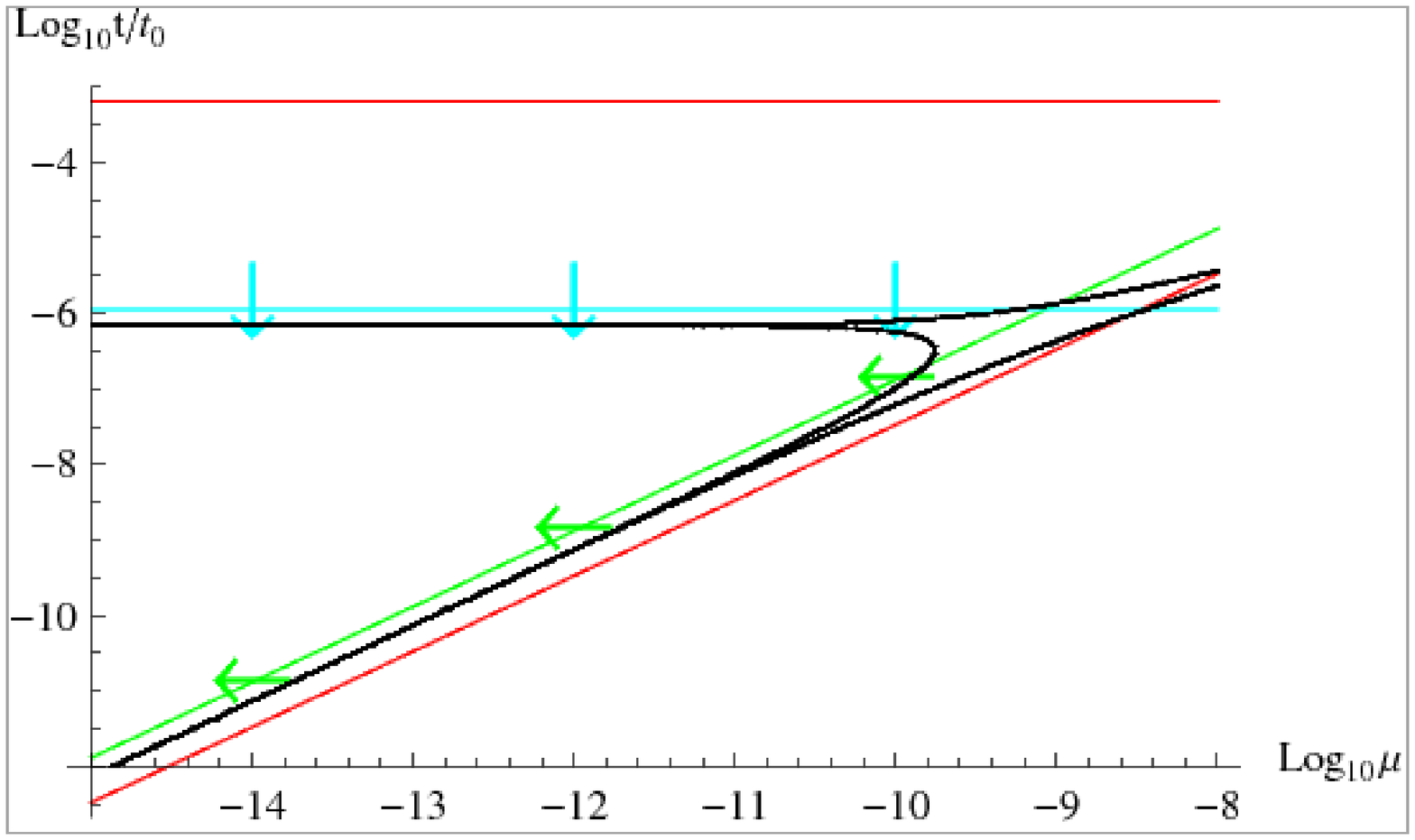}
\caption{\label{fig-summary-ti-mu-numerical2} Same as
\ref{fig-summary-ti-mu-numerical} except for physical
radius 10 kpc. The red lines (damping to the characteristic
rotation velocity and lifetime) are unchanged. All the other lines move. Capture
requires earlier formation and retention allows larger
$G \mu$.
} }

\clearpage

\section{Monte-Carlo Dynamics}
\label{appendix-monte-carlo}

The probability that the particle is bound with position $\vec x$,
momentum $\vec v$ and length $l$ today
\be
\frac{dP_{bnd}}{d{\vec x}d{\vec v}d l} = \frac{1}{\tilde{V}}\int d{\vec x'} d{\vec v'} d{l'} \frac{dP}{d\vec v'} \frac{dP}{dl'} 
\delta^3({\vec x} - {\vec X}) \delta^3({\vec v} - {\vec V}) \delta(l-L) \theta_{AB} \left({\vec x},{\vec v} \right)
\ee
where $\vec X$, $\vec V$ and $L$ are the formal time-dependent solutions for
initial position $\vec x'$, momentum $\vec v'$ and length $l'$ at time $t'$.
The integration is over the distribution of position and momentum
while $l'=l_i$ and $t'=t_i$ are fixed.

There are some significant computational simplifications. First, only
the properties of the bound particles with $l>0$ are of interest.
To avoid sampling initial conditions which imply $\theta_B=0$ and/or
loops that have evaporated use the
following approximations.  In flat FRW space, directly integrate the
motion and length of a particle with given initial momentum from $t_i$
to $t_0$. The particle must live to the current epoch and
be able to enter the spherical volume
defined by today's turn-around radius or there is no chance of it
being accreted.  This limits the minimum length $l'$ as well as
the maximum distance and the range of
angles between $\hat x$ and $\hat v$ that need to be sampled. For
acceptable $l'$ the
Monte-Carlo points are weighted by the ratio of the volume actually
sampled to the total volume.

For efficiency order the kinematic integrals (1) magnitude of $v'$, (2)
magnitude of $x'$, (3) direction $\hat x'$ and (4) direction $\hat
p'$. First, sample $v'$ according to $dP/dv'$ (weight is 1). Second,
calculate $x_{max}$, the maximum initial displacement from the
perturbation center that a particle can have and still reach the
turn-around volume by today, and sample $x'$ from the volume associated
with $x_{max}$ (weight is $4 \pi x_{max}^3/3 \tilde{V}$). Third, sample
$\hat x'$ in the full sphere (weight is 1). Fourth, sample $\hat v'$
from the maximum angular range that allows motion from
$x'$ to reach the turn-around volume by today (weight is ratio of the
angular extent sampled to the full $4 \pi$ extent).

To generate the final distribution, weight each bound particle
according to the sampling above. In the end, ignore
the orbital phase of the particle, i.e. marginalize the two-dimensional
distribution to give the distribution of semi-major axes
\be
\frac{dP_{bnd}}{dr dl} = \int d{\vec x}d{\vec v} \frac{dP_{bnd}}{d{\vec x}d{\vec v}d{l}} \delta(r - R)
\ee
where $R=R({\vec x},{\vec v},t_0)$ is the formal expression for the
semi-major axis in terms of the current phase space coordinates.

The second simplification is to recognize that once a particle is
bound to the perturbation the physical dimensions of its orbit are
fixed if the rocket effect is ignored. Therefore, its unnecessary to
integrate all particles from $t_i$ to $t_0$. Stop once a particle has
passed back and forth through the perturbation center several times
and been captured.  The error (i.e. the characteristic size of the
change in $r$ in the future due to the growth of the perturbation) can
be inferred by the variation in the relative heights of the peaks in
figure \ref{fig-bou}. To the extent that the orbit is exactly radial,
one could extrapolate from the last calculated $r$ to the asymptotic
one. In this work, the integration after $N_c$ bounces ($N_c=3$-$8$)
is halted and the value of $r$ at that time is adopted. A check that
the final results are insensitive to $N_c$ is made.

The third simplification (used in \S \ref{sec-loop-clustering} and
\ref{sec-halo-profile}) is to ignore the rocket effect before capture
and apply the retention and lifetime criteria at the current
epoch. Retention is more stringent than capture so little error is
made. This allows the same simulation to be used for different choices
of $\mu$.

A fourth simplification is to work directly with the cumulative
distribution rather than the differential one. It is ultimately
necessary to record only the weight and the semi-major axis of the
bound particles generated by Monte-Carlo sampling. A sort of the
semi-major axis of the final sample allows simple construction of the
cumulative probability distribution from the weights.

\bibliographystyle{JHEP.bst}
\bibliography{MyLibPlus}

\end{document}